\documentstyle[12pt,epsfig]{article}
\newlength{\dinwidth}
\newlength{\dinmargin}
\newcommand{\ba}{\begin{array}}
\newcommand{\ea}{\end{array}}
\newcommand{\beq}{\begin{equation}}
\newcommand{\eeq}{\end{equation}}
\newcommand{\bea}{\begin{eqnarray}}
\newcommand{\eea}{\end{eqnarray}}
\newcommand{\tr}{\mbox{Tr}}

\newcommand{\bean}{\begin{eqnarray*}}
\newcommand{\eean}{\end{eqnarray*}}

\def\bce{\begin{center}}
\def\ece{\end{center}}

\def\nonu{\nonumber}

\def\pa{\partial}

\def\La{\Lambda}

\def\si{\sigma}

\def\S{{\bf S}}

\def\eps6{{\displaystyle \mathop{\epsilon}^{6}}{}}

\def\nab6{{\displaystyle \mathop{\nabla}^{6}}{}}
\def\to{\rightarrow}

\begin{document}
\thispagestyle{empty} \addtocounter{page}{-1}
\begin{flushright}
%NSF-ITP 99-098\\
%KIAS-P03016 \\
TIT-HEP-497 \\
{\tt hep-th/0306068}\\
revised, Sept., 2003\\
\end{flushright}

\vspace*{1.3cm} \centerline{\Large \bf Phases of} \vskip0.3cm
\centerline{ \Large \bf ${\cal N}=1$  $SO(N_c)$ Gauge Theories with 
Flavors} 
%\vskip0.3cm \centerline{ \Large \bf
% via
%Matrix Model}
%\vskip0.3cm
%\centerline{\Large \bf and Matrix Model }
\vspace*{1.5cm}
\centerline{{\bf Changhyun Ahn}$^1$, {\bf  Bo Feng}$^2$  and 
{\bf Yutaka Ookouchi}$^3$}
\vspace*{1.0cm} \centerline{\it $^1$Department of Physics,
Kyungpook National University, Daegu 702-701, Korea}
\vspace*{0.2cm} \centerline{\it $^2$Institute for Advanced Study,
 Olden Lane, Princeton, NJ 08540, USA}
\vspace*{0.2cm} \centerline{\it $^3$Department of Physics,
 Tokyo Institute of
Technology, Tokyo 152-8511, Japan}

\vspace*{0.8cm} \centerline{\tt
ahn@knu.ac.kr,  \qquad fengb@ias.edu,}
\vspace{0.2cm}
\centerline{\tt
ookouchi@th.phys.titech.ac.jp}

 \vskip2cm

\centerline{\bf Abstract}
\vspace*{0.5cm}

We studied the phase structures 
of ${\cal N}=1$ supersymmetric $SO(N_c)$ gauge theory
with $N_f$ flavors in the vector representation as we deformed
the ${\cal N}=2$ supersymmetric QCD 
by adding the superpotential of
arbitrary polynomial for the adjoint chiral scalar field. 
%These
%phases are characterized by Chebyshev branch or Special branch
%for $SO(N_i)$ factor and baryonic or non-baryonic $r$ branch for
%$U(N_i)$ factor.
Using weak and strong coupling analyses, we determined
the most general factorization forms for various breaking patterns.
%for example, the  three different breaking
%patterns of quartic superpotential: $SO(N_c) \rightarrow SO(N_0) \times  
%U(N_1)$, 
%$SO(N_c) \rightarrow SO(N_c)$ and $SO(N_c) \rightarrow U([N_c/2])$.
%We distinguish between  the case with massive flavors   
%and the case with massless flavors 
%which will behave differently because they are
%charged under the $U(N)$ and $SO(N)$ at IR respectively.
We observed all kinds of smooth transitions for quartic superpotential. 
%like $SO(N_c)\leftrightarrow SO(M_0)\times U(M_1)$,
%$SO(N_0)\times U(N_1)\leftrightarrow SO(M_0)\times U(M_1)$
%and $U([N_c/2])\leftrightarrow SO(M_0)\times U(M_1)$. 
%We
%develop also other techniques which are used for the study
%of phase structures, like the general multiplication map and
%addition map.
% which can relate the $SO(2N_c)$ gauge group
%to $SO(2N_c+1)$ gauge group.
%Furthermore, as a by-product, we discuss the matrix model 
%curve and a generalized Konishi anomaly
%equation for the case with flavors. 

\baselineskip=18pt
\newpage
\renewcommand{\theequation}{\arabic{section}\mbox{.}\arabic{equation}}

%\baselineskip 30pt

%%%%%%%%%%%%%%%%%%%%%%%%%%%%%%%%%%%%%%%%%%%%%%%%%%%%%%%%%%%%%%%%%%%%%%%%%
\section{Introduction and summary}
\setcounter{equation}{0}
%%%%%%%%%%%%%%%%%%%%%%%%%%%%%%%%%%%%%%%%%%%%%%%%%%%%%%%%%%%%%%%%%%%%%%%%%

\indent

The ${\cal N}=1$ supersymmetric gauge theories in four dimensions
have  rich structures and one obtains nonperturbative aspects by studying
the holomorphic effective superpotential which determines 
the quantum moduli space.  
A large class of interesting gauge theories can be obtained from the choice
of geometries in which D5-branes wrap partially over the nontrivial cycles 
\cite{vafa,civ,ckv,cfikv}.
As far as the effective superpotential is concerned, the geometry in which
the four dimensional gauge theories are realized on the worldvolume of 
D5-branes wrapping around $\S^2$ is replaced by  a dual geometry in which
D5-branes are replaced by RR fluxes and the $\S^2$ by $\S^3$.
These RR fluxes provide the effective superpotential which corresponds to
one of four dimensional gauge theories on the D5-branes 
\cite{gukov1,gukov2,tv}.   
This equivalence has been tested for several different models in 
different contexts \cite{eot,fo1,cv,feng1,ookouchi,dot1}. 

A new recipe for the computation of the effective superpotential
was proposed by Dijkgraaf and Vafa \cite{dv1,dv2,dv3} through the  
free energies in a certain matrix model.
This matrix model analysis could be interpreted within purely field 
theoretical point of view without a string theory \cite{cdsw}. 
Based on  the anomalous Ward identity of a generalized Konishi 
anomaly,
a powerful machinery for the nonperturbative aspects was obtained.
Moreover, a new kind of duality where one can transit several vacua 
with different broken gauge groups continuously and holomorphically by 
changing the parameters of the superpotential was found in \cite{csw}.
The extension of \cite{csw} to the ${\cal N}=1$ supersymmetric gauge
theories with the gauge group $SO/USp$ was found in \cite{ao} in which there 
were no flavors and the phase
structures of these theories, a matrix model curve and a generalized Konishi
anomaly equation, were obtained. 
More recently \cite{bfhn}, 
by adding the flavors in the fundamental representation 
to the theory of \cite{csw}, the vacuum structures in 
classically and quantum mechanically  were described and 
an addition map as well as 
multiplication map were developed.

In this paper, 
we study the phase structures 
of ${\cal N}=1$ supersymmetric $SO(N_c)$ gauge theory
with $N_f$ flavors in the vector representation by deforming 
the ${\cal N}=2$ supersymmetric QCD with the superpotential of
arbitrary polynomial for adjoint chiral scalar field, by applying the 
methods in \cite{csw,ao,bfhn}. These kinds of study were initiated
by Cachazo, Seiberg and Witten in \cite{csw}, where a kind of new
duality was found. 
This paper is a generalization of \cite{ao}
to the case with flavors. We found that with flavors the phase
structure is richer and that more interesting dualities show up.     
We list some partial papers \cite{Ferr}-\cite{itoh} 
on the recent works, along the line  of 
\cite{dv1,dv2,dv3}. 

In section 2,
we describe the classical moduli space of
${\cal N}=2$ SQCD deformed to ${\cal N}=1$ theory by adding the 
superpotential $W(\Phi)$ (\ref{treesup1}). The gauge group $SO(N_c)$ will
break to $SO(N_0)\times \prod_{j=1}^n U(N_j)$ with $N_0+\sum_{j=1}^n 
2N_j=N_c$
by choosing the adjoint chiral field $\Phi$ to be the root of
$W'(x)$  or $\pm m_i$, the mass parameters. To have pure Coulomb branch
where no factor $U(N_j)$ is higgsed, we restrict ourselves to
the case $W^{\prime}(\pm m_i)=0$. For each factor with some effective
massless flavors, 
there exists a rich structure
of Higgs branches, characterized by an integer $r_i$, 
that meets the Coulomb
branch along the submanifold called the root of the $r$-th Higgs branch.

In section 3.1, we discuss the quantum moduli space of $SO(N)$ by both the
weak and strong coupling analyses.
When the difference between the roots of $W^{\prime}(x)$ is much 
larger than
${\cal N}=2$ dynamical scale $\La$ (in other words, in the weak coupling
region), the adjoint scalar field $\Phi$ can be integrated out and it gives
a low energy effective ${\cal N}=1$ superpotential. Under this condition,
one can suppress higher order terms except the quadratic piece 
in the superpotential  (\ref{treesup1}). 
Then the effective superpotential 
consists of the classical part plus nonperturbative part. 
%We summarize the
%quantum theory with the mass deformation for $\Phi$ in the weak coupling
%analysis for the regions of various number of flavors. 
There exist 
two groups of solutions, i.e., Chebyshev vacua and Special vacua,
 according to the unbroken flavor symmetry, meson matrix $M$ 
(\ref{classical-part}) and  various phases of 
vacua. 
 
At the scales below
the  ${\cal N}=1$ scale $\La_1$ (when the roots of
$W^{\prime}(x)$ are almost the same), 
the strong coupling analysis is relevant.     
We need to  look for the special points where some number of 
magnetic monopoles (mutually local or non local)  
 become massless, on the 
submanifold of the Coulomb branch of  ${\cal N}=2$ $SO(N)$ which
is not lifted by the ${\cal N}=1$ deformation. The conditions for
these special points are translated into a
particular factorization form
of the corresponding Seiberg-Witten curve. 
We discuss the characters of these curves
at the Chebyshev branch or the Special branch, especially, the 
power of factor $t=x^2$ and the number of single roots.  

In  section 3.2, 
combining the quantum moduli space of $SO(N)$ group with the
quantum moduli space of $U(N)$ group studied in \cite{bfhn},
we give the most general factorization forms of
the curves with the proper number of single roots and double roots
for various symmetry breaking patterns, which generalize the
results in \cite{fo1,cv,feng1,ookouchi}.
 From the point of view
of the geometry, these various breaking patterns correspond to
the various distributions of wrapping D5-branes among the 
roots of $W'(x)$,
i.e.,  
some of the roots do not have wrapping D5-branes. 
We also summarize
the counting of vacua for various phases which will be used heavily
in the examples in sections 4 and 5. For $U(N_i)$ without flavors, 
the number of vacua is given by the Witten index $w(N_i)=N_i$.
For $U(N_i)$ with flavors, it is given by 
(\ref{vacua}) (for more details, see \cite{bfhn}). 
For $SO(N_i)$ without flavors, it is also given by the
Witten index and finally for $SO(N_i)$ with $M_i$ flavors, 
the number of vacua is given by
$(N_i-M_i-2)$ for the Chebyshev vacua 
and one for Special vacua.

In section 4, we analyze the simplest nontrivial examples for quartic 
tree level superpotential with {\it massive} flavors for $SO(4)$, $SO(5)$,
and $SO(6)$  gauge groups
\footnote{ 
Although for the pure case \cite{ao}, the description for $SO(7)$ and
$ SO(8)$ gauge 
theories was given, in this paper we will skip it because on the one hand, 
it will give rise to the complicated solutions and it is hard to analyze
and on the other hand, we expect one cannot see any new interesting 
phenomena.}. 
For these cases, there are two
values which can be chosen by $\Phi$ and we have the following
breaking patterns: $SO(N_c)\to SO(N_0)\times \widehat{U(N_1)}$,
$SO(N_c)\to SO(N_c)$, and $SO(N_c)\to \widehat{U([N_c/2])}$. Depending on
both the $SO(N)$ factor at the Chebyshev branch or the Special branch
and the $\widehat{U(N)}$ factor at the various $r$ baryonic or non-baryonic
branches, the curve takes a different factorization form. By solving
the factorization of the curve, we find various phases predicted by
weak and strong coupling analyses that lead to the matches of
 the counting of vacua.
We find also three interesting smooth transitions among these three
breaking patterns:  $SO(N_c)\leftrightarrow SO(M_0)\times 
\widehat{U(M_1)}$,
$SO(N_0)\times \widehat{U(N_1)} \leftrightarrow SO(M_0)\times 
\widehat{U(M_1)}$,
and $U([N_c/2])\leftrightarrow SO(M_0)\times \widehat{U(M_1)}$. 
Especially
the smooth transitions of  $SO(N_c)\leftrightarrow SO(M_0)\times 
\widehat{U(M_1)}$
and 
$U([N_c/2])\leftrightarrow SO(M_0)\times \widehat{U(M_1)}$ cannot be found
if we do not consider the breaking patterns 
$SO(N_c)\to \widehat{U([N_c/2])}$
and $SO(N_c)\to SO(N_c)$. We also discuss  carefully how the
smooth transition arises when we tune the parameters of
the deformed superpotential.
The phase
structures for various gauge groups have been summarized
in the Tables which can be found in sections 4 and 5.

In section 5,  we move to the {\it massless} flavors
with quartic deformed superpotential for $SO(N_c)$ 
where $N_c=4,5,6$ and $7$. In this case, contrary to section 5,
at the IR limit, the $\widehat{SO(N_0)}$ factor has massless 
flavors instead of a $U(N_i)$ factor having effective massless
flavors as in section 5. Because of this difference, new features
arise. For example, there is Special branch for general $SO(N_c)$ with
massless flavors while for massive case, only $SO(2)$, $SO(3)$, and
$SO(4)$ gauge groups have the Special branch. Also with massless
flavors, there are no smooth transitions as in 
$\widehat{SO(N_c)} \leftrightarrow  
\widehat{SO(N_0)}
\times U(N_1)$ and $\widehat{SO(N_c)} 
\leftrightarrow U([N_c/2])$. Furthermore, for 
the smooth transition 
$\widehat{SO(N_0)} \times U(N_1)\leftrightarrow 
\widehat{SO(M_0)} \times U(M_1)$
in the Special branch,
we have $M_0=(2N_f-N_0+4)$, which is the relationship between
$\widehat{SO(N_0)}$ and $\widehat{SO(M_0)}$ 
to be the {\sl Seiberg dual} pair. 
In fact, the smooth transition in this case may be rooted in
the Seiberg duality. In this section, we  also give the
general discussion for the possible smooth transition
in the case of massless flavors
\footnote{In fact, with a little
modification, the result can also be applied to the massive flavors.}.

In Appendix A,
by using the ${\cal N}=2$ curve together with monopole constraints we are 
interested in and applying the contour integral formula, we derive  
the matrix model curve (\ref{matrixcurveform}) 
for deformed superpotential with an arbitrary degree
and the relationship (\ref{newmatrix}) (or the most general
expression (\ref{general-relationship}))
between the matrix model curve (i.e., the 
single-root part of the factorized Seiberg-Witten curve)
and the deformed superpotential $W'(x)$. The formula  
(\ref{general-relationship}) 
will be used to determine the number of
vacua for fixed tree level superpotential. Using these results,
we have also checked  the generalized Konishi anomaly equation 
for our gauge theory with flavors (\ref{Konequation}).

In Appendix B,
in order to understand the vacua of different theories, 
we discuss both the addition map and multiplication map. 
The addition map with {\it massive} flavors relates
the vacua of $SO(N_c)$ gauge theory with $N_f$ flavors 
in the $r$-th branch 
to those of $SO(N_c^{\prime})$ gauge theory with $N_f^{\prime}$ flavors 
in the $r^{\prime}$-branch where both $N_c$ and $N_c^{\prime}$ are 
even or odd. This phenomenon is the same exactly  as the one in the 
$U(N_c)$ gauge theory with flavors. 
However, the addition map with {\it massless} flavors can relate
$SO(2N_c)$ gauge theory to $SO(2N_c+1)$ gauge theory.  
For the multiplication map, we present the most general form.
Through this multiplication map, one can obtain the unknown 
factorization of the gauge group with higher rank 
from the known factorization of the gauge 
group with lower rank. In this derivation, the properties of Chebyshev 
polynomials are used. In several specific examples, we demonstrate 
the general results. Some interesting points of these 
general multiplication maps are:
(1) We can map the results of $SO(2M)$ to $SO(2N+1)$ and vice
versa; (2) We can map the results without massless flavors to 
the case with some massless flavors. This is a somewhat unusual result.

One can also study 
the phase structures 
of ${\cal N}=1$ supersymmetric $USp(2N_c)$ gauge theory
with $N_f$ flavors in the fundamental representation 
with equal footing described in this paper. However,  
due to the  length of the paper, 
we will publish our results in a separate paper.

Finally let us mention a few interesting directions to be done
in the future:

%\begin{itemize}
%\item (1). 
$\bullet$ 
In this paper, we restrict ourselves to the case
$W'(\pm m)=0$. It would be interesting to study the general
case without this constraint $W'(\pm m)=0$ along the line
of \cite{ca}. For the general case, the parameter space is bigger
and we expect to have more smooth transitions in this bigger space.

$\bullet$ We have discussed some indices which distinguish 
between different phases. 
%However, we do not exhaust the complete list
%of these indices. 
It will be useful to have more concrete
discussion of phases with the complete list of 
indices, along the line of \cite{csw}.

$\bullet$ In this paper, we have discussed the phase structure with
adjoint chiral field $\Phi$. It would be interesting to discuss
the phase structure with other representations, especially the
second rank tensor representation. The descriptions with 
these representations
have attracted some attention recently 
\cite{aus,ald,kra,Aganagic:2003db}.

%\end{itemize}

%%%%%%%%%%%%%%%%%%%%%%%%%%%%%%%%%%%%%%%%%%%%%%%%%%%%%%%%%%%%%%%%%%%%%%%%%%%%%
\section{The classical moduli space of $SO(N_c)$ supersymmetric QCD}
\setcounter{equation}{0}
%%%%%%%%%%%%%%%%%%%%%%%%%%%%%%%%%%%%%%%%%%%%%%%%%%%%%%%%%%%%%%%%%%%%%%%%%%%%%

\indent

In this section, we will discuss the classical moduli space of
$SO(N_c)$ with flavors and the ${\cal N}=1$ deformation (\ref{treesup1})
so that the total superpotential will be (\ref{SO-W-111}). 
Although the classical picture will be modified by quantum corrections,
it does give some useful information about the quantum moduli,
especially in the weak coupling region.
 
%%in certain situation, the classical analysis can be used to find out 
%%information that is also valid in the full quantum theory.
%%The useful information on the number of vacua (???) or
%%symmetry breaking patterns can be obtained from a 
%%semiclassical analysis. 

%%%%%%%%%%%%%%%%%%%%%%%%%%%%%%%%%%%%%%%%%%%%%%%%%%%
%%\subsection{The classical vacua of $SO(N_c)$ supersymmetric QCD}
%%%%%%%%%%%%%%%%%%%%%%%%%%%%%%%%%%%%%%%%%%%%%%%%%%

Let us consider an ${\cal 
N}=1$ supersymmetric $SO(N_c)$
gauge theory with $N_f$ flavors of quark $Q_a^i(i=1,2, \cdots, 2N_f, 
a=1,2, \cdots, N_c)$ in the vector representation. 
The tree level superpotential of the theory 
is obtained from ${\cal N}=2$ SQCD by adding the 
arbitrary polynomial of the adjoint scalar $\Phi_{ab}$
belonging to the ${\cal N}=2$ vector multiplet: 
\bea
\label{SO-W-111}
W_{tree}(\Phi,Q)=
\sqrt{2} Q_a^i \Phi_{ab} Q_b^j J_{ij}+\sqrt{2} m_{ij} Q_a^i Q_a^j
+\sum_{s=1}^{k+1}\frac{g_{2s}}{2s}\mbox{Tr}\Phi^{2s}
%\sum_{s=1}^{k+1} g_{2s} u_{2s}
\eea 
where the first two terms come from the ${\cal N}=2$ theory and the third
term, $W(\Phi)$, can be described  as a small perturbation of
${\cal N}=2$ $SO(N_c)$ gauge theory 
\cite{Konishi-SO,ty,kty,as,aps,hanany,hms,dkp,aot,eot,feng1}
\begin{eqnarray}
W(\Phi)=\sum_{s=1}^{k+1}\frac{g_{2s}}{2s}\mbox{Tr}\Phi^{2s}\equiv
\sum_{s=1}^{k+1} g_{2s} u_{2s}, \qquad u_{2s} \equiv \frac{1}{2s}
\mbox{Tr} \Phi^{2s}
\label{treesup1}
%\nonu
\end{eqnarray}
where $\Phi_{ab}$ is 
an adjoint scalar chiral superfield that plays the role of
a deformation breaking ${\cal N}=2$ supersymmetry to ${\cal N}=1$ 
supersymmetry. Note that in \cite{Konishi-SO} the coefficient for
the mass term of quark in the tree superpotential is different from 
$\sqrt{2}$ used here, but it can be absorbed in the mass matrix and 
the only quadratic mass deformation for $\Phi$ 
with $g_2=\mu$ (other parameters
are vanishing) was considered in 
\cite{Konishi-SO}.
The $J_{ij}$ is
the symplectic metric used to raise and lower the $SO(N_c)$ flavor indices
while $m_{ij}$  is a quark mass matrix,
and they are defined as
\bea
%\label{SO-J-m}
J=\left(\begin{array}{cc} 0 & 1 \\ -1 & 0 \end{array} \right)
\otimes {\bf I}_{N_f\times N_f}, \qquad
m=\left(\begin{array}{cc} 0 & 1 \\ 1 & 0 \end{array} \right)
\otimes \mbox{diag} (m_1, \cdots, m_{N_f})~
\label{jandm}
\eea
where ${\bf I}_{N_f\times N_f}$ is the $N_f\times N_f$ identity matrix.
The ${\cal N}=2$ theory without $W(\Phi)$
is asymptotically free for $N_f<N_c-2$  
(and generates 
${\cal N}=2$ strong-coupling scale $\La$), conformal for
$N_f=N_c-2$ (scale invariant) and infrared (IR) free for $N_f>N_c-2$.

The classical vacuum structure, the zeroes of the scalar potential,
can be obtained by solving 
D-terms and F-terms.
We summarize the following results from
the mechanism of adjoint vevs as follows:

1) The eigenvalues of $\Phi$, $\pm i \phi_i$,  
can only be  the roots of $W'(x)$
or $\pm m_i$; thus, the  gauge group $SO(N_c)$ with $N_f$ flavors 
is broken to 
the product of blocks with or without the effective massless flavors.
Among these blocks at most one
block is $SO(\mu_{n+1})$  and others, $U(\mu_i)$ where $i=1,2, 
\cdots, n$. 
%Each factor which has the effective massless
%flavors can have various Higgs branches.

2)  If $\phi_i=\pm i m_i$ but $W'(\pm m_i)\neq 0$ ($W'(x)$ has
some roots which are not equal to $\pm m_i$), the 
corresponding gauge symmetry of that block will be completely higgsed;
Because of this, we will restrict the following discussions to
the case of $W'(\pm m)= 0$ for which there are richer structures.

3)  For each block, if it has  ``effective'' massless flavors, there
are various Higgs branches classified by some integers $r$.

%%%%%%%%%%%%%%%%%%%%%%%%%%%%%%%%%%%%%%%%%%%%%%%%%%%%%%%
\section{The quantum moduli space  of $SO(N_c)$ supersymmetric QCD}
\setcounter{equation}{0}
%%%%%%%%%%%%%%%%%%%%%%%%%%%%%%%%%%%%%%%%%%%%%%%%%%%%%%%

\indent

From the above section we see that the gauge group $SO(N_c)$ is
broken to the product of various $U(N_i)$ with at most
one $SO(N)$. To get the quantum moduli space we need first
to understand the quantum moduli for various factors. 
The quantum theory of $U(N_i)$ with effective {\it massless}
flavors has been discussed in \cite{aps1,Konishi-SU,bfhn}
and we will not repeat it here. Here we will focus on the
quantum moduli of $SO(N)$ 
\footnote{The number of colors $N$ 
%(which is equal to 
%$N_0$ in previous section and we simply write as $N$ for simplicity) 
is less than or equal to $N_c$ and it 
 can be $2, 4, \cdots, N_c$ for $N_c$ even
and $3, 5, \cdots, N_c$ for $N_c$ odd.} with {\it massless} flavors
discussed in \cite{aps,Konishi-SO}.

%%%%%%%%%%%%%%%%%%%%%%%%%%%%%%%%%%%%%%%%%%%%%
\subsection{The quantum theory of $SO(N)$  supersymmetric QCD}
%%%%%%%%%%%%%%%%%%%%%%%%%%%%%%%%%%%%%%%%%%%%%

\indent

The quantum theory of   $SO(N)$ with mass deformation $\frac{1}{2}
\mu\tr\Phi^2$
has been presented in \cite{aps,Konishi-SO} for both 
weak and strong coupling analyses. 
%We will summarize their work with brief derivations.
Notice that when $N\leq 4$, it is not a simple Lie group, so we will
restrict our discussion 
to the case $N\geq 5$ \footnote{For  $N\geq 5$ the Witten
index $w(N)$ of $SO(N)$ gauge group is $(N-2)$ while for $N\leq 4$
the Witten indices are $1,1,1,2,4$ for $N=0,1,2,3,4$ respectively.
We will give some explanations why the Witten index is different
for $N\leq 4$.}.  

%%%%%%%%%%%%%%%%%%%%%%%%%%%%%%%%
\subsubsection{The weak coupling analysis}
%%%%%%%%%%%%%%%%%%%%%%%%%%%%%%

\indent 

Special corners in parameter space will place ${\cal N}=1$
vacua in regions where the gauge symmetry breaking scale is much larger
than the ${\cal N}=2$ dynamical scale, $\La$. 
Then the gauge coupling is small.
When the  mass $\mu$ for adjoint scalar $\Phi$  is larger
compared with the dynamical scale $\La$, the adjoint scalar
$\Phi$ can be integrated out to lead to a low energy effective
${\cal N}=1$ superpotential from which one can recover the 
original theory as in \cite{is}.
The weak coupling analysis is valid when the difference between the
roots of the polynomial $W'(x)$ are much bigger than $\La$.
Under this condition we can only consider the quadratic part 
of the effective superpotential
since the higher powers of it will be suppressed by $\mu$.
Then the relevant superpotential is 
\bea
W_{tree}(\Phi,Q)=
\sqrt{2} Q_a^i \Phi_{ab} Q_b^j J_{ij}+
\frac{\mu}{2}\mbox{Tr}\Phi^{2}.
\nonu
\eea
Integrating out $\Phi$ will give 
\bea  \label{classical-part}
W_{tree}(\Phi,Q)=
-\frac{1}{2\mu} \mbox{Tr} \left(M J M J \right), 
\qquad M^{ij} = Q^i \cdot Q^j.
\eea
Then the effective superpotential will consist of 
the above classical part (\ref{classical-part}) 
plus nonperturbative effects. 
To find   ${\cal N}=1$ vacua, the effective superpotential
should be minimized.  We can study the 
vacuum structure, the number of vacua, and  global symmetry
breakings according to the range of the number of flavors $N_f$.

We summarize two groups of solutions as follows: 

1) The first group has 
non-degenerated meson matrix $M$. They are all in a confining phase
with unbroken flavor symmetry $U(N_f)$ and the number of vacua
is given by $(N-N_f-2)$. As we will show,
they are given by {\it Chebyshev point} in the corresponding 
Seiberg-Witten(SW)-curve.
Its position in the 
${\cal N}=2$ moduli space is located by the roots of 
the Chebyshev polynomial; the light degrees of freedom are mutually 
nonlocal and the theory flows to an 
interacting ${\cal N}=2$ superconformal theory.  
Since the superconformal theory is  nontrivial, 
one does not have a local effective Lagrangian description 
for those theories. The symmetry breaking pattern is obtained 
by the analysis at large $\mu >> \La $ where there exists an 
effective description of the theory by integrating out the 
adjoint scalar $\Phi$. The dynamical condensation of mesons
breaks the $USp(2N_f)$ flavor symmetry to $U(N_f)$.

2) The second group exists
only when $2N_f\geq N-4$ \footnote{From this condition we see
that when $N_f=0$, it is satisfied only when $N\leq 4$. This
explains why the Witten index of $SO(N)$ without flavors
is different for $N\leq 4$ because there are extra contributions
from this second group.
These special cases are also related to the fact that
there is a factor $x^4$ (or $x^2$)
in the Seiberg-Witten curve as we will see in the later examples.}
 with unbroken flavor symmetry $USp(2N_f)$.
Unlike the first group, the vacua can be in any of Higgs, Coulomb, or
confining phases. They are given by the Special point in the SW-curve.
In the {\it Special point} observed in
\cite{aps}, the gauge symmetry is enhanced to IR free $
SO(\widetilde{N})=SO(2N_f-N+4)$ which is the Seiberg dual 
gauge group \cite{is}. 
The full $USp(2N_f)$ 
global symmetry remains unchanged since there are no meson 
condensates and no dynamical symmetry breaking occurs. 
These vacua are in the non-Abelian free magnetic phase.

%%%%%%%%%%%%%%%%%%%%%%%%%%%%%%%%%%%%%%%%%%%%
\subsubsection{The strong coupling analysis}
%%%%%%%%%%%%%%%%%%%%%%%%%%%%%%%%%%%%%%%%%%%

\indent

When $W(\Phi)$ is very small compared with  dynamical
scale $\La$, the ${\cal N}=1$ quantum theory 
can be considered as a small perturbation of a strongly coupled 
${\cal N}=2$ gauge theory without $W(\Phi)$. In this region
of parameters, we can use the Seiberg-Witten curve  in which
all the nontrivial dynamicses are  already encoded. 
 The ${\cal N}=2$ 
Seiberg-Witten curve derivation in the
context of the matrix model was found in \cite{ahnnam1,ow}.  
Turning on the perturbed superpotential lifts most points 
on the Coulomb branch except some points (the Higgs  
roots) where a certain number of mutually local or nonlocal monopoles 
becomes massless.
  
The strong coupling analysis has been done in \cite{aps,Konishi-SO}.
Let us recall that the curve of $SO(N)$ is given by
\footnote{Notice that
there is a factor $(t-m_k^2)$ instead of $(t+m_k^2)$. 
The reason is as follows.
The $SU(N)$ case has a factor 
$(x+m_k)$ for $\mbox{det} (x+m)$ in the second term
for diagonalized mass matrix $m$. 
For $USp(2N)$ group, the 
masses are given by $i \si_2 \otimes \mbox{diag}(m_1, m_2, \cdots, m_{N_f})$ 
so $\mbox{det} (x+m)$ contains a factor $(x^2+m_k^2)$. For $SO(N)$
case, the  masses are given by (\ref{jandm}) 
so $\mbox{det} (x+m)$ has a factor $(x^2-m_k^2)$.}
\bea
%\label{SO-SW-gen}
y^2=  \prod_{j=1}^{[N/2]} (t-\phi_j^2)^2- 4 \Lambda^{2(N-2-N_f)}
t^{1+\epsilon} \prod_{k=1}^{N_f} (t-m_k^2)
\nonu
\eea
where $t=x^2$ and $\epsilon=0$ for $N$ odd and $\epsilon=1$ for
$N$ even \cite{soevencurve,sooddcurve}. 
For our massless case, it is reduced to
\bea
y^2 &= &  \prod_{j=1}^{N/2} (t-\phi_j^2)^2-4 \Lambda^{2(N-2-N_f)}
t^{N_f+2}, \qquad \mbox{for}  \;\;\; N \;\;\; \mbox{even}, 
%\label{SO-SW-massless-even} 
\nonu \\
y^2 &= &  \prod_{j=1}^{(N-1)/2} (t-\phi_j^2)^2-4 \Lambda^{2(N-2-N_f)}
t^{N_f+1}, \qquad \mbox{for} \;\;\;  N \;\;\; \mbox{odd}. 
%\label{SO-SW-massless-odd}
\nonu \eea
By setting $r$ $\phi_j$'s to vanish,  the curve is
\bea
y^2= t^{2r} \left[\prod_{j=1}^{([N/2]-r}(t-\phi_j^2)^2
- 4 \Lambda^{2(N-2-N_f)} t^{1+\epsilon+N_f-2r} \right]
\label{curver}
\eea 
which has a factor $t$ with some power.
The results in  \cite{aps,Konishi-SO} are that there
are unlifted vacua under the mass deformation $\frac{1}{2} \mu\tr \Phi^2$
only when the power of factor $t$ in the curve $y^2$ is some
particular numbers, i.e., 
\bea
\widetilde{N}_c=(2N_f-N+4)/(2N_f-N+3)
\nonu
\eea 
for $N$ is even/odd 
or 
\bea
(N_f+3)/(N_f+1)/(N_f+2)/(N_f+2)
\nonu 
\eea
for $(N,N_f)=(e,e)/(e,o)/(o,e)/(o,o)$  
where we denote $e$ by an even  number and
$o$ by an odd number. The first case  \cite{aps,Konishi-SO}
is the character of the Special branch which in some sense
corresponds to the baryonic branch of $U(N)$ theory. The
second case  indicates the Chebyshev branch where the
low energy effective ${\cal N}=2$  theory is a nontrivial
conformal theory.

The reason for the above conclusion is the following. For general $r$,
it corresponds to the trivial superconformal theory classified as
 class 1 in \cite{Hori}. To have the vacua, we must have enough
massless monopoles which can happen if and only if 
$r=\widetilde{N}_c=2N_f-N+4$ when $N$ is even. 
It is easy to show that there are
%$(2N_f-N+4)$ 
$(N-N_f-2)$
mutual local monopoles by noticing that from (\ref{curver})
(here we use the example where $N$  is even)
\bea
y^2 &= &  t^{2N_f-N+4} \left( P^2_{N-N_f-2}(t)-4\Lambda^{2(N-2-N_f)}
t^{N-2-N_f} \right) \nonu \\
 &= &  t^{2N_f-N+4} \left( t^{N-2-N_f}-\Lambda^{2(N-2-N_f)} 
\right)^2
\nonu
\eea
where to have a perfect square form in the bracket
implying the maximal degeneracy of the Riemann surface, 
we have chosen $P_{N-N_f-2}(t)=t^{N-2-N_f}+
\Lambda^{2(N-2-N_f)}$ \cite{aps}. This particular
case is called the ``Special branch'' where the
curve $y^2$ is a square form, which is a typical
character of this branch.  
The Special branch, in fact,  gives the second
group of the solution analyzed in the weak coupling region.

Besides the trivial superconformal fixed point, we can obtain a
non-trivial superconformal fixed point in IR by 
having a proper power of $t$ in the curve,
 where mutually non-local monopoles
are massless, and call it the ``Chebyshev branch''. 
To see these nontrivial fixed points, 
let us assume that  both $N_f$ and $N$ are even. The curve
is given by $y^2= t^{N_f+2}\left( P^2_{(N-N_f-2)/2}(t)-4
\Lambda^{2(N-N_f-2)} \right)$. If we take
the characteristic function as 
\bea
P_{(N-N_f-2)/2}(t)=2 (\eta \Lambda)^{(N-N_f-2)}
{\cal T}_{(N-N_f-2)} \left({\sqrt{t}\over 2\eta \Lambda} \right)
\nonu
\eea
with $\eta^{2(N-N_f-2)}=1$, we have the curve
\footnote{
Here we have used a useful relation between the Chebyshev polynomials
${\cal T}_K^2(x)-1 = (x^2-1) 
{\cal U}_{K-1}^2(x)$. They are defined as ${\cal T}_K(x)=\cos(Kx)$ and
${\cal U}_{K-1}(x)={1\over K}{\partial{\cal T}_K(x) \over \partial
x}$. }
\bea
y^2 & = & t^{N_f+2} 4 (\eta \Lambda)^{2(N-N_f-2)} \left[
{\cal T}^2_{(N-N_f-2)} \left({\sqrt{t}\over 2\eta \Lambda} \right)-1 
\right] \nonu 
\\
& = &t^{N_f+2} 4 (\eta \Lambda)^{2(N-N_f-2)} 
\left[ \left({\sqrt{t}\over 2\eta \Lambda} \right)^2-1 \right]
{\cal U}^2_{(N-N_f-3)} \left({\sqrt{t}\over 2\eta \Lambda} \right).
\nonu
\eea
Notice that although $\eta^{2(N-N_f-2)}=1$, the characteristic function
$P_{(N-N_f-2)/2}(t)$
is a function of $t$ and $\eta^2$, so we will have only
the $(N-N_f-2)$ solution  predicted from the weak coupling
analysis. Notice also that 
except a factor $t^{N_f+2}$, there are  single roots 
at $t=0$ and $t=4 (\eta \Lambda)^2$ from the factor 
$\left[ \left({\sqrt{t}\over 2\eta \Lambda} \right)^2-1 \right]
{\cal U}^2_{(N-N_f-3)} \left({\sqrt{t}\over 2\eta \Lambda} \right)
$, so finally we have a factor $t^{N_f+3}$ for this case 
\footnote{The second kind of Chebyshev polynomial ${\cal U_{K}}(x)$ 
contains a factor $x$ for odd degree $K$. }. 
Notice that $(N_f+3)$ is an odd number which will be the
property of the Chebyshev branch in the following discussions.
 Similar arguments can be made for
other three cases. The Chebyshev branch discussed here will
give the first group of the solutions in the weak coupling region.

%%%%%%%%%%%%%%%%%%%%%%%%%%%%%%%%%%%%%%%%%%%%%%%%%%%%%%%%%%%%
\subsection{The factorized form of  a hyperelliptic curve}
%%%%%%%%%%%%%%%%%%%%%%%%%%%%%%%%%%%%%%%%%%%%%%%%%%%%%%%%%%%%%

\indent

Now combining the proper factorization form of the $SO(N)$ curve 
with the massless flavors described in the 
previous subsection with the  proper factorization form  of 
the $U(N_i)$ curve with the flavors in  \cite{bfhn}, we can describe
the general curve form. The basic idea is that first we 
need to have the proper prefactor (like $t^p$ for $SO(N)$ part and
$(t-m^2)^{2r}$ for the $U(N_i)$ part at the $r$-th branch. 
More details about the power $p$ will be presented in a later section.).
After factorizing out these prefactors, we require the
remaining curve to have the proper number of double roots
and single roots, which will fix the form of factorization 
eventually.

The key part of the above procedure is the number of single
roots. The basic rule is the following. For every $U(N_i)$ 
factor in the non-baryonic branch, we have {\it two} single roots
while there is {\it no} single root in the baryonic branch. 
For the possible $SO(N)$ factor, if it is at
the Chebyshev branch, we have {\it two} single roots where one
of them is at the origin; i.e., we have a factor $t$ for that single
root. However, if it is at the Special branch, there is
{\it no} single root for the block  $SO(N)$. Adding
the single roots together for all blocks, we obtain the
final number of the single roots. If there are $2n$
single roots, we will have a factor $F_{2n}(t)$ in
the curve. Furthermore, as we will discuss soon,
the polynomial $F_{2n}(t)$ has some relationship
with the deformed superpotential $W^{\prime}(x)$. One key point
of all the above discussions is that the Special branch
(or the Baryonic branch) will have {\it one} more double root
than the Chebyshev branch (or the Non-baryonic branch)
in the factorization form. This fact will be used repeatedly 
in the following study.

To demonstrate our idea, we consider the following examples
in which $SO(2N_c)$ group is broken into the following three cases:
\bea
SO(2N_c)\to SO(2N_0)\times U(N_1), \qquad SO(2N_c)\to U(N_c), 
\qquad SO(2N_c)\to SO(2N_c),
\nonu
\eea 
where the generalization to multiple blocks
will be trivial and straightforward
\footnote{For the breaking pattern  
$SO(2N_c)\to U(N_c)$, the eigenvalues of
$\Phi$ are the same and are nonzero while for $SO(2N_c)\to SO(2N_c)$,
all of those are vanishing.}. 
Similar analysis 
corresponding to $SO(2N_c+1)$ can be done with 
no difficulty. 

For the broken pattern 
$SO(2N_c)\to SO(2N_0)\times U(N_1)$, by counting the number of the single
roots, we derive the following four possible curves
\bea 
\label{example-1}
y^2 & = & t F_3(t) H^2_{N_c-2}(t), \\
y^2 & = &  F_2(t)  H^2_{N_c-1}(t), \label{example-2} \\
y^2 & = &  t F_1(t) H^2_{N_c-1}(t), \label{example-3} \\
y^2 & = & H^2_{N_c}(t). \label{example-4}
\eea 
Curve (\ref{example-1}) is  
for $SO(2N_0)$ at the Chebyshev branch and $U(N_1)$ at the
non-baryonic branch. The reason is that first we need have
four single roots. 
Secondly, because  $SO(2N_0)$ is at
the Chebyshev branch, one of the four single roots must be
at the origin $t=0$ and finally $F_4(t)=t F_3(t)$. 
Curve (\ref{example-2}) is  
for $SO(2N_0)$ at the Special branch and $U(N_1)$ at the
non-baryonic branch. Factor $F_2(t)$ with two single
roots will record the information of $U(1)\subset U(N_1)$.
Curve (\ref{example-3}) is 
for $SO(2N_0)$ at the Chebyshev branch and $U(N_1)$ at the
baryonic branch. Since $SO(2N_0)$ is at the Chebyshev branch 
we should have $F_2(t)=tF_1(t)$. Finally
curve (\ref{example-4}) is
for $SO(2N_0)$ at the  Special branch and $U(N_1)$ at the
baryonic branch, where no single root is required.
 Notice that the function $H_{p}(t)$ will have  a proper
number of $(t-m^2)$ or $t$ to count the prefactor
for various branches.

For the broken pattern $SO(2N_c)\to U(N_c)$, we have
\bea 
\label{example-5}
y^2 & = &  F_2(t)  H^2_{N_c-1}(t),  \\
y^2 & = & H^2_{N_c}(t), \label{example-6}
\eea
where curve (\ref{example-5}) is for $U(N_c)$ at the non-baryonic
branch and curve (\ref{example-6}), for $U(N_c)$ at the baryonic
branch. Again, function $H_{p}(t)$ has a proper number of $(t-m^2)$
to count the prefactor for various $r$-th branches. It is also
interesting to see that the curve form (\ref{example-5})
is identical to the (\ref{example-2}), which as we will
show later, provides the possibility for smooth transition
between these two breaking patterns. 

For  the broken pattern $SO(2N_c)\to SO(2N_c)$, we have
\bea 
\label{example-7}
y^2 & = &  t F_1(t) H^2_{N_c-1}(t),  \\
y^2 & = & H^2_{N_c}(t), \label{example-8}
\eea 
where curve (\ref{example-7}) is for $SO(2N_c)$ at the
Chebyshev branch and curve (\ref{example-8}),
for $SO(2N_c)$ at the Special branch. The function $H_{p}(t)$ will
have a proper number of factor $t$ to count the 
prefactor required by Special or Chebyshev branch. It is
noteworthy to notice that although both curves (\ref{example-3})
and (\ref{example-7}) appear to be identical, 
they can be distinguished
by a factor $H_{p}(t)$ where different powers of $t$ and $(t-m^2)$
can arise. For example, there may be 
a factor $(t-m^2)$ in  $H_{p}(t)$ of the curve (\ref{example-3}),
but it does not exist in  $H_{p}(t)$ of the curve (\ref{example-7}).

Before ending this subsection, let us give the rules for the
counting of vacua. The total number of vacua is the product
of the number of vacua of various blocks. For each block,
there are four cases: $U(N_i)$ without flavors, $U(N_i)$ with
flavors, $SO(N)$ without flavors, and $SO(N)$ with flavors.
For $U(N_i)$ without flavors, the counting is given by the Witten
index $w(N_i)=N_i$. 
For $U(N_i)$ with $M_i$ flavors, the result has been
given in \cite{bfhn} and is summarized as follows:
\begin{eqnarray}
\mbox{The number of vacua}&=&\left\{				
\begin{array}{lll} 
2N_i-M_i  & r < M_i/2, & M_i \leq N_i,   \\
N_i-M_i/2 & r = M_i/2, & M_i \leq N_i,  \\
2N_i-M_i  & r \geq M_i-N_i, & M_i \geq N_i+1,  \\
N_i- r & r < M_i-N_i, & M_i \geq N_i+1,  \\
1 & r =N_i -1(\mbox{nonbaryonic}), &  \\				
1 & r=M_i-N_i, N_i(\mbox{baryonic}). 
\end{array} \right.
\label{vacua} 
\end{eqnarray}
For $SO(N)$ without flavors, the counting is given by the Witten
index $w(N)=(N-2)$ for $N \geq 5$ and $4,2,1,1,1$ for $N=4,3,2,1,0$
(notice that in previous subsection we have given an explanation of
the Witten index for $N\leq 4$).
For $SO(N)$ with $M$ flavors, there are $(N-M-2)$ vacua
from the Chebyshev branch and one vacuum from the Special
branch (if it exists).

%%%%%%%%%%%%%%%%%%%%%%%%%%%%%%%%%%%%%%%%%%%%%%%%%%%%%%%%%%%%%%%%%%%%%%%%%%%%%
\section{Quartic superpotential with massive flavors}
\setcounter{equation}{0}
%%%%%%%%%%%%%%%%%%%%%%%%%%%%%%%%%%%%%%%%%%%%%%%%%%%%%%%%%%%%%%%%%%%%%%%%%%%%%

\indent

In this section we will deal with and analyze the explicit examples 
with quartic tree level superpotential of degree 4 (that is, $k=1$ from
[\ref{treesup1}]) satisfying
$W^{\prime}(\pm m)=0$,
\begin{eqnarray}
W^{\prime}(x)=x(x^2-m^2)
\label{extree}
\end{eqnarray}
where we do not have a constant term in $x$ because 
the superpotential $W(x)$ is a function of $x^2$ with 
a constant term;
therefore, after differentiating this with respect to $x$, 
superpotential $W(x)$ will give rise to
the above expression (\ref{extree}).
In these examples the gauge group $SO(N_c)$ 
can break into three cases: two blocks with 
\footnote{As in \cite{bfhn}, 
we use the notation for hat in 
$\widehat{U(N_1)}$ to denote a gauge theory with flavors 
charged under the $U(N_1)$ group.}
\bea
SO(N_c)\to 
SO(N_0)\times \widehat{U(N_1)},\qquad  N_c=N_0+2N_1
\nonu
\eea
under the semiclassical limit 
$\Lambda \to 0$ which we will call the {\it non-degenerated} case and
one block with 
\bea
SO(N_c)\to SO(N_c) \qquad \mbox{ or} 
\qquad
SO(N_c)\to  \widehat{U([N_c/2])}
\nonu
\eea
which we will call the {\it degenerated} case. 
Since for the sake of simplicity,
we consider the case with equal masses of flavors, 
we can write the Seiberg-Witten curves for these 
gauge theories as follows: 
\bea
y^2&=&P^2_{2N_c}(x)-4x^4\Lambda^{4N_c-2N_f-4}(x^2-m^2)^{N_f}\qquad 
\mbox{for }\ SO(2N_c), \nonu \\
y^2&=&P^2_{2N_c}(x)-4x^2\Lambda^{4N_c-2N_f-2}(x^2-m^2)^{N_f}\qquad 
\mbox{for }\ SO(2N_c+1),
\nonu
\eea
where  $m$ in these curves is the same as the one in 
(\ref{extree})
\footnote{For gauge theories with massless flavors we will 
study in next section in which the tree level superpotential 
(\ref{massless-W}) 
is 
different from (\ref{extree}) and all the flavors must be charged 
under $SO(N_0)$, not $U(N_1)$. For degenerated case, 
$SO(N_c)\to SO(N_c)$, the flavors are charged under the factor 
$SO(N_c)$. }. 
%In addition, we study the examples where all $N_f$ flavors 
%are massive and charged only under $\widehat{U(N_1)}$ in 
%the semiclassical limit, not $SO(N_0)$ factor. 
%Thus we should have the condition $W^{\prime}(-m)=0$ in the superpotential. 
%Therefore we assumed the quartic tree level superpotential as 
%(\ref{extree}). 
On the $r$-th branch these Seiberg-Witten curves 
are factorized as follows, based on the discussion of previous section 3
and \cite{aps1,bfhn},
\bea
(x^2-m^2)^{2r}&&\!\!\!\!\!\!\!\!\!\! \left[P^2_{2(N_c-r)}(x)-4x^4
\Lambda^{4N_c-2N_f-4}(x^2-m^2)^{N_f-2r} \right] \qquad \mbox{for }
\ SO(2N_c), 
\label{42swso2n}  
 \\
(x^2-m^2)^{2r}&&\!\!\!\!\!\!\!\!\!\! \left[P^2_{2(N_c-r)}(x)-4x^2
\Lambda^{4N_c-2N_f-2}(x^2-m^2)^{N_f-2r} \right] \qquad \mbox{for }
\ SO(2N_c+1), 
\label{42swso2n+1} 
\eea
where the $(x^2-m^2)^{2r}$ is the prefactor for that particular branch. For
the expressions in the bracket we need to count 
the proper number of single roots
and double roots as given in previous section 3. Furthermore,
there are {\it two} special $r$'s 
which give the baryonic branch of $\widehat{U(N_1)}$ factor 
\footnote{This holds also for the degenerated case:
$SO(N_c)\to  \widehat{U([N_c/2])}$. In this case, $N_1$ becomes
$N_c/2$.}
\bea
r=N_1, \qquad \mbox{or} \qquad r=N_f-N_1. 
\label{conr}
\eea
If the $r$ does not satisfy this condition, it is called a nonbaryonic
branch.

For convenience we list the proper factorization form of the curve
in various situations. 
Note that the roots for $W^{\prime}(x)$ are $x=0$ and $x=\pm m$
from (\ref{extree}). 
For the non-baryonic branch of $\widehat{U(N_1)}$ factor 
 with some D5-branes wrapping
around the origin $x=0$, 
the curve corresponding to (\ref{example-1}) is 
\bea
P^2_{2(N_c-r)}-4x^4\Lambda^{4N_c-2N_f-4}(x^2-m^2)^{N_f-2r}
\!\!\!\!&=&\!\!\!\!x^2 H^2_{2N_c-2r-4}(x) F_{6}(x) \quad  \mbox{for} 
\ SO(2N_c), 
\label{42swso2nnb}  
\\
P^2_{2(N_c-r)}-4x^2\Lambda^{4N_c-2N_f-2}(x^2-m^2)^{N_f-2r}
\!\!\!\!&=&\!\!\!\!x^2 H^2_{2N_c-2r-4}(x) F_{6}(x) \quad  \mbox{for}
\ SO(2N_c+1). 
\label{42swso2n+1nb} 
\eea
For the baryonic branch of $\widehat{U(N_1)}$ factor 
 with some D5-branes wrapping around the origin $x=0$ (or
for the degenerated case where all D5-branes are wrapping around 
the origin $x=0$, the curve has the same form: (\ref{example-7})),
the curve corresponding to (\ref{example-3}) is
\bea
P^2_{2(N_c-r)}-4x^4\Lambda^{4N_c-2N_f-4}(x^2-m^2)^{N_f-2r}\!\!\!\!&=
&\!\!\!\!x^2 H^2_{2N_c-2r-2}(x) F_{2}(x) \quad \mbox{for} \
 SO(2N_c), 
\label{42swso2nb}  
\\ 
P^2_{2(N_c-r)}-4x^2\Lambda^{4N_c-2N_f-2}(x^2-m^2)^{N_f-2r}\!\!\!\!&=
&\!\!\!\!x^2 H^2_{2N_c-2r-2}(x) F_{2}(x) \quad \mbox{for} \
SO(2N_c+1). 
\label{42swso2n+1b}
\eea
For the non-baryonic branch of $\widehat{U(N_1)}$ factor 
 (or for degenerated case with all D5-branes 
wrapping around the $x=\pm m$, the curve is the same: (\ref{example-5})) 
the curve corresponding to (\ref{example-2}) is
\bea
P^2_{2(N_c-r)}-4x^4\Lambda^{4N_c-2N_f-4}(x^2-m^2)^{N_f-2r}\!\!\!\!&=
&\!\!\!\! H^2_{2N_c-2r-2}(x) F_{4}(x) \quad \mbox{for} \
 SO(2N_c), 
\label{42swso2nb--1}  
%\nonu 
\\ 
P^2_{2(N_c-r)}-4x^2\Lambda^{4N_c-2N_f-2}(x^2-m^2)^{N_f-2r}\!\!\!\!&=
&\!\!\!\! H^2_{2N_c-2r-2}(x) F_{4}(x) \quad \mbox{for} \ SO(2N_c+1). 
\label{42swso2n+1b--1}
%\nonu
\eea
Finally for the baryonic branch of $\widehat{U(N_1)}$ factor 
(or for degenerated case with all D5-branes wrapping 
around the $x=\pm m$, the curve takes the same form: 
(\ref{example-6}) or 
(\ref{example-8})) 
the curve corresponding to (\ref{example-4}) is
\bea
P^2_{2(N_c-r)}-4x^4\Lambda^{4N_c-2N_f-4}(x^2-m^2)^{N_f-2r}\!\!\!\!&=
&\!\!\!\! H^2_{2N_c-2r}(x) \qquad \mbox{for }
\ SO(2N_c),  
\label{42swso2nb--2}  
%\nonu
\\ 
P^2_{2(N_c-r)}-4x^2\Lambda^{4N_c-2N_f-2}(x^2-m^2)^{N_f-2r}\!\!\!\!&=
&\!\!\!\! H^2_{2N_c-2r}(x) \qquad \mbox{for }
\ SO(2N_c+1). 
\label{42swso2n+1b--2}
%\nonu
\eea
In fact, these curves have been given in section 3.2 where we
used the fact that $t=x^2$ and $r=0$ 
\footnote{ Moreover, the curves corresponding to
(\ref{example-5}), (\ref{example-6}), (\ref{example-7}) and 
(\ref{example-8}) can be written similarly
and we will describe them in the following discussions when we need to 
explain each case.}.

Next we want to clarify the number of flavors $N_f$ that we will study 
below. For the exponent of $\Lambda$ in (\ref{42swso2n}) and 
(\ref{42swso2n+1}) to be positive, we will concentrate on $SO(2N_c)$ 
gauge theories with the condition $N_f<2N_c-2$ and 
$SO(2N_c+1)$ gauge theories with 
the condition $N_f<2N_c-1$, 
which are the conditions for the theories to be 
asymptotically free.  Thus taking into account of $r=\mbox{min} 
\left(N_c,\frac{N_f}{2} \right)$, we will study the 
examples with the condition
\bea
r\le \frac{N_f}{2}.
\label{rcondition}
\eea
Finally let us  emphasize the  comments on the relation between 
function $F_6(x)$ appearing in (\ref{42swso2nnb}) and (\ref{42swso2n+1nb}) 
and 
$W^{\prime}(x)$ from (\ref{extree}). 
According to  relation (\ref{newmatrix}), we have 
to take care of the second term of (\ref{newmatrix}) on the left hand side
when 
the number of flavors $N_f$ is greater than $(2N_c-4)$ for $SO(2N_c)$ 
gauge theories and $(2N_c-3)$ for $SO(2N_c+1)$ gauge theories. 
This is a kind of new phenomenon that did not appear in the pure case 
\cite{ao}.
For 
example, on the non-baryonic branch of $\widehat{U(N_1)}$ factor 
for $SO(4)$ with $N_f=1$, $SO(5)$ 
with $N_f=2$, $SO(6)$ with $N_f=3$ and $SO(7)$ with $N_f=4$ cases, 
we must be attentive to  
these modifications due to the presence of flavors.
%For the case with $SO(8)$ gauge theory with $N_f=5$ flavors,
%we do have similar thing but in this paper, we will not consider this 
%particular case.

Now we are ready to deal with the explicit examples for $SO(N_c)$ 
gauge theory with massive flavors where $N_c=4,5,6,7$.
The number of flavors $N_f$ is restricted to $N_f<N_c-2$ and the 
index $r$ satisfies (\ref{rcondition}). Therefore,
 for given number of colors
$N_c$, the quantities $N_f$ and $r$ are fixed.

%%%%%%%%%%%%%%%%%%%%%%%%%%%%%%%%%%%%%%%%%
\subsection{ $SO(4)$ case}
%%%%%%%%%%%%%%%%%%%%%%%%%%%%%%%%%%%%%%%%%

\indent 

For this gauge theory we will discuss the number of flavors 
$N_f=0,1$  cases. 
There are three breaking patterns $SO(4)\to SO(2)\times 
\widehat{U(1)}$,
$SO(4)\to SO(4)$, and 
$SO(4)\to \widehat{U(2)}$. 

%%%%%%%%%%%%%%%%%%%%%%%%%
\subsubsection{$N_f=0$}
%%%%%%%%%%%%%%%%%%%%%%%%%

\noindent

%%%%%%%%%%%%%%%%%%%%%%%%
{\bf 1. Non-degenerated case}
%%%%%%%%%%%%%%%%%%%%%%%%

\indent

In \cite{ao}, the case with 
$N_f=0$  was discussed. Since the factorization problem 
was trivial, 
the characteristic function $P_{4}(x)$ could be represented as
\bea
P_4(x)=x^2(x^2-v^2).
\nonu
\eea
From this $P_4(x)$, we could obtain a tree level superpotential and 
a deformed function $f_2(x)$,
\bea
W^{\prime}(x)=x(x^2-v^2), \qquad f_2(x)=-4x^2\Lambda^4.
\nonu
\eea
There is only one vacuum for a given $W^{\prime}(x)$. To satisfy the 
condition $W^{\prime}(\pm m)=0$ discussed above, we should have $v^2=m^2$. 
Under the semiclassical limit $\Lambda\to 0$, the gauge group $SO(4)$ 
breaks into $SO(2)\times U(1)$. Notice that the Witten index $w(2)$ 
of $SO(2)$ is one and therefore the  total number of vacua becomes one from
the weak coupling analysis. We realize that the number of vacua from 
the strong coupling analysis 
coincides with the one from the weak coupling analysis. 

In fact, by factoring out $x^2$ common to 
$W^{\prime}(x)^2$ and $f_2(x)$, 
the full curve can be written as 
\bea
y^2= x^2 \left( W^{\prime}(x)^2 + f_2(x) \right)=
  [(x^2-v^2)^2-4 \La^4] (x^2)^2 \equiv F_4(x) H^2_2(x),
\nonu
\eea
which indicates that the $SO(2)$ is at the Special branch
where the power of $x^2$ is two (even) in the above curve and that $U(1)$
at the nonbaryonic $r=0$ branch because the condition 
(\ref{conr}) is not satisfied. 
See also  (\ref{42swso2nb--1}).
%This
%is exactly the result in the weak coupling analysis in 3.1.1 when we discuss
%the quantum moduli space for the range $2N_f=N-2$. 
%This phenomena
%will happen for all the examples later and we will not repeat them. 

%%%%%%%%%%%%%%%%%%%%%%%%
{\bf 2. Degenerated case}
%%%%%%%%%%%%%%%%%%%%%%%%

\indent

The above calculations for nondegenerated case 
assume that every root of $W'(x)$ has at least
one D5-brane wrapping around it. However, there are the situations
where some roots do {\it not} have wrapping D5-branes around them. 
For our example, there
are two cases; one is that there is no D5-brane around zero ($x=0$) 
and the other is that there is no D5-brane around $x=\pm m$.
For both cases, the curve corresponding to (\ref{example-5}) 
should be factorized with  factor $F_4(x)$ as
\bea
y^2=P^2_4(x)-4 \Lambda^4  x^4= F_4(x) H^2_2(x),
\nonu
\eea
where $F_4(x)$ can have a factor $x^2$ further depending 
on whether the $SO(N)$ factor
is at the Chebyshev branch or not. See also (\ref{example-7}).
Writing $P_4(x)=(x^4-s_1 x^2+s_2)$, $H_2(x)=(x^2+a)$ and $F_4(x)=(x^4
+b x^2+c)$, we have the following solutions
\bean
b=2(a\mp 2\Lambda^2), \qquad c=a^2, \qquad 
s_1=-2a\pm 2\Lambda^2, \qquad s_2=a^2.
\nonu
\eean
If $\Lambda \to 0$, but $a\neq 0$, we get a symmetry breaking 
$SO(4)\to U(2)$ where $U(2)$ is at the nonbaryonic $r=0$ branch. If
the classical limit goes to $\Lambda,a \to 0$ where
$F_4(x)$ contains the $x^2$ factor, we get 
$SO(4)\to SO(4)$. It is obvious that there is no smooth 
interpolation between these vacua because there exists a U(1) at the
IR for the former symmetry breaking while there exists no $U(1)$ for
the latter symmetry breaking.

To count  the vacua, we need to determine the value of $a$. If
there are D5-branes wrapping around the $x=\pm m$, we get $b=-2m^2$. Putting
it back to the expression for $b$, we found two values for $a$
(each sign gives one solution for $a$); thus,
there exist two vacua for $SO(4)\to U(2)$. 
If all D5-branes wrap around the origin $x=0$,
there are two cases to be distinguished. The first case is
that the $SO(4)$ factor is 
at Chebyshev vacua, so we need a factor $x^2$ in the
curve which determines  the value of $c=0$ (so $a=0$ too). There are
two vacua corresponding to the $\pm$ sign. It is noteworthy  that
the curve is $y^2=(x^2)^3(x^2+b)$ with the power of $x^2$ being three. 
This 
is the character of $SO(4)$ without flavors at the Chebyshev branch.
More detailed discussion of the power of $t=x^2$ can be found in
the next section.
The second case is that
the $SO(4)$ factor is 
at the Special vacua, so $F_4(x)$ should be a complete square
form. This determines the value of $a$ as 
$a=\Lambda^2$ if $b=2(a- 2\Lambda^2)$
or $a=-\Lambda^2$ if $b=2(a+2\Lambda^2)$. Although it seems that we
have two solutions, in fact, the curve is the same for both cases and
presumably we should count them as one.
 
However, it is known that for $SO(4)=SU(2)_L\times SU(2)_R$, we should have
$2\times 2=4$ vacua. Among these four vacua, two of them have
the relations 
$S_L=S_R$ and $S_L+S_R \neq 0$, so they correspond to the above
two solutions at the Chebyshev vacua. 
The other two have the relation $S_L+S_R=0$, but we found only one solution
for the curve at the Special vacua. Our understanding is that
since $S_L+S_R=0$ for these two vacua, they correspond
to the same point in the ${\cal N}=2$ curve. Later we will meet same
problem again and again, where from the curve we find only one
solution for $SO(4)$ at the Special branch, but it should be counted 
as two vacua.

%%%%%%%%%%%%%%%%%%%%%%%%
\subsubsection{$N_f=1$}
%%%%%%%%%%%%%%%%%%%%%%%%%%%

In this case, from the condition $r\le \frac{N_f}{2}$ we have only the $r=0$
branch. For the breaking pattern $SO(4)\to SO(2)\times 
\widehat{U(1)}$, the $r=0$ 
branch can be a non-baryonic or baryonic branch. 
%because 
%there exists only one relation $r=N_f-N_1$($r \ne N_1$)
On the other hand, for the breaking pattern 
$SO(4)\to \widehat{U(2)}$, the  $r=0$ is non-baryonic since $r \ne 
N_f-N_1$ and $r \ne N_1$.

%%%%%%%%%%%%%%%%%%%%%%%%%%%%%%%%%%%%%%%%%
$\bullet$ {\bf   Baryonic $r=0$ branch}
%%%%%%%%%%%%%%%%%%%%%%%%%%%%%%%%%%%%%%%%

\noindent
Since we are considering a baryonic branch, we have an {\it extra} 
massless monopole (that is, one more double root) 
and from the relation (\ref{42swso2nb}) 
the following factorization problem arises,
\bea
y^2= P^2_4(x)-4x^4\Lambda^{2}(x^2-m^2)=x^2 H^2_2(x) F_2(x).
\nonu
\eea
To be consistent  on both sides, 
the characteristic function $P_4(x)$ on the left hand side must
have an $x^2$ factor, i.e., $P_4(x)=x^2(x^2-a^2)$ due to the presence of
an $x^2$ factor in the right hand side. In addition, we assume 
$H_2(x)=(x^2-b^2)$. Assuming that $b\neq 0$ (the $b=0$ case will correspond
to $SO(4)\to SO(4)$ and will be discussed later),
we must have $F_2(x)=x^2$, 
 $a^2=(m^2-\Lambda^2)$, and $b^2=(m^2+\Lambda^2)$; thus,
the characteristic function and $F_2(x)$ behave  as 
\bea
P_4(x)=x^2\left( x^2+\Lambda^2-m^2 \right), \qquad F_2(x)=x^2.
\nonu
\eea
Under the semiclassical limit $\Lambda\to 0$, the 
gauge group $SO(4)$ breaks 
into $SO(2)\times \widehat{U(1)}$. In this case, we  have only one vacuum. 
This number matches the one obtained from the 
weak coupling analysis, 
because both $SO(2)$ and $\widehat{U(1)}$ with $N_f=1$ theories have only 
one vacuum. 
Notice also that 
for this special case, the curve $y^2$ has  a total square form with a factor
$x^4$ implying that $SO(2)$ is at Special branch.

%%%%%%%%%%%%%%%%%%%%%%%%%%%%%%%%%%%%%%%%%%%%
$\bullet$ {\bf   Non-baryonic $r=0$ branch}
%%%%%%%%%%%%%%%%%%%%%%%%%%%%%%%%%%%%%%%%%%%

%%%%%%%%%%%%%%%%%%%%%%%%
{\bf 1. Non-degenerated case}
%%%%%%%%%%%%%%%%%%%%%%%%

\indent

In this case the factorization problem
(\ref{42swso2nnb})
 becomes as
\bea
P_4^2(x)-4x^4\Lambda^2\left(x^2-m^2 \right)=x^2 F_6(x).
\nonu
\eea
This equation is easily solved if we assume that $P_4(x)=x^2P_2(x)
\equiv x^2(x^2-A)$ by recognizing that there is a factor $x^2$ on the 
right hand side
and finally one obtains
\bea
F_6(x)=x^2\left[(x^2-A)^2-4\Lambda^2(x^2-m^2) \right].
\nonu
\eea
Taking into account of (\ref{newmatrix}) we have $m^2=A$. 
Under the semiclassical limit $\Lambda \to 0$, the characteristic 
function goes to $P_4(x)=x^2(x^2-m^2)$, which means that the gauge 
group $SO(4)$ breaks into $SO(2)\times \widehat{U(1)}$. We  
have only one vacuum, which matches the counting obtained by general
analysis given in previous section (weak coupling analysis).
The curve $y^2$ has  a total square form with a factor
$x^4$ implying that $SO(2)$ is at Special branch.

%%%%%%%%%%%%%%%%%%%%%%%%%%%%%
{\bf 2. Degenerated case}
%%%%%%%%%%%%%%%%%%%%%%%%%%%%%%

\indent

The curve should be factorized as, together with a term $F_4(x)$ on the 
right hand side as before,
\bea
y^2=P^2_4(x) -4x^4\Lambda^{2}(x^2-m^2)=F_4(x) H^2_2(x).
\nonu
\eea 
Parameterized by $P_4(x)=(x^4-s_1 x^2+s_2)$, $H_2(x)=(x^2+a)$ and 
$F_4(x)=(x^4+b x^2+c)$ we have the following relations
\bea
s_1 & = & \frac{-2a - b - 4{\La}^2}{2}, \nonu \\
s_2 & = & \frac{4c + 4a\left( b - 4{\La}^2 \right)
        - \left( b + 4{\La}^2 \right)^2 - 
    16{\La}^2 m^2}{8}, \nonu \\
c & = & -\left( \frac{32 a^2 {\La}^2 + 
      \left( b + 4 {\La}^2 \right)^3 + 
      16 {\La}^2
       \left( b + 4 {\La}^2 \right) m^2 + 
      a \left( -2 b^2 + 16 b {\La}^2 + 
         32 {\La}^2
          \left( 3 {\La}^2 + m^2 \right) 
         \right) }{8 a - 
      4 \left( b + 4 {\La}^2 \right) } \right), \nonu \\
0 & = & a \left( 4 a^2 - 4 a
      \left( b - 5 {\La}^2 \right)  + 
     \left( b + 4 {\La}^2 \right)^2 \right) 
    \nonu \\
& & +  \left( \left( -2 a + b \right)^2 + 
     8 \left( 4 a + b \right) {\La}^2 + 
     16 {\La}^4 \right) m^2 + 
  16 {\La}^2 m^4.
\nonu
\eea
To see the limit where there is D5-brane wrapping around $x =\pm m$,
we put \footnote{
To discuss the smooth transition, we take $a$ to be free parameter
while the mass $m$ is determined by $a$, i.e., $m$ will change as
the $a$ does. This is the philosophy used in \cite{csw,bfhn}
(see, for example, the equation (6.30) in \cite{bfhn}) which
is very convenient for this purpose. We can also use the 
direct method which will be demonstrated later by one simple
example. To count the vacua, we really need to fix $m$ and
 then solve $a$.} 
$b=-2 m^2-4 \La^4$ by the relationship of $F_4(x)$ and $(x^2-m^2)$
\footnote{If we use $t=x^2$, the curve is effectively the one for 
$U(2)$ with three flavors
where two of them are massless and one is massive.
Then using
the relation $F_4(x)+4 \Lambda^2 x^2\sim (x^2-m^2)^2+O(x^0)$, 
we can read off
$F_4(x)=x^4-(2m^2+4 \Lambda^2) x^2+m^4+d$.}
and get
\bea
0= a^2 \left( a + 9 {\La}^2 \right)  + 
  3 a \left( a + 4 {\La}^2 \right) m^2 + 
  \left( 3 a + 4 {\La}^2 \right) m^4 + m^6.
\nonu
\eea
There are two limits we can take: (1) $\Lambda \to 0$, but
$a\to \mbox{constant}$. In this limit, all three solutions provide
$m^2\to -a$
and it is $SO(4)\to \widehat{U(2)}$; (2) $\Lambda \to 0$, $a\to 0$
which gives $SO(4)\to SO(4)$. However,
from the above equation by keeping the last term $m^6$, we see
that all of the solutions give
$m^2\to 0$ in this limit. Since we require $m^2$ to be 
nonzero (that is, we are considering massive flavors),
this limit is inconsistent with our assumption and should not
be taken. In fact, 
as we have seen, the non-baryonic $r=0$ branch of
$SO(2)\times \widehat{U(1)}$ theory will have a factor $x^4$
while for $\widehat{U(2)}$ theory,
the curve does not have any factor $x^2$,
 so we should not
expect any smooth transition between them.
To count  the number of  vacua, 
we fix $m$ and then solve $a$. There exist three
solutions. They match the counting, 
$(2N-N_f)=(2\times 2-1)=3$, of $\widehat{U(2)}$ with one flavor.

If all D5-branes wrap around the origin $x=0$, there are two cases to
be considered. The first one is when the extra $x^2$ factor
occurs in $F_4(x)$: $F_4(x)=x^2(x^2+d)$ or $c=0$. In
fact, it is the same curve of the baryonic $r=0$ branch, but with
$H_2(x)=x^2$. There are two solutions with the expression
$P_4(x)=x^2(x^2\pm 2\Lambda^2)$ by $\pm$ sign providing  
a breaking 
$SO(4)\to SO(4)$ under the semiclassical limit $\La \to 0$.
These two vacua are the Chebyshev vacua of $SO(4)$ gauge theory (notice that
the counting becomes $N-N_f-2=4-0-2=2$) with an overall factor 
$x^6$ in the curve. The second one is that
 function $F_4(x)$ should be
a complete square form. There are two solutions. One of them is
$P_4(x)=x^2(x^2-[m^2-\Lambda^2])$ 
which gives a breaking $SO(4)\to SO(2)\times
\widehat{U(1)}$ under the semiclassical limit $\La \to 0$ 
(it is the baryonic $r=0$ branch found above)
and other, $P_4(x)=x^4+\Lambda^2 x^2-m^2 \Lambda^2$
which provides  a breaking 
 $SO(4)\to SO(4)$ under the semiclassical limit $\La \to 0$
and $SO(4)$ is at the Special branch. 
%and is the one we are looking for here.
In other words, among the four vacua of $SO(4)$, two of them are found
in the Chebyshev vacua and two of them correspond to the same 
point in the Special vacua.

Now we summarize what we have obtained in Table \ref{tableso4}
by specifying the flavors $N_f$, symmetry breaking patterns, 
various branches, the exponent of $t=x^2$ in the curve, $U(1)$ at the
IR, the number of vacua, and the possibility  of smooth transition.
It turned out that the number of vacua is exactly the same 
as from the 
weak coupling analysis .

\begin{table} 
\begin{center}
\begin{tabular}{|c|c|c|c|c|c|c|} \hline
 $N_f$  &  Group  & Branch &  Power of $t(=x^2)$ &  $U(1)$ & 
Number of vacua & Connection
 \\ \hline
 0 & $SO(2)\times U(1)$ & $(S,0_{NB})$ & $t^2$ & 1 &  1  & \\
\cline{2-7}  
 & $SO(4)$  & $(C)$ &  $t^3$ & 0 & 2 & \\\cline{3-7}
  &  & $(S)$ &  $t^0$ & 0 & 2 & \\\cline{2-7}
& $U(2)$ & $(0_{NB})$ & $t^0$ & 1 & 2 & \\ \hline 
1 &  $SO(2)\times \widehat{U(1)}$ & $(S,0_{NB})$ & $t^2$ & 1 &  1  & 
\\ \cline{3-7}
& & $(S,0_{B})$ & $t^2$ & 0 &  1  & \\ \cline{2-7}
  & $SO(4)$  & $(C)$ &  $t^3$ & 0 & 2 & \\\cline{3-7}
  &  & $(S)$ &  $t^0$ & 0 & 2 & \\\cline{2-7}
& $\widehat{U(2)}$ & $(0_{NB})$ & $t^0$ & 1 & 3 & \\ \hline     
\end{tabular} 
\end{center}
\caption{\sl The summary of the phase structure of the $SO(4)$ gauge group. 
Here we use $C/S$ for Chebyshev or Special branch for $SO(N_i)$ factor
and $r_{NB}/r_{B}$ for the $r$-th non-baryonic or $r$-th baryonic
branch. In this table, we list the power of $t$ and the $U(1)$ which is
present in the nonbaryonic branch, at the
IR. They are the indices to see whether two phases could have smooth
transition. In the case of  massive flavors, the power of $t$ is
$3$ for $SO(N_i)$ for $N_i$ even and 
$1$ for $SO(N_i)$ for $N_i$ odd at the Chebyshev branch.
For massive flavors, only $SO(N_i)$ where $N_i=2,3,4$ 
can be at the Special branch
and the power is $2,0,0$ for $N_i=2,3,4$. For this table, we see that
for both $N_f=0$ and $N_f=1$, there are {\it no} two phases having the same
factor $t^p$ where $p$ is some number and 
the number of $U(1)$, so there exists no  smooth
transition between them.}
\label{tableso4}
\end{table} 

%%%%%%%%%%%%%%%%%%%%%%%%%%%%%%%%%%%%%%%%%
\subsection{SO(5) case}
%%%%%%%%%%%%%%%%%%%%%%%%%%%%%%%%%%%%%%%%%%%

\indent

For this gauge theory we will discuss the number of flavors 
$N_f=0, 1, 2$  cases. 
There are three breaking patterns $SO(5)\to SO(3)\times 
\widehat{U(1)}$,
$SO(5)\to SO(5)$ and 
$SO(5)\to \widehat{U(2)}$. 

%%%%%%%%%%%%%%%%%%%%%%%%%
\subsubsection{$N_f=0$}
%%%%%%%%%%%%%%%%%%%%%%%%%%%%%

\noindent

%%%%%%%%%%%%%%%%%%%%%%%%
{\bf 1. Non-degenerated case}
%%%%%%%%%%%%%%%%%%%%%%%%

\indent

The theory with $N_f=0$ was discussed in \cite{ao}. Since the 
factorization problem was trivial, the
characteristic function was given by,
\bea
P_4(x)=x^2(x^2-l^2).
\nonu
\eea
To satisfy the condition for $W^{\prime}(\pm m)=0$, 
we must have the relation, 
$l^2=m^2$. Under the semiclassical limit $\Lambda\to 0$, the gauge group
$SO(5)$
breaks into $SO(3)\times U(1)$ where $SO(3)$ is at the 
Chebyshev branch with a factor $x^2$ in the curve
while $U(1)$ is at the nonbaryonic branch. 
However, since the Witten index
$w(3)=2$, we 
expect to find {\it two} solutions
instead of one. 
In fact, another solution will be found in the degenerated case
\footnote{For the two vacua of $SO(3)$, one is at the Chebyshev branch
having factor $t=x^2$ and another, at the Special branch with
factor $t^0$. 
However, since the Special branch has extra double roots
just like the one in the degenerated case, they have the same
factorization form of the curve and for simplicity we
use the degenerated case to include all of them.}.

%%%%%%%%%%%%%%%%%%%%%%%%%
{\bf 2. Degenerated case}
%%%%%%%%%%%%%%%%%%%%%%%

\indent

For D5-branes wrapping around only one root of $W'(x)$ (either the origin or 
$x=\pm m$),  the curve
is
\bea
y^2=P^2_2(t)-4 \Lambda^6 t=F_2(t) H^2_1(t), \qquad t \equiv x^2.
\nonu
\eea
First let us discuss a case in which there are D5-branes wrapping around
$\pm m$.
%For the case where D5-branes are wrapping around the origin $x=0$, 
%the function $F_2(t)$ should possess a factor $t$ (\ref{42swso2n+1b}).
Parameterizing $P_2(t)=(t^2-s_1 t+s_2)$, $F_2(t)=(t-a)^2+b$ and
$H_1(t)=(t+c)$, we found only one combination $(s_1,s_2,a,c)$ as a 
function of the parameter $b$ (it is easy to see that $b\neq 0$)
\bea
s_1 & = & {b^2\over 8  \Lambda^6}-{4  \Lambda^6 \over b}, \qquad
s_2 = {b\over 256} \left(64+{b^3\over  \Lambda^{12}} \right), \nonu \\
a & = & {b^2\over 16  \Lambda^6}-{4  \Lambda^6 \over b}, \qquad
c={- b^2 \over 16  \Lambda^6}.
\nonu
\eea 
Now we can discuss the smooth interpolation. 
First using the relationship of $F_2(t)$ and $(t-m^2)$ we 
get $a=m^2$. If $\Lambda\to 0$, but
$b^2/\Lambda^6\equiv \beta \neq 0$, $a=m^2\to \beta/16$ and
$P_2(t)\to (t-{\beta \over 16})^2$ 
so the breaking pattern is $SO(5)\to U(2)$.  If $\Lambda\to 0$, but
$\Lambda^6/b\equiv \gamma \neq 0$, $a=m^2\to -4 \gamma$ and
$P_2(t) \to t(t+4 \gamma)$,
the breaking pattern is $SO(5)\rightarrow 
SO(3)\times U(1)$ where $SO(3)$ is at the Special branch. 
From this we see that $U(2)$ is smoothly connected to 
$SO(3)\times U(1)$.  
It is noteworthy that at the Special point of $SO(3)$ without
flavor, the power of $t$ is given by $(2N_f-N+3)=(0-3+3)=0$
according to the argument in section 3.1.2.
\footnote{
The proper power of $t$ at the Chebyshev branch and Special branch
will be summarized in the next section, where more details
can be found.}
Because of this, both $SO(3)\times U(1)$ and $U(2)$ can have
the same form of curve.

Now let us count the number of  vacua. 
To do so, we need to fix the mass $m^2$
and solve 
$b$. Using the relation $a=m^2$, 
we find three solutions for $b$.
It is obvious that two of them correspond to the limit
$b^2/\Lambda^6\equiv \beta \neq 0$ which gives the two vacua of
$U(2)$ and the third one, to the limit
$\Lambda^6/b\equiv \gamma \neq 0$ which gives the vacua of
$SO(3)\times U(1)$. The third one is the missing one for 
$SO(3)$ mentioned in the  non-degenerated case.

The above discussion shows the smooth transition clearly. However, to
deepen our understanding, we use another method by solving 
the $b$ in terms of the $m^2$. One of these three solutions will turn out 
\bea
b=\frac{-4{\left( \frac{2}{3} \right) }^{\frac{1}{3}}
     {\Lambda}^6m^2}{{\left( -9
         {\Lambda}^{12} + 
        {\sqrt{3}}{\sqrt{{\Lambda}^{18}
             \left( 27{\Lambda}^6 - 4m^6 \right)
               }} \right) }^{\frac{1}{3}}} - 
  2{\left( \frac{2}{3} \right) }^{\frac{2}{3}}
   {\left( -9{\Lambda}^{12} + 
       {\sqrt{3}}{\sqrt{{\Lambda}^{18}
            \left( 27{\Lambda}^6 - 4m^6 \right) 
            }} \right) }^{\frac{1}{3}}.
\nonu
\eea
Let us consider the following expression in the above
\bea 
  {\left( -9{\Lambda}^{12} + 
       {\sqrt{3}}{\sqrt{{\Lambda}^{18}
            \left( 27{\Lambda}^6 - 4m^6 \right) 
            }} \right) }^{\frac{1}{3}} 
 & = &  \Lambda^3 {\left( -9{\Lambda}^{3}+ {\sqrt{3}}
{\sqrt{\left( 27{\Lambda}^6 - 4m^6 \right) 
            }} \right) }^{\frac{1}{3}} \nonumber \\
& = & 12^{1\over 6} \Lambda^3 m \omega_6,~~~~\omega_6^6=-1,~~~
\Lambda\rightarrow 0
%\label{expression-1}
\nonu
\eea
where we have taken the limit $\Lambda\rightarrow 0$ in the last line,
but carefully kept the phase factor $\omega_6$. The phase factor
comes from the expression 
\bea
(27{\Lambda}^6 - 4 m^6)^{1\over 6}=\left(4 (m^2-\alpha_1)
(m^2-\alpha_2)(m^2-\alpha_3) \right)^{1\over 6}
\nonu
\eea 
where three roots $\alpha_i$ are the order $\La^2$. If we 
start from very
large $m^2$, surround only one root once and go back to
large $m^2$, there will be a phase factor $e^{2\pi i\over 6}=\eta$
where $\eta^6=1$. Thus depending on the path, in general  we  have
\bea
b\sim -{4\over \sqrt{3}} \Lambda^3 m [ \omega_6 \eta^k+\omega_6^{-1}
\eta^{-k}].
\nonu
\eea
For general $k$, we find $[ \omega_6 \eta^k+\omega_6^{-1}
\eta^{-k}]\neq 0$ and $b\sim \Lambda^3$. This case gives
the breaking pattern  $SO(5)\to U(2)$
in the limit $\Lambda\to 0$. However, when $k=1$, we have
$ (\omega_6 \eta)^2=\eta^3=-1$ (where we have used 
$\omega_6^2=\eta$) and $[ \omega_6 \eta^k+\omega_6^{-1}
\eta^{-k}]= 0$. This case will give rise to $b\sim \Lambda^6$ 
and the breaking pattern $SO(5)\to SO(3)\times U(1)$.
From this analysis we see that starting from $k=0$ and 
surrounding
one root once, we make a smooth transition from one phase to
another phase.

The above analysis demonstrates
explicitly the origin of the smooth transition among  the phases
in the parameter space of deformed superpotential. However, in
most cases it is hard to make such a detailed analysis. It is
more convenient to use $b$ as a free parameter while $m^2$  
is determined by $b$ when our focus is just the smooth transition.
Later we mainly use this method.
To count the number of vacua, we really need to fix $m^2$ and find the
number of solutions of $b$. Although we cannot solve this 
exactly most of the time, 
the counting can be done easily by observing the
degree of the equation and  various limits we take.

Finally we consider a case in which all D5-branes wrap around the 
origin $x=0$.
In this case,
we need to have a factor $t$ in $F_2(t)$, or $a^2+b=0$ using the above
results
\footnote{
Notice that for this case, $a,b$ are fixed by this equation
$a^2+b=0$, so there is no free parameter in the curve to be adjusted.
This explains why the above smooth transition, where $b$ is a free
parameter, does not include this phase.}.
There are three solutions $b=-4 \Lambda^4, 2(1\pm i \sqrt{3}) \Lambda^4$
which give three vacua with the 
breaking pattern $SO(5)\to SO(5)$ where
$SO(5)$ is at the Chebyshev branch. The
counting goes like  $(N-2)=3$ which is consistent with the one 
from both the strong 
and weak coupling analyses. 

%%%%%%%%%%%%%%%%%%%%%%%
\subsubsection{$N_f=1$}
%%%%%%%%%%%%%%%%%%%%%%%

\indent

In this case, there is only one nondegenerated  
breaking pattern $SO(5)\rightarrow SO(3)
\times \widehat{U(1)}$ where we can have both non-baryonic and
baryonic $r=0$ branch due to the fact $r = N_f-N_1$ and $r \ne N_1$. 
Besides that, we have two degenerated
breaking patterns $SO(5)\rightarrow SO(5)$ and $SO(5)\rightarrow
\widehat{U(2)}$ where only the non-baryonic $r=0$ branch exists
because the relations  $r \neq N_f-N_1$ and $r \ne N_1$ hold.  

%%%%%%%%%%%%%%%%%%%%%%%%%%%%%%%%%%%%%%%%%%%%%%%%%%%%%
%{\bf The case $SO(5)\rightarrow SO(3)\times \widehat{U(1)}$}

%%%%%%%%%%%%%%%%%%%%%%%%%%%%%%%%%%%%%%%5
$\bullet$ {\bf  Baryonic $r=0$ branch}
%%%%%%%%%%%%%%%%%%%%%%%%%%%%%%%%%%%%%%%

\noindent
The curve should be factorized together with a function $F_2(x)$ as
\bea
y^2= P^2_4(x)-4x^2\Lambda^{4}(x^2-m^2)=x^2 H^2_2(x) F_2(x).
\nonu
\eea
Setting $H_2(x)=(x^2-b)$, $F_2(x)=x^2-a$ and $P_4(x)=x^2(x^2-s_1)$, 
we get
\bea 
\label{puzzle}
s_1=b-{2 \Lambda^4 m^2 \over b^2}, 
\qquad a=-{4 \Lambda^4 m^2 \over b^2}, \qquad
~~-b^4+b^3 m^2+ \Lambda^4 m^4=0.
\eea
From (\ref{puzzle}) we see two limits: (1) If $\La \to 0$ and 
$b\to \mbox{constant}$, by keeping the first and second terms we find
$m^2\to b$ and the gauge group is broken to 
$SO(5) \to SO(3)\times \widehat{U(1)}$
where $SO(3)$ is at the Chebyshev branch and 
$\widehat{U(1)}$, the baryonic $r=0$ branch; (2) If $\La \to 0$,
but $b\sim \La^{4/3}$, by keeping the second and the third
terms we find $m^2\to b^3/\La^4\sim \mbox{finite}$ and the gauge
group is broken to $SO(5) \to SO(5)$.
Thus, we see a smooth transition between these two phases.
This is also consistent with the indices. Both phases have
the same $t^1$ and zero $U(1)$ factor at IR.

Now we can count the vacua by solving $b$ for fixed $m^2$.
 Clearly there
are four solutions. Keeping the first and second terms
we get one solution $b\sim -m^2$ which has the classical
limit $SO(3)\times \widehat{U(1)}$ where $SO(3)$ is at
the Chebyshev branch. Keeping the second and third terms
we get three solutions $b\sim \Lambda^{4/3}$ which give
the three vacua of $SO(5)$ and are consistent with 
the counting  $N-2=3$ from the weak coupling analysis.

To show clearly the rightness of the above result, we use the
$m^2$ to solve $b$ directly. One of these four solutions
will be 
\bea
\Delta &= & -9 \Lambda^4 m^8-\sqrt{3} \Lambda^4 m^6\sqrt{256 \Lambda^4
+27 m^4}, \nonu
%\label{Delta}
\\
\Sigma &= & {\Delta^{1\over 3} \over 2^{1\over 3} 3^{2\over 3}}
+{m^4\over 4}-{ 4 2^{1\over 3} 3^{-1\over 3} \Lambda^4 m^4
\over  \Delta^{1\over 3}}, \nonu 
%\label{Sigma} 
\\
b & = & {m^2\over 4}-{1\over 2} \sqrt{\Sigma}-{1\over 2}
\sqrt{-\Sigma + {3 m^4\over 4}-{m^6 \over 4 \sqrt{\Sigma}}}. \nonu
%\label{final-b}
\eea
Let us consider the classical limit $\Lambda\to 0$. First we have
$\Delta \leq \Lambda^4$. Using this factor and considering the
phase factor  as we did for the $SO(5)$ with $N_f=0$,
we have $\sqrt{\Sigma}\sim \eta {m^2\over 2}$ with
$\eta^2=1$. Using this, we finally have
\bea
b\sim {m^4\over 4} \left(1-\eta-\omega \sqrt{2(1-\eta)} 
\right),~~~\omega^2=1
\nonu
\eea
with two phase factors $\eta, \omega$ depending on the path of
$m^2$.   For $\eta=1$ we have 
$b\sim 0$ at the classical limit. For $\eta=-1$, $\omega=1$
again $b\sim 0$. However, for $\eta=-1$, $\omega=-1$ we have
$b\sim m^2$. The limit of $b\sim 0$ gives the breaking 
$SO(5)\to SO(5)$ while the limit $b\sim m^2$ gives the breaking 
$SO(5)\to SO(3)\times \widehat{U(1)}$. These calculations
confirm our conclusions derived by the simpler method.

For another vacuum of $SO(3)\times \widehat{U(1)}$ where $SO(3)$ is at
the Special branch, we just need to 
choose $P_4(x)=x^2(x^2-m^2)+ \Lambda^4$ so that 
$y^2=(x^2(x^2-m^2)+ \Lambda^4)^2$ is a total square form.

%%%%%%%%%%%%%%%%%%%%%%%%%%%%%%%%%%%%%%%%%%
$\bullet$ {\bf  Non-baryonic $r=0$ branch}
%%%%%%%%%%%%%%%%%%%%%%%%%%%%%%%%%%%%%%%%%

%%%%%%%%%%%%%%%%%%%%%%%%
{\bf 1. Non-degenerated case}
%%%%%%%%%%%%%%%%%%%%%%%%

\noindent
The curve should be
\bea
y^2= P^2_4(x)-4x^2\Lambda^{4}(x^2-m^2)=x^2 F_6(x).
\nonu
\eea
It is easy to solve and it turns out 
$P_4(x)=x^2(x^2-s)$ and $F_6(x)=x^2[(x^2-s)^2-
4 \Lambda^{4}(x^2-m^2)]$. To determine the parameter $s$, by noticing that
for $N_f=1$, $F_6(x)=W'(x)^2+a x^2 +b$, we get $s=m^2$. 
Here we found only {\it one} solution
for $SO(5)\rightarrow SO(3)\times \widehat{U(1)}$ where $SO(3)$ is at
the Chebyshev branch with a factor $x^2$ in the curve. Another one
will be given from the Special vacua  of $SO(3)$ in the degenerated case.

%%%%%%%%%%%%%%%%%%%%%%%%%
{\bf 2. Degenerated case}
%%%%%%%%%%%%%%%%%%%%%%%%%%%

\noindent
For the degenerated case where D5-branes wrap around only one root,
the curve should be the form with a factor $F_4(x)$
\bea
y^2= P^2_4(x)-4x^2\Lambda^{4}(x^2-m^2)=F_4(x) H^2_2(x).
\nonu
\eea
Parameterizing as $P_4(x)=(x^4-s_1 x^2+s_2)$,  $H_2(x)=(x^2+a)$, and
$F_4(x)=(x^2-b)^2+c$, we found the following solutions
\bea
s_1 & = & -a+b,~~~~s_2={-2ab+c+4\La^4 \over 2},~~~c={4\La^4 (a-b+m^4)
\over a+b}, \nonu \\
0 & = & (a^2+a m^2) ( (a+b)^2-4\La^4)-\La^4 m^4.
\nonu
\eea
Using the relationship between $F_4(x)$ and $(x^2-m^2)$, we
get $b=m^2$. Putting it into the last equation above we 
get
\bea \label{SO(5)-m-1}
0 & = &(a+m^2)^3 a- 4\La^4 a( a+m^2)-\La^4 m^4.
\eea
This equation (\ref{SO(5)-m-1}) has three solutions of $m^2$. If $\La\to 0$,
but $a\to \mbox{constant}$, 
keeping the first and third terms we have all
three $m^2$ going to $-a$ and $(a+m)^2\sim \La^{4/3}$. This
limit gives $SO(5)\to \widehat{U(2)}$. If $\La\to 0$ and
$a\to 0$ but $a\sim \La^4$, keeping the first and 
third terms we get 
one $m^2\to \La^4/a\neq 0$ (another two solutions of $m^2$ go to 
$0$). This limit gives 
$SO(5)\to SO(3)\times \widehat{U(1)}$ where $SO(3)$ is at the
Special branch. From this discussion we see
that at least one solution of $m^2$ gives a smooth transition 
between these two phases.

To count the vacua, notice that for fixed $m^2$ we have four
solutions $a$ from  (\ref{SO(5)-m-1}). Keeping the first and
third terms we see that there are three solutions $a\to -m^2$ and
one solution $a\to 0$; i.e., there are three vacua for
$SO(5)\to \widehat{U(2)}$ by counting $2N-N_f=4-1=3$
and one vacuum for
$SO(5)\to SO(3)\times \widehat{U(1)}$ where $SO(3)$ is at the
Special branch.

%%%%%%%%%%%%%%%%%%%%%%%%%
\subsubsection{$N_f=2$}
%%%%%%%%%%%%%%%%%%%%%%%

In this case we have the following situations. For $SO(5)\rightarrow SO(3)
\times \widehat{U(1)}$, we can have the $r=0$ non-baryonic branch 
with two vacua and
the $r=1$ baryonic branch with two vacua. For $SO(5)\to SO(5)$,
there are three vacua, while for  $SO(5) \to \widehat{U(2)}$
we can have the $r=0$ non-baryonic branch with two vacua, the $r=0$ baryonic
branch with one vacuum, and the $r=1$ non-baryonic branch with one vacuum.
We will study these vacuum structures in detail.

%%%%%%%%%%%%%%%%%%%%%%%%%%%%%%%%%%%%%%%%%
%{\bf The case of $SO(5)\rightarrow SO(3)
%\times \widehat{U(1)}$}
%%%%%%%%%%%%%%%%%%%%%%%%%%%%

%%%%%%%%%%%%%%%%%%%%%%%%%%%%%%%%%%%%%%%%%
$\bullet$ {\bf  Baryonic $r=1$ branch}
%%%%%%%%%%%%%%%%%%%%%%%%%%%%%%%%%%%%%%

\noindent
As previously discussed, on $r=1$ branch, the Seiberg-Witten curve is 
factorized as follows:
\bea
y^2=\left(x^2-m^2 \right)^2 \left[P^2_2(x)-4x^2\Lambda^2 \right].
\nonu
\eea
In addition to this factor, we have the following factorization problems,
\begin{eqnarray}
P^2_2(x)-4x^2\Lambda^2=x^2F_2(x).
\nonu
\end{eqnarray}
The solution for this equation is easily obtained. $P_2(x)=x^2$
because this characteristic function should have an $x^2$ factor in order 
for both sides to be consistent with each other. 
Thus, we have $P_4(x)=x^2\left(x^2-m^2 \right)$, which means that 
the gauge group $SO(5)$ breaks into $SO(3)\times \widehat{U(1)}$. 
For fixed $m$, there is only one vacuum which comes from the 
Chebyshev point with a factor $x^2$ in the curve. To get another vacuum,
we must choose $P_2(x)=(x^2+\Lambda^2)$ , so that
\bea
P^2_2(x)-4x^2\Lambda^2=(x^2-\Lambda^2)^2.
\nonu
\eea
This vacuum comes from the Special point. Thus, we get all of two vacua
for $r=1$ baryonic branch.

%%%%%%%%%%%%%%%%%%%%%%%%%%%%%%%%%%%%%%%%%
$\bullet$ {\bf  Non-baryonic $r=1$ branch}
%%%%%%%%%%%%%%%%%%%%%%%%%%%%%%%%%%%%%%

\noindent
In this case, we need use the relationship of 
$F_4(x)=P_2(x)^2-4 x^2 \La^2$ and $(x^2-m^2)^2$ so that
$P_2(x)=(x^2-m^2)$. There is only one solution which gives
the one vacuum of $SO(5)\to \widehat{U(2)}$ at the $r=1$ non-baryonic
branch.

%%%%%%%%%%%%%%%%%%%%%%%%%%%%%%%%%%%%%%%%%%%%
$\bullet$ {\bf  Non-baryonic $r=0$ branch}
%%%%%%%%%%%%%%%%%%%%%%%%%%%%%%%%%%%%%%%%%%%%%%

%%%%%%%%%%%%%%%%%%%%%%%%
{\bf 1. Non-degenerated case}
%%%%%%%%%%%%%%%%%%%%%%%%

\noindent
In this case, from (\ref{42swso2n+1nb}) we have the following 
factorization problem,
\bea
P^2_4(x)-4x^2\Lambda^2 (x^2-m^2)^2=x^2 F_6(x).
\nonu
\eea
To satisfy this equation we must have $x^2$ factor in $P_4(x)$.
 Thus,  assuming that $P_4(x)=x^2(x^2-a^2)$, 
we can obtain $F_6(x)$ as follows:
\bea
F_6(x)=x^2(x^2-a^2)^2-4\Lambda^2(x^2-m^2)^2.
\nonu
\eea
At this point we want to determine $W^{\prime}(x)$ from $F_6(x)$. 
As already discussed above, we should take into account  the 
second term of (\ref{newmatrix}) 
\footnote{In this case, we are 
considering the tree level superpotential (\ref{extree}), thus 
$g_{2n+2}$ is $1$. In addition, since the equation (\ref{newmatrix}) 
is the result for $SO(2N_c)$ theory, we have to extend the result 
to $SO(2N_c+1)$ theories by replacing $x^2\Lambda^{4N_c-2N_f-4}$ in 
the second term to $\Lambda^{4N_c-2N_f-2}$.}.
\begin{eqnarray}
F_6(x)+4\Lambda^2 x^2 \left(x^2-m^2 \right)=W^{\prime}(x)^2+{\cal O}(x^2).
\nonu
\end{eqnarray}
From this equation, we can read off  $W^{\prime}(x)=x^2(x^2-a^2)$. 
Thus, according to  the condition $W^{\prime}(\pm m)=0$, we identify 
$a^2$ with $m^2$. Finally the characteristic function becomes
\begin{eqnarray}
P_4(x)=x^2\left(x^2-m^2 \right),
\nonu
\end{eqnarray}
which means that the gauge group $SO(5)$ breaks into $SO(3)\times 
\widehat{U(1)}$ where $SO(3)$ is at the Chebyshev branch with a factor 
$x^2$ in the curve. 
For fixed $m$, there is only one vacuum. 
We will find another vacuum coming from the degenerated case.

%%%%%%%%%%%%%%%%%%%%%%%%
{\bf 2. Degenerated case}
%%%%%%%%%%%%%%%%%%%%%%%%%%%

\noindent
Now let us consider the degenerated case with the curve (note the factor
$F_4(x)$),
\bea
y^2=P^2_4(x)-4x^2\Lambda^2 (x^2-m^2)^2= F_4(x) H^2_2(x).
\nonu
\eea
Writing $P_4(x)=(x^4-s_1 x^2+s_2)$,  $H_2(x)=(x^2+a)$ and
$F_4(x)=x^4+b x^2+c$ we found 
\bea
s_1 & = & \frac{-2 a - b - 4 {\La}^2}{2}, \nonu \\
s_2 & = & \frac{4 c + 4 a \left( b - 4 {\La}^2 \right)
        - \left( b + 4 {\La}^2 \right)^2 - 
    32 {\La}^2 m^2}{8}, \nonu \\
c & = & -\left( \frac{\frac{32 a^2 {\La}^2 - 
         2 a \left( b - 12 {\La}^2 \right) 
          \left( b + 4 {\La}^2 \right)  + 
         \left( b + 4 {\La}^2 \right)^3}{8}
       + 4{\La}^2
       \left( 2 a + b + 4 {\La}^2 \right) 
       m^2 + 4 {\La}^2 m^4}{a - \frac{b}{2} - 
      2 {\La}^2} \right), \nonu \\
0 & = & a \left( 4 a^2 - 4 a
      \left( b - 5 {\La}^2 \right)  + 
     \left( b + 4 {\La}^2 \right)^2 \right) 
   + 24 a {\La}^2 m^2 + 
  4 {\La}^2 m^4.
\nonu
\eea
Putting $b=-2m^2-4\La^2$ we can solve
\footnote{In this case,
$c=-4 a \La^2+4 \La^2 m^2+m^4$ is very simple compared with
the above messy expression. }
\bea
0 = 4 a^2 \left( a + 9 {\La}^2 \right)  + 
  8 a \left( a + 3 {\La}^2 \right) m^2 + 
  4 \left( a + {\La}^2 \right) m^4 \Longrightarrow
m^2 & = & {-a^{3/2} \pm 3i a \La \over \sqrt{a}\mp i\La}.
\nonu 
\eea
At the limit $\La\to 0$ and $a\to \mbox{constant}$, 
we get both $m^2\to -a$
and $SO(5)\to \widehat{U(2)}$. At the limit  $\La\to 0$, but
$\sqrt{a}\to i\La+ \alpha \La^3$, we have $m^2\neq 0$ and
$SO(5)\to SO(3) \times \widehat{U(1)}$
where $SO(3)$ is at the Special branch. Thus, there is smooth
transition between these two phases.
To count the vacua, notice that for fixed $m^2$, we have
three solutions for $a$. Setting $\La=0$, the equation 
becomes $4 a (a+m^2)^2=0$. Thus we have two vacua for 
 $\widehat{U(2)}$ and one vacuum for $SO(3) \times \widehat{U(1)}$.

Next we discuss the case in which 
all D5-branes wrap around the origin $x=0$; 
thus,  $F_4(x)$ should have a
factor $x^2$. Writing $P_2(t)=t(t-s_1)$, $H_1(t)=t+a$ and 
$F_2(t)=t(t+b)$ (where we have used $t=x^2$ for simplicity), we 
get $s_1=-a-2\La^2+{2\La^2 m^4 \over a^2}$, $b=-{4\La^2 m^4 \over a^2}$
and 
\bean
0= a^2 \left( a + {\La}^2 \right)  - 
  2 a {\La}^2 m^2 + {\La}^2 m^4,~~~
\mbox{or}~~ m^2=\pm i {a^{3/2}\over \La}+a.
\eean
There is only one limit $a\sim \La^{2/3}$ to get finite $m^2$. For fixed
$m^2$, there are three solutions of $a$ which give the 
three wanted vacua of $SO(5)\to SO(5)$ where
$SO(5)$ is at the Chebyshev branch with a factor $x^2=t$ 
in the curve.

Finally we want to get the $r=0$ baryonic branch of $\widehat{U(2)}$.
To achieve this, we require  $y^2$ to be square. 
See also (\ref{42swso2n+1b--2}).
There are
two solutions. One of them is $P_2(t)=(t-m^2)(t+\La^2)$ (for
simplicity we have used $t=x^2$) which 
gives the $r=1$ baryonic branch of $SO(3)\times \widehat{U(1)}$.
Another is $P_2(t)=(t-m^2)^2+\La^2 t$ which is the one we want.
It is easy to see that the latter solution gives
$y^2=[(t-m^2)^2-\La^2 t]^2\neq H_1(t)^2 F_1(t)^2$, which explains why
we could not get it in the previous paragraph. In other words,
for the baryonic branch, the relationship 
$F_4(x)+4 \Lambda^2 x^2= ({W'(x) \over x})^2+
{\cal O}(x^0)$ fails.

Now we summarize the results in Table \ref{tableso5}
by specifying the flavors $N_f$, symmetry breaking patterns, 
various branches, the exponent of $t=x^2$ in the curve, $U(1)$ at the
IR, the number of vacua, and the possibility  of smooth transition.
It turned out that the number of vacua is exactly the same 
as from the 
weak coupling analysis.
\begin{table}
\begin{center}
\begin{tabular}{|c|c|c|c|c|c|c|} 
\hline
 $N_f$  &  Group  & Branch &  Power of $t(=x^2)$ &  $U(1)$ & 
Number of vacua & Connection
 \nonu \\ 
\hline
 0 & $SO(3)\times U(1)$ & $(C,0_{NB})$ & $t^1$ & 1 &  1  & 
\nonu \\ \cline{3-7} 
& &  $(S,0_{NB})$ & $t^0$ & 1 &  1  & A \nonu \\ \cline{2-7}  
 & $SO(5)$  & $(C)$ &  $t^1$ & 0 & 3 & \nonu  \\\cline{2-7}
& $U(2)$ & $(0_{NB})$ & $t^0$ & 1 & 2 & A \nonu \\ \hline 
1 &  $SO(3)\times \widehat{U(1)}$ & $(C,0_{NB})$ & $t^1$ & 1 &  1  & 
\nonu \\ \cline{3-7}
& & $(S,0_{NB})$ & $t^0$ & 1 &  1  & B \nonu \\ \cline{3-7}
& & $(C,0_{B})$ & $t^1$ & 0 &  1  & D \nonu \\ \cline{3-7}
& & $(S,0_{B})$ & $t^0$ & 0 &  1  & \nonu \\ \cline{2-7}
  & $SO(5)$  & $(C)$ &  $t^1$ & 0 & 3 & D \nonu \\ \cline{2-7}
& $\widehat{U(2)}$ & $(0_{NB})$ & $t^0$ & 1 & 3 & B \nonu \\ \hline 
2 &  $SO(3)\times \widehat{U(1)}$ & $(C,0_{NB})$ & $t^1$ & 1 &  1  & 
\nonu \\ \cline{3-7}    
& & $(S,0_{NB})$ & $t^0$ & 1 &  1  & C \nonu \\ \cline{3-7}
& & $(C,1_{B})$ & $t^1$ & 0 &  1  & \nonu \\ \cline{3-7}
& & $(S,1_{B})$ & $t^0$ & 0 &  1  & \nonu \\ \cline{2-7}
 & $SO(5)$  & $(C)$ &  $t^1$ & 0 & 3 & \nonu \\ \cline{2-7}
& $\widehat{U(2)}$ & $(0_{NB})$ & $t^0$ & 1 & 2 & C \nonu \\ \cline{3-7}
& &  $(0_{B})$ & $t^0$ & 0 & 1 &  \nonu \\ \cline{3-7}
& &  $(1_{NB})$ & $t^0$ & 1 & 1 & \nonu \\ \hline 
\end{tabular}
\end{center} 
\caption{\sl The summary of the phase structure of $SO(5)$ gauge group. 
There are three pairs of smoothly connected phases indicated by
the letters $A,B,C$. That is, the two branches denoted by $A, B$, or $C$
are smoothly connected with each other.
 For $N_f=2$, since we can have $r=0,1$, the $r$ is also
an index for phases and should be preserved under the smooth transition. 
One application is that 
although for $N_f=2$, $SO(3)\times \widehat{U(1)}$ at $(S,1_{B})$
and $\widehat{U(2)}$ at $(0_{B})$ have the same $t^0$ and zero $U(1)$,
they can {\it not} be smoothly connected because the index 
$r$ is different.}
\label{tableso5}
\end{table} 

%%%%%%%%%%%%%%%%%%%%%%%%%%%%%%%%%%%%%%%%%%%
\subsection{SO(6) case}
%%%%%%%%%%%%%%%%%%%%%%%%%%%%%%%%%%%%%%%%%%%%

\indent 

The next example is $SO(6)$ gauge theory, which is more interesting. 
For this gauge theory we will discuss the number of flavors 
$N_f=0,1,2,3$  cases. 
There are three breaking patterns $SO(6)\to SO(4)\times 
\widehat{U(1)}$, $SO(6)\to SO(2)\times \widehat{U(2)}$,
$SO(6)\to SO(6)$, and 
$SO(6)\to \widehat{U(3)}$. 

%%%%%%%%%%%%%%%%%%%%%%%
\subsubsection{$N_f=0$}
%%%%%%%%%%%%%%%%%%%%%%

\noindent

%%%%%%%%%%%%%%%%%%%%%%%%
{\bf 1. Non-degenerated case}
%%%%%%%%%%%%%%%%%%%%%%%%

\indent

In \cite{ao} the theory with $N_f=0$  was discussed intensively. For the 
breaking pattern $SO(6)\to SO(2)\times U(2)$ where $SO(2)$ is at
the Special branch, there are two vacua 
which are confining vacua. These confining vacua are constructed from 
the Coulomb branch with the breaking pattern $SO(4)\to SO(2)\times U(1)$ 
by the multiplication map. In addition to these vacua, there are two vacua
with the breaking pattern $SO(6)\to SO(4)\times U(1)$. These
two vacua come from the Chebyshev branch of $SO(4)$ factor, and
we will find another two vacua from the Special branch of $SO(4)$
in the degenerated case.

%%%%%%%%%%%%%%%%%%%%%%%%
{\bf 2. Degenerated case}
%%%%%%%%%%%%%%%%%%%

\indent

To see all the vacua including the one with the breaking pattern 
$SO(6)\to U(3)$, it is necessary to 
consider the degenerated case. The 
factorization problem becomes 
\bea
P^2_6(x)-4x^4\Lambda^{8}=H^2_4(x) F_4(x).
\nonu
\eea
After solving this factorization problem, we obtain the following 
solutions,
\bea
P_6(x)&=&x^6-\epsilon \frac{3G^4+\Lambda^8}{G\Lambda^4}x^4+
\left(G^2-\frac{3G^6}{\Lambda^8} \right)x^2+\epsilon 
\frac{G(G^4-\Lambda^8)^2}{\Lambda^{12}}, \nonu \\
F_4(x)&=&\left(x^2-\epsilon \frac{G^4+\Lambda^8}{G\Lambda^4} 
\right)^2+4G^2, 
\label{423sol1}
\eea
where $\epsilon^2 \equiv -1$. Note that this use of $\epsilon$ 
is slightly different from previous one. If we remember the relation 
$F_4(x)\equiv \frac{W^{\prime}(x)^2}{x^2}+d$, we can see the relation,
\bea
\Delta=\epsilon \frac{G^4+\Lambda^8}{G\Lambda^4},\qquad \mbox{where}
\qquad  W^{\prime}(x)\equiv x\left(x^2-\Delta \right). 
\label{423cond1}
\eea
In this case we can take two semiclassical limits with $\Lambda\to 0$:

1. $G\to 0$ with fixed $\frac{\epsilon G^3}{\Lambda^4}\equiv v$: 
Under these limits, since the characteristic function goes to $P_6(x)
\to \left(x^2-v \right)^3$, the gauge group $SO(6)$ breaks into $U(3)$. 
From the condition (\ref{423cond1}), $\Delta =\epsilon 
\frac{G^3}{\Lambda^4}$, we obtain three values for $\epsilon G=
\left(-\Delta \Lambda^4 \right)^{\frac{1}{3}}$. Thus substituting 
(\ref{423sol1}), we can see three vacua, which agree with the number 
obtained by weak coupling analysis, 
because pure $U(3)$ theory has three vacua.

2. $G\to 0$ with fixed $\frac{\epsilon \Lambda^4}{G}\equiv w$: 
Under this limit since the characteristic function behaves as $P_6(x)
\to x^4(x^2-w)$, the gauge group $SO(6)$ breaks into $SO(4)\times U(1)$.
From condition (\ref{423cond1}), $\Delta=\epsilon \frac{\Lambda^4}{G}$,
we obtain two values for $G$. These two vacua of  $SO(4)\times U(1)$
are the missing two vacua mentioned in the non-degenerated case.
Adding all the contributions from the 
degenerated case and nondegenerated case, 
the result  agrees with the one from the 
weak coupling analysis, because the Witten index for pure $SO(4)$ 
gauge theory has four vacua.

Notice that from the above analysis we get a smooth transition between
$U(3)$ and $SO(4)\times U(1)$ at the Special branch. The reason 
is that the power of $t=x^2$ of $SO(4)$ at  the Special branch
is $(2N_f-N+4)=0$, which is the same as the case of $U(3)$. Similar
phenomena have been observed in the previous subsection of the $SO(5)$ 
gauge group.

Finally, to get the breaking $SO(6)\to SO(6)$ (see also 
(\ref{example-7})), we require a $x^2$
factor in $F_4(x)$ which can be obtained when $G^4-\Lambda^8=0$. There
are four solutions for $G$. Combining the $\epsilon$, it seems
we will get eight solutions. However,  careful study shows that there
are, in fact, only four solutions which match the counting 
$(N-2)=4$ of the Witten index for $SO(6)$ gauge theory where
$SO(6)$ is at the Chebyshev branch.

%%%%%%%%%%%%%%%%%%%%%%%
\subsubsection{$N_f=1$}
%%%%%%%%%%%%%%%%%%%%%%%%

\indent

In this example, there is  only the $r=0$ branch. For the breaking 
pattern $SO(6)\to SO(4)\times \widehat{U(1)}$, this $r=0$ branch 
can be non-baryonic or baryonic due to the fact $r=N_1$ 
while for the breaking pattern 
$SO(6)\to SO(2)\times \widehat{U(2)}$, this branch is non-baryonic
because both relations $r \neq N_f-N_1$ and $r \neq N_1$ hold in this case. 
In addition, for the breaking pattern $SO(6)\to \widehat{U(3)}$ it 
is a non-baryonic branch because  $r \neq N_f-N_1$ and $r \neq N_1$. 
For $SO(6)\to SO(6)$, only the Chebyshev branch exists with four
vacua.

%%%%%%%%%%%%%%%%%%%%%%%%%%%%%%%%%%%%%%%%%%%%5
$\bullet$ {\bf  Non-baryonic $r=0$ branch}
%%%%%%%%%%%%%%%%%%%%%%%%%%%%%%%%%%%%%%%%%%5

%%%%%%%%%%%%%%%%%%%%%%%%
{\bf 1. Non-degenerated case}
%%%%%%%%%%%%%%%%%%%%%%%%

\indent

Since this branch is non-baryonic  we have the following 
factorization problem,
\bea
P^2_{6}(x)-4x^4\Lambda^{6}(x^2-m^2)=x^2H^2_{2}(x)F_6(x).
\nonu
\eea
To satisfy this equation, the characteristic function must have 
an $x^2$ factor, $P_6(x)=x^2P_4(x)$. Then one of $H_2(x)$ and $F_6(x)$
on the right hand side 
must have a factor $x^2$. If the factor $x^2$ is contained in the function
$F_6(x)$, it
belongs, in fact, to the degenerated case, so we consider the
case $H_2(x)=x^2$. 
Parameterizing $P_6(x)=x^2(x^4-s_1 x^2+s_2)$ and  $F_6(x)=x^6+a x^4+b 
x^2+c$, we found the solutions:
\bean
s_1=-{a\over 2}, \qquad s_2=\mp 2i \Lambda^3 m, \qquad
b={a^2\over 4}\mp 
4i \Lambda^3 m, \qquad c=-4 \La^6\mp 2i a\Lambda^3 m.
\nonu
\eean 
Obviously the classical limit is $SO(6)\to SO(4)\times
\widehat{U(1)}$. By the relationship of $W'(x)$ and $F_6(x)$ we
find $a=-2m^2$, so there are two vacua of $ SO(4)\times\widehat{U(1)}$
which come from the Chebyshev branch of the $SO(4)$ factor where there is a
factor $x^6$ in the curve.

%%%%%%%%%%%%%%%%%%%%%%%%%
{\bf 2. Degenerated case}
%%%%%%%%%%%%%%%%%%%%

\indent

Next we go to the degenerated case. The factorization problem of 
this case is given by
\bea
P^2_6(x)-4x^4\Lambda^6(x^2-m^2)=H^2_4(x) F_4(x).
\nonu
\eea
Parameterizing $P_3(t)=t^3-s_1 t^2+s_2 t+s_3$, $H_2(t)=t^2-at+b$,
$F_2(t)=t^2-2 m^2 t+m^4 +c$ where we have used the relationship
between $F_2(t)$ and $(t-m^2)^2$ to simplify 
the calculation \footnote{
The logic of the calculation is following. First we 
factorize the curve with some parameters. Then we establish the
relationship between these parameters and $m^2$. This is the
method used in $SO(4)$ and $SO(5)$. However, we can use the
relationship of $m^2$ and $F_4(x)$ to solve some parameters
firstly, then to solve the factorization form. The latter method
is more convenient for our purpose.}.
The solution is
\bea
b & = & \frac{c}{4} - \frac{4 a {\La}^6}{c} + 
  a m^2 - m^4, \nonu \\
a & = & \frac{2 c \left( 4 c {\La}^6
       \left( c^3 - 64 {\La}^{12} \right)  + 
      c^2 \left( c^3 + 64 {\La}^{12} \right)
         m^2 + 8 {\La}^6
       \left( 3 c^3 + 64 {\La}^{12} \right) 
       m^4 \right) }{\left( c^3 + 
      64 {\La}^{12} \right) 
    \left( c^3 - 64 {\La}^{12} + 
      32 c {\La}^6 m^2 \right) }, \nonu \\
0 & = &
  \left( \left( c^3 - 64 {\La}^{12} \right)^
     2 - 1024 c^2 {\La}^{12} m^4 - 
    1024 c {\La}^{12} m^8 \right). 
\label{blabla}
\eea
From the last equation in (\ref{blabla}), 
we see that at $\La\to 0$ we must have
$c\to 0$. Therefore the solution of $m^2$ is roughly
\bea
m^8\sim {{\left( c^3 - 64 {\La}^{12} \right) }^
     2 \over  1024 c {\La}^{12}}.
\nonu
\eea
The first limit is that $c\sim \La^{2.4}$, so that all  
four solutions provide $m^2\neq 0$
and $a\sim 2m^2$ and $b\sim m^4$. This limit gives 
$SO(6)\to \widehat{U(3)}$. The second limit is that 
$c\sim \alpha \La^{12}$, so that all four solutions give
  $m^2\neq 0$,
$a\sim {-\alpha \over 4} \La^6\sim 0$, $b\sim 0$. This limit gives
$SO(6)\to SO(4)\times \widehat{U(1)}$ where $SO(4)$ is at the
Special branch. Thus, we have a smooth transition between these
two phases.

To count the vacua, for fixed $m^2$ there are six solutions of $c$.
Keeping the first and third terms of (\ref{blabla}) 
we have five $c\sim \La^{2.4}$,
which is consistent with the counting $(2N-N_f)=6-1=5$ of 
$\widehat{U(3)}$. Keeping the second and third terms of (\ref{blabla})
we have one solution of $c\sim \La^{12}$, which gives the one
vacuum of $SO(4)\times \widehat{U(1)}$ where $SO(4)$ is at the
Special branch
\footnote{Notice that although the Special branch of
$SO(4)$ should have two vacua, but the counting from the curve
gives always one. We have met this situation many times already.}.

Now we consider other special cases. The first one is that $b=0$ so
that $H_2(t)=t(t-a)$. There are three solutions: $a=-\omega_3 \La^2+
m^2$, $c=-4 \omega_3^{-1} \La^4$ with $\omega_3^3=-1$. They are the
three vacua of 
 $SO(6)\to SO(2)\times \widehat{U(2)}$ by counting number 
$(2N-N_f)=4-1=3$.

The next one is that $F_2(t)=t(t+a)$ and
$H_2(t)=t(t+b)$. The solutions are given by  
\bea \label{ttt}
b={a\over 2}+{16 \La^6 \over a^2}, \qquad
a^6 + 4096 \La^{12}+ a^2 \La^6 (128a+256 m^2)=0.
\eea
From the (\ref{ttt}) we see that $a\to 0$ in the limit $\La\to 0$.
There are two limits we can take. If $a\sim 4\La^{3/2}$,
we have $m^2\sim -1$, $b\sim 0$ which gives $SO(6)\to SO(6)$. 
 If $a\sim \La^3$, we have $m^2\sim -16$, $b\sim 16$ which gives
$SO(6)\to SO(4)\times \widehat{U(1)}$ where $SO(4)$ is at 
Chebyshev branch and $\widehat{U(1)}$ is at the $r=0$ baryonic branch.

From these considerations we see that there is a smooth transition 
between $SO(6)\to SO(6)$ and $SO(6)\to SO(4)\times \widehat{U(1)}$.
Indices for these two phases are the same: $t^3$ and non IR $U(1)$.
It is interesting to compare them with the baryonic $r=0$ branch of
$SO(5)$ with $N_f=1$, where we do not find a smooth transition
between $SO(5)\to SO(5)$ and $SO(5)\to SO(3)\times \widehat{U(1)}$.
For $SO(5)$, these two phases come from different solutions of $m^2$
while for $SO(6)$ there is only one solution for $m^2$. This
different behavior of $m^2$ as a function of parameters 
explains the different phase structure, but the physics behinds
these calculations is still unclear.

To count the vacua at fixed $m^2$,
keeping the first and the third terms 
of the second equation (\ref{ttt})
we find four solutions 
$a\sim m^{1/2}\La^{3/2}$ which give the limit
$SO(6)\to SO(6)$. Keeping the second and the third terms  
we find two solutions  $a\sim \pm 4i\La^3/m$ and $b\sim -m^2$.
These two solutions give the breaking pattern 
$SO(6)\to SO(4)\times \widehat{U(1)}$ where $SO(4)$ is at 
Chebyshev branch and $\widehat{U(1)}$ is at the $r=0$ baryonic branch.

%%%%%%%%%%%%%%%%%%%%%%%%%%%%%%%%%%%%%%
$\bullet$ {\bf  Baryonic $r=0$ branch}
%%%%%%%%%%%%%%%%%%%%%%%%%%%%%%%%%%%%%%

\noindent
Since this branch is baryonic, from the relation 
(\ref{42swso2nb}) we have the following 
factorization problem,
\begin{eqnarray}
P^2_{3}(t)-4t^2\Lambda^{6}(t-m^2)=t H^2_{2}(t) F_1(t). 
%\label{42fac}
\nonu
\end{eqnarray}
In fact, for the $SO(4)$ at the Chebyshev branch, we need
to have $H_2(t)=t(t+b)$. This is exactly the same problem 
discussed above.

To get the missing vacua, we just need to choose 
$P_3(t)=t^3-m^2 t^2+\La^6$, so that $y^2=(t^3-m^2 t^2-\La^6)^2$.
This gives the $r=0$ baryonic branch of 
$SO(6)\to SO(4)\times \widehat{U(1)}$ where $SO(4)$ factor 
is at the Special branch.

%%%%%%%%%%%%%%%%%%%%%%%
\subsubsection{$N_f=2$}
%%%%%%%%%%%%%%%%%%%%%%%

\indent

In this case, we have two branches $r=0,1$. For the breaking 
pattern $SO(6)\to SO(4)\times \widehat{U(1)}$, the $r=1$ branch is 
baryonic because $r=N_f-N_1=N_1$
while the $r=0$ branch is non-baryonic. For the breaking 
pattern $SO(6)\to SO(2)\times \widehat{U(2)}$ the $r=1$ branch is 
non-baryonic while the $r=0$ branch can be non-baryonic or baryonic because 
$r=N_f-N_1$. 
Moreover, for the breaking pattern $SO(6)\to \widehat{U(3)}$ both 
branches of $r=0,1$ are non-baryonic. 
For the breaking pattern, $SO(6)\to SO(6)$, there is only Chebyshev branch.

%%%%%%%%%%%%%%%%%%%%%%%%%%%%%%%%%%%%%%%%%
$\bullet$ {\bf  Baryonic $r=1$ branch }
%%%%%%%%%%%%%%%%%%%%%%%%%%%%%%%%%%%%%%%%%%

\noindent
%As discussed above in $r=1$ branch, 
The Seiberg-Witten curve 
is factorized as,
\bea
y^2=\left(x^2-m^2 \right)^2 \left[P^2_4(x)-4x^4\Lambda^{4} \right]. 
\label{42factsw}
%\nonu
\eea
In addition to this factor, we have the following factorization problem
with a factor $F_2(x)$,
\bea
y^2= P^2_4(x)-4x^4\Lambda^{4}=x^2 H^2_2(x) F_2(x).
\nonu
\eea
To satisfy this constraint, the characteristic function $P_4(x)$ must have 
$x^2$ factor, i.e. $P_4(x)=x^2(x^2-a^2)$. In addition, we assume that 
$H_2(x)=(x^2-b^2)$. After solving the factorization problem we obtain 
the following results, 
\bea
P_4(x)=x^2(x^2-2\eta \Lambda^2), \qquad F_2(x)=(x^2-4\eta \Lambda^2),
\nonu
\eea
where $\eta$ is $2$-nd roots of unity. Thus taking into account 
the $(x^2-m^2)$ factor, we obtain 
\begin{eqnarray}
P_6(x)=x^2(x^2-m^2)(x^2-2\eta \Lambda^2). 
%\label{9maycon3}
\nonu
\end{eqnarray}
Under the semiclassical limit, $\Lambda \to 0$, the characteristic function 
becomes $P_6(x)\to x^4 (x^2-m^2)$, which means that the gauge group 
$SO(6)$ breaks into $SO(4)\times \widehat{U(1)}$. Since there is $\eta$ 
we have two vacua for fixed $m$ 
which come from the Chebyshev branch of $SO(4)$ factor. 

In addition to this solution, we have other solutions from  the following 
case,
\begin{eqnarray}
P_4^2(x)-4x^4\Lambda^4=H_4^2(x).
\nonu
\end{eqnarray}
If we parameterize $H_4(x)=(x^2-a)(x^2-b)$, 
we have two solutions $a\to -\Lambda^2,b\to \Lambda^2$ or 
$a\to \Lambda^2, b\to -\Lambda^2$. Both cases lead to 
the conclusion that $P_4(x)$ becomes $P_4(x)=x^4+\Lambda^4$. 
As above, these solutions also exhibit the 
breaking pattern $SO(6)\to SO(4)\times \widehat{U(1)}$. 
However, these come from the Special branch of $SO(4)$. 
Therefore, we have four solutions from the strong coupling analysis. 
This number agrees with the one 
from the previous analysis of the weak coupling. 
Although $SO(4)$ gauge theory has 
four vacua, $\widehat{U(1)}$ with $N_f=2$ gauge theory has 
only one vacuum.

%%%%%%%%%%%%%%%%%%%%%%%%%%%%%%%%%%%%%%%%%%%%%
$\bullet$ {\bf  Non-baryonic $r=1$ branch }
%%%%%%%%%%%%%%%%%%%%%%%%%%%%%%%%%%%%%%%%%

%%%%%%%%%%%%%%%%%%%%%%%%
{\bf 1. Non-degenerated case}
%%%%%%%%%%%%%%%%%%%%%%%%

\indent

As in the previous case, we can factorize the Seiberg-Witten curve 
as (\ref{42factsw}). In this case, since the branch is non-baryonic, 
we have different factorization problem,
\bea
y^2= P^2_4(x)-4x^4\Lambda^{4}=x^2F_6(x)
\nonu
\eea
in which the factor $(x^2-m^2)^2$ has been factorized out. 
To solve this equation, we assume that $P_4(x)=x^2(x^2-a^2)$. 
After inserting this relation, we find that $F_6(x)$ must have 
an $x^2$ factor. However, this implies the degenerated case. 
Therefore, we have no new solution for this non-degenerated case.

%%%%%%%%%%%%%%%%%%%%%%%%%%%%%%
{\bf 2. Degenerated case}
%%%%%%%%%%%%%%%%%%%%%%%%%%%

\noindent
Next we will consider 
the degenerated case. The factorization problem 
 is given by 
\bea
P^2_4(x)-4x^4\Lambda^4=H^2_2(x) F_4(x).
\nonu
\eea
Solving this factorization problem, 
we obtain two kinds of solutions. One is given by
\bea
P_4(x)&=&\left(x^2-a \right)^2+2\eta \Lambda^2 x^2, \nonu \\
F_4(x)&=&\left(x^2-a+2\eta \Lambda^2 \right)^2+4a 
\eta \Lambda^2-4\Lambda^4,
\nonu
\eea
where $\eta$ is the $2$-nd root of unity. Taking into account that 
$W^{\prime}(\pm m)=0$, we have one constraint,
\bea
m^2=a+2\eta \Lambda^2.
\nonu
\eea
In this case, we can take only one semiclassical limit, $\Lambda\to 0$. 
Under the limit, 
the characteristic function goes to $P_6(x)\to (x^2-m^2)^3$, 
which means that the gauge group $SO(6)$ breaks into $\widehat{U(3)}$. 
Thus, we have two vacua from the strong coupling approach. 
This number of
vacua agrees with the weak coupling analysis 
given in (\ref{vacua}). 
The $\widehat{U(3)}$ theory with $N_f=2$ and $r=1$ has 
two vacua. 

The other solutions are derived 
from $H_2(x)=x^2$. 
Inserting $P_4(x)=x^2(x^2-a^2)$ into the factorization problem, we obtain 
\bea
F_4(x)=(x^2-a^2)^2-4\Lambda^4 \equiv 
\left(\frac{W^{\prime}(x)}{x} \right)^2+d.
\nonu
\eea
The condition $W^{\prime}(\pm m)=0$ leads to the solution $a^2=m^2$. 
After all, we 
obtain the trivial solution,
\bea
P_6(x)=x^2(x^2-m^2)^2.
\nonu
\eea
The gauge group $SO(6)$ breaks into $SO(2)\times \widehat{U(2)}$. 
Thus, we obtained only one vacuum from the strong coupling analysis 
which comes from the Special branch of $SO(2)$. This 
number matches the one from the weak coupling analysis. The 
$\widehat{U(2)}$ 
with $N_f=2$ gauge theory has only one vacuum from (\ref{vacua}). 
Since $SO(2)$ theory has only one vacuum, the total number of vacua is 
only one, which agrees with the results of the strong coupling analysis.

%%%%%%%%%%%%%%%%%%%%%%%%%%%%%%%%%%%%%%%%%%%%%%
$\bullet$ {\bf  Non-baryonic $r=0$ branch }
%%%%%%%%%%%%%%%%%%%%%%%%%%%%%%%%%%%%%%%%%%%%%%

%%%%%%%%%%%%%%%%%%%%%%%%
{\bf 1. Non-degenerated case}
%%%%%%%%%%%%%%%%%%%%%%%%

\noindent
Next we consider a non-baryonic $r=0$ branch. From (\ref{42swso2nnb}) we 
have the following factorization problem,
\begin{eqnarray}
P^2_6(x)-4x^4 \Lambda^4(x^2-m^2)^2=x^2 H^2_2(x) F_6(x). 
\label{42fac6nb}
%\nonu
\end{eqnarray}
Assuming that $P_6(x)=x^2P_4(x)$, we can rewrite (\ref{42fac6nb}) as 
follows:
\bea
x^2\left(P_4(x)-2\Lambda^2(x^2-m^2) \right) \left(P_4(x)+2
\Lambda^2(x^2-m^2) \right)=H^2_2(x) F_6(x).
\nonu
\eea
If we include the additional factor  $x^2$  in a function $F_6(x)$ 
it goes to the degenerated case. 
Here we want to discuss the non-degenerated case so we 
consider the $H_2(x)=x^2$ case. The solution is given by
\bea
P_6(x)&=&x^2 \left[x^4+Ax^2-2\eta \Lambda^2 m^2 \right], \nonu \\
F_6(x)&=&x^2 \left(x^2+A \right)^2-4\left(\Lambda^4+\eta \Lambda^2 
m^2 \right)\left(x^2+\frac{A-2\eta \Lambda^2}{m^2+\eta \Lambda^2}
m^2 \right),
\nonu
\eea
where $\eta$ is the $2$-nd root of unity. Taking into account that
$W^{\prime}(\pm m)=0$ we have the condition $m^2=-A$. 
Under the semiclassical 
limit, $\Lambda \to 0$, the characteristic function goes to 
$P_6(x)\to x^4(x^2-m^2)$, 
which means that the gauge group $SO(6)$ breaks 
into $SO(4)\times \widehat{U(1)}$. 
These vacua come from the Chebyshev branch of the $SO(4)$ factor.
%What about the counting of vacua?

%%%%%%%%%%%%%%%%%%%%%%%%%%
{\bf 2. Degenerated case}
%%%%%%%%%%%%%%%%%%%%%%%%%%

Now we consider  the degenerated $r=0$ branch. Using $t=x^2$,
the curve should be factorized as
\bea
P^2_3(t) - 4\Lambda^{4} t^2 (t-m^2)^2= H^2_2(t) F_2(t).
\nonu
\eea
We have two cases to be discussed for this $r=0$ branch. The first
case is $H_2(t)=tH_1(t)$ which constrains to $P_3(t)=tP_2(t)$. Then we
have the following solution
\bean
P_2(t)-2 \eta \Lambda^{2}(t-m^2) &= &H^2_1(t)=(t-a)^2, \\
P_2(t)+2 \eta \Lambda^{2}(t-m^2) &= & (t-a)^2+ 4 \eta 
\Lambda^{2}(t-m^2),
\nonu
\eean
with $\eta^2=1$. There are two situations we need to consider for
this kind of solution. The first is that there are D5-branes wrapping
around the $x=\pm m$; then we have
$$
a=m-2 \eta \Lambda^{2},
$$
by the relationship of $F_2(t)$ and $(t-m^2)^2$. It gives
$SO(6)\to SO(2)\times \widehat{U(2)}$ at the classical
limit where $SO(2)$ is at the Special branch. There are 
two vacua corresponding to $\eta=\pm 1$, which is consistent with
the counting that $U(2)$ with two flavors at $r=0$ non-baryonic
branch has two vacua. The second situation is that $F_2(t)$ has
a further factor $t$  which leads to
$$
a^2-4\eta \Lambda^{2} m^2=0~.
$$
It gives $SO(6)\to SO(6)$ at the classical limits. There are
four  solutions, which is consistent with the counting that
$SO(6)$ at the Chebyshev branch has a Witten index $(N-2)=4$. 
Notice that because of the second situation, $F_2(t)$ has a
factor $t$, and the first situation is not smoothly connected to the
second situation.

Having considered the case that $H_2(t)=tH_1(t)$, we move to
the case that there is no factor $t$ in the curve $y^2$. Then
$P_3(t)-2\eta  \Lambda^{2}t(t-m^2)$ has only one double root for
every $\eta$, or
\bea
P_3(t)+ 2\Lambda^2 t(t-m^2) &= & (t-a-b)^2(t+c), \nonu \\
P_3(t)- 2\Lambda^2 t(t-m^2) &= & (t-a+b)^2(t+d), \nonu 
\eea
where
\bea
c & = & 2b+2\Lambda^2-{a(b+2\Lambda^2)\over b}+{\Lambda^2 m^2 \over b},
\nonu \\
d & = & -2b-2\Lambda^2-{a(b+2\Lambda^2)\over b}+{\Lambda^2 m^2 \over b},
\nonu \\
0 & = & b^2(b+\Lambda^2)+a \Lambda^2(m^2-a).
\nonu 
\eea
Using the relationship of $F_2(t)=(t+c)(t+d)$ and $(t-m^2)$,
we get
\bea 
{ 2ab (b+2\Lambda^2)-2 b \Lambda^2 m^2 \over b^2}=2m^2~, 
\nonu
\eea
from which we can solve 
\bea
a={ (b+\Lambda^2) m^2 \over b+2\Lambda^2}, \qquad
c=2(b+\Lambda^2)-m^2,~~~d=-2(b+\Lambda^2)-m^2
\nonu
\eea
and
\bea 
\label{SO6-Nf=2-r=0}
\left(b+\Lambda^2 \right) 
\left(b^2+{\Lambda^4 m^4 \over (b+2\Lambda^2)^2} \right)=0.
\eea
Forgetting about 
the factor $\left(b+\Lambda^2 \right)$ we found that
there is only one limit $b\sim \La$ to get finite $m^2$. In this
limit  $m^4\sim -1$,
$a\sim m^2$, $c,d\sim -m^2$, so it  gives $SO(6)\to \widehat{U(3)}$.
For fixed $m^2$, there are four solutions of $b$ which give
the four vacua of $\widehat{U(3)}$ by counting $(2N-N_f)=6-2=4$.

However, where is the breaking $SO(6)\to SO(4)\times  \widehat{U(1)}$?
It comes from $\left(b+\Lambda^2 \right)=0$, so that $a=0$ and $c=d=-m^2$.
In fact, the  curve has the same form as the curve of the $r=1$ branch.
Similar phenomena have been
observed in \cite{bfhn}; for example, the $U(2)$ with $N_f=2$.
Since this non-baryonic $r=0$ branch is coincident
with the   baryonic $r=1$ branch in this particular example,
we do not have a smooth transition from $\widehat{U(3)}$
to $SO(4)\times  \widehat{U(1)}$. It is very interesting to
recall that there is smooth transition for $N_f=0,1$.

%%%%%%%%%%%%%%%%%%%%%%%%%%%%%%%%%%%%%%%%%%
$\bullet$ {\bf  Baryonic $r=0$ branch }
%%%%%%%%%%%%%%%%%%%%%%%%%%%%%%%%%%%%%%%%%%%%

For the baryonic branch, the curve should be
\bea
P^2_3(t)-4 \La^4 t^2(t-m^2)^2=H^2_3(t).
\nonu
\eea
Parameterizing $H_3(t)=t^3-a t^2+b t+c$, there are two
solutions. The first one is $a=2m^2$, $b=m^4-\La^4$, and
$c=0$ which gives $r=0$ baryonic branch of $SO(2)\times
\widehat{U(2)}$. The $SO(2)$ is at the
Special branch with a factor $t^2$ in the curve. The second solution
is $a=m^2$, $b=-\La^4$, and $c=-\La^4 m^2$ which
gives the $r=1$ baryonic branch of 
$SO(4)\times \widehat{U(1)}$ discussed before.

%%%%%%%%%%%%%%%%%%%%%%%%
\subsubsection{$N_f=3$}
%%%%%%%%%%%%%%%%%%%%%%%%

In this case, some of the branches can be described by
the addition map we have discussed in previous section.

%\subsection{The $N_f=3$ case}
 
%The phase structure can be found in Table \ref{tableso6}.
 
%%%%%%%%%%%%%%%%%%%%%%%%%%%%%%%%%%%%%%%%%%%%
$\bullet$ {\bf  Non-baryonic $r=1$ branch}
%%%%%%%%%%%%%%%%%%%%%%%%%%%%%%%%%%%%%%%%%%%
 
In this case, the curve is simplified as
\bea
y^2=(x^2-m^2)^2 \left[P^2_4(x) -4x^4\Lambda^2(x^2-m^2)\right],
\nonu
\eea
where the term in the bracket is, in fact, the curve of
$SO(4)$ with $N_f=1$. This is just the familiar application
of addition maps. Using the corresponding  result of $SO(4)$,
we  find three vacua for $SO(6)\to \widehat{U(3)}$ and
one vacua for $SO(6)\to SO(2) \times \widehat{U(2)}$. There is no
smooth transition in this case.

%%%%%%%%%%%%%%%%%%%%%%%%%%%%%%%%%%%%%%%
$\bullet$ {\bf  Baryonic $r=1$ branch}
%%%%%%%%%%%%%%%%%%%%%%%%%%%%%%%%%%%%%%%%%
 
Again, we can use the addition map to reduce the problem to
the $r=0$ baryonic branch of $SO(4)$ with $N_f=1$. From there,
we get one vacuum for $SO(6)\to SO(2) \times \widehat{U(2)}$,
two vacua for $SO(6)\to SO(4) \times \widehat{U(1)}$ where
$SO(4)$ is at the Chebyshev branch, and two vacua 
for $SO(6)\to SO(4) \times \widehat{U(1)}$ where
$SO(4)$ is at the Special branch.
 
%%%%%%%%%%%%%%%%%%%%%%%%%%%%%%%%%%%%%%%%%%%%
$\bullet$ {\bf  Non-baryonic $r=0$ branch}
%%%%%%%%%%%%%%%%%%%%%%%%%%%%%%%%%%%%%%%%%%%%
 
%%%%%%%%%%%%%%%%%%%%%%%%
{\bf 1. Non-degenerated case}
%%%%%%%%%%%%%%%%%%%%%%%%
 
\noindent
In this case, we have the following factorization problem,
\bea
P^2_6(x)-4x^4\Lambda^2(x^2-m^2)^3=x^2 H^2_{2}(x) F_6(x).
\nonu
\eea
To satisfy this equation, 
we must have a $x^2$ factor in $P_6(x)=x^2P_4(x)$. 
Let us assume that
\bean
P_4(x)&=& x^4+Ax^2+B, \quad H_2(x)= x^2-C,  \\
F_6(x)&=&\left[ x(x^2-D) \right]^2+G \left(x^2+F \right). 
\eean
After solving the factorization problem, we have three kinds of solutions. 
However,
 two of them have an $x^2$ factor in $F_6(x)$, which means that this 
becomes the 
degenerated case. Therefore, in the non-degenerated case, 
we have only the solution below.
  
The third solution is given by 
\bean
P_6(x)&=&x^2\left[x^4+Ax^2-2i\eta \Lambda m^3 \right],\qquad 
G=4\left(A\Lambda^2-\Lambda^4+3\Lambda m^2-i\eta \Lambda m^3 \right), \nonu 
 \\
F&=&\frac{m^3\left(A-3i\eta \Lambda m \right)}{iA\eta 
\Lambda-i\eta \Lambda^3+3i\eta \Lambda m^2+m^3},\qquad D=-A+2\Lambda^2.
\eean
Thus taking into account 
 the relation $W^{\prime}(x)$ and $F_6(x)$, we have 
$m^2=-A$. 
Under the semiclassical limit, the characteristic function 
behaves as $P_6(x)=x^4(x^2-m^2)$, which means that the 
classical group $SO(6)$ breaks into 
$SO(4)\times \widehat{U(1)}$.
Thus, we have obtained two vacua from the 
analysis of strong coupling, which comes from the Chebyshev branch of 
the $SO(4)$ factor.
 
%%%%%%%%%%%%%%%%%%%%%%%%%
{\bf 2. Degenerated case}
%%%%%%%%%%%%%%%%%%%%%%%%

For this case, the curve should be factorized as
\bea
P^2_3(t)- 4 \Lambda^2 t^2 (t-m^2)^3= F_2(t) H^2_2(t).
\nonu
\eea
There are two cases we should discuss. The first case is
that $H_2(t)=tH_1(t)$. After dividing out the $t^2$, at two
sides we parameterize $P_2(t)=t^2-s_1 t+s_2$, $H_1(t)=(t-a)$,
and $F_2(t)=t^2+b t+c$. Solving $s_1,s_2,c$ in terms of $a$ and $b$ we get
the following constraint
\bea
\La^2 \left( a - m^2 \right)^3
   \left( 4 a^2 + 4 a \left( b - 
        5 {\Lambda}^2 \right)  + 
     \left( b + 4 {\Lambda}^2 \right)^2 + 
     36 \La^2 m^2 \right)=0.
\nonu
\eea
At the classical limit, we have $b=-2 a^2$. If there are D5-branes wrapping
around the $x=\pm m$, we will have $b\sim -2m^2$, thus giving
$SO(6)\to SO(2)\times \widehat{U(2)}$. For given $b$, there are 
two solutions for $a$ which are consistent with the counting
of $U(2)$ with $N_f=3$ at the $r=0$ non-baryonic branch. 
If there are no D5-branes wrapping around the $x=\pm m$, 
we need to have an extra $t$
factor in $F_2(t)$; i.e., $c=0$. There are four solutions having
$a,b\to 0$ which give the four vacua of $SO(6)\to SO(6)$. 

Now we move to the second case in which
there is no factor $t$ in $H_2(t)$. This case will give
the $r=0$ non-baryonic branch of $SO(6)\to \widehat{U(3)}$ and
the $r=0$ non-baryonic branch of $SO(6)\to SO(4)\times \widehat{U(1)}$
where $SO(4)$ is at the Special branch. Although the indices for these
two phases are the same, it is not clear whether they 
are smoothly connected or not.
To demonstrate the smooth
transition, we must use the analytic expression, which is too complex
to be solved. To count the number of vacua
we can use a numerical method. Since we are unable to solve it,
we will not discuss if any further.

%%%%%%%%%%%%%%%%%%%%%%%%%%%%%%%%%%%%%%%%%%
$\bullet$ {\bf  Baryonic $r=0$ branch }
%%%%%%%%%%%%%%%%%%%%%%%%%%%%%%%%%%%%%%%%%%%%

This can happen only for $SO(6)\to \widehat{U(3)}$ and the
corresponding curve should be factorized as
\bea
P^2_3(t)- 4 \Lambda^2 t^2 (t-m^2)^3= H^2_3(t). 
\nonu
\eea
There are three solutions for $H_3(t)=t^3-a t^2+b t-c$. 
The first one is $a=2m^2+\Lambda^2$, $b=m^2(m^2+\Lambda^2)$,
and $c=0$ which gives the $r=1$ baryonic branch of $SO(2)\times
\widehat{U(2)}$. The second one is 
$a=m^2+\Lambda^2$, $b=2\Lambda^2 m^2$, and $c= \Lambda^2 m^4$
which gives the $r=1$ baryonic branch of $SO(4)\times
\widehat{U(1)}$. The third one is $a=3 m^2+\Lambda^2$, $b=3m^4$, and
$c=m^6$ which gives the baryonic $r=0$ branch of $\widehat{U(3)}$
we are looking for.

The results are summarized in Table \ref{tableso6}
by specifying the flavors $N_f$, symmetry breaking patterns, 
various branches, the exponent of $t=x^2$ in the curve, $U(1)$ at the
IR, the number of vacua, and the possibility  of a smooth transition.
It turns out that the number of vacua is exactly the same 
as from the 
weak coupling analysis.
 
\begin{table}
\begin{center}
\begin{tabular}{|c|c|c|c|c|c|c|} \hline
$N_f$  &  Group  & Branch &  Power of
$t(=x^2)$ &  $U(1)$ & Number of vacua & Connection
\\ \hline
0 & $SO(2)\times U(2)$ & $(S,0_{NB})$ & $t^2$ & 1 &  2  & \\ \cline{2-7} 
& $SO(4)\times U(1)$ & $(C,0_{NB})$ &  $t^3$ & 1 &  2  & \\ \cline{3-7}
& & $(S,0_{NB})$ &  $t^0$ & 1 &  2  & A \\ \cline{2-7}
& $SO(6)$ &  $(C)$ &  $t^3$ & 0 & 4 & \\ \cline{2-7}
& $U(3)$  &  $(0_{NB})$ & $t^0$ & 1 & 3 & A
\\ \hline 
1 & $SO(2)\times \widehat{U(2)}$ 
& $(S,0_{NB})$ & $t^2$ & 1 &  3  & \\ \cline{2-7}
& $SO(4)\times \widehat{U(1)}$ 
& $(C,0_{NB})$ &  $t^3$ & 1 &  2  & \\ \cline{3-7}
& & $(S,0_{NB})$ &  $t^0$ & 1 &  2  &  B\\ \cline{3-7}
& & $(C,0_{B})$ &  $t^3$ & 0 &  2  & C \\ \cline{3-7}
& & $(S,0_{B})$ &  $t^0$ & 0 &  2  &  \\ \cline{2-7}
& $SO(6)$ &  $(C)$ &  $t^3$ & 0 & 4 & C\\ \cline{2-7}
& $\widehat{U(3)}$  &  $(0_{NB})$ & $t^0$ & 1 & 5 & B
\\ \hline 
2 & $SO(2)\times \widehat{U(2)}$ 
& $(S,1_{NB})$ & $t^2$ & 1 &  1  &  \\ \cline{3-7}
& & $(S,0_{B})$ &  $t^2$ & 0 &  1  &  \\ \cline{3-7}
& & $(S,0_{NB})$ &  $t^2$ & 1 &  2  &  \\  \cline{2-7}
& $SO(4)\times \widehat{U(1)}$ 
& $(C,1_{B})$ &  $t^3$ & 0 &  2  &  \\ \cline{3-7}
& & $(S,1_{B})$ &  $t^0$ & 0 &  2  &  \\ \cline{3-7}
& & $(C,0_{NB})$ &  $t^3$ & 1 &  2  &  \\ \cline{3-7}
& & $(S,0_{NB})$ &  $t^0$ & 1 &  2  &  \\ \cline{2-7}
& $SO(6)$ &  $(C)$ &  $t^3$ & 0 & 4 & \\ \cline{2-7}
& $\widehat{U(3)}$  &  $(1_{NB})$ & $t^0$ & 1 & 2 &  \\ \cline{3-7}
&  &  $(0_{NB})$ & $t^0$ & 1 & 4 &  
\\ \hline 
3 & $SO(2)\times \widehat{U(2)}$ 
& $(S,1_{NB})$ & $t^2$ & 1 &  1  &  \\ \cline{3-7}
& & $(S,1_{B})$ &  $t^2$ & 0 &  1  &  \\ \cline{3-7}
& & $(S,0_{NB})$ &  $t^2$ & 1 &  2  &  \\  \cline{2-7}
& $SO(4)\times \widehat{U(1)}$ & 
$(C,1_{B})$ &  $t^3$ & 0 &  2  &  \\ \cline{3-7}
& & $(S,1_{B})$ &  $t^0$ & 0 &  2  &  \\ \cline{3-7}
& & $(C,0_{NB})$ &  $t^3$ & 1 &  2  &  \\\cline{3-7}
& & $(S,0_{NB})$ &  $t^0$ & 1 &  2  & * \\  \cline{2-7}
& $SO(6)$ &  $(C)$ &  $t^3$ & 0 & 4 & \\ \cline{2-7}
& $\widehat{U(3)}$  &  $(1_{NB})$ & $t^0$ & 1 & 3 &  \\ \cline{3-7}
& &  $(0_{NB})$ & $t^0$ & 1 & 3 &  * \\ \cline{3-7}
& &  $(0_{B})$ & $t^0$ & 0 & 1 & 
\\ \hline 
\end{tabular}
\end{center} 
\caption{\sl The summary of phase structure of $SO(6)$ gauge group. 
Here again we use  capital letters to indicate  the three phases 
having smooth transitions. The * means that it is  not clear whether
they are smoothly connected or not.}
\label{tableso6}
\end{table}

For the $SO(7)$ case, we refer to  version one in the hep-th archive 
for detail. 
In this paper 
we only summarize the results in Table \ref{tableso7} and
Table \ref{tableso7-1}
by specifying the flavors $N_f$, symmetry breaking patterns, 
various branches, the exponent of $t=x^2$ in the curve, $U(1)$ at the
IR, the number of vacua, and the possibility  of a smooth transition.
It turned out that the number of vacua is exactly the same 
as from the 
weak coupling analysis.

\begin{table}
\begin{center}
\begin{tabular}{|c|c|c|c|c|c|c|} \hline
 $N_f$  &  Group  & branch &  Power of $t(=x^2)$ 
&  $U(1)$ & Number of vacua & Connection
 \\ \hline
 0 & $SO(3)\times U(2)$ & $(S,0_{NB})$ & $t^0$ & 1 &  2  
& A \\ \cline{3-7} 
& & $(C,0_{NB})$ &  $t^1$ & 1 &  2  & B \\ \cline{2-7}
& $SO(5)\times U(1)$ & $(C,0_{NB})$ &  $t^1$ & 1 &  3  & 
B \\ \cline{2-7}
& $SO(7)$ &  $(C)$ &  $t^1$ & 0 & 5 & \\ \cline{2-7}
& $U(3)$  &  $(0_{NB})$ & $t^0$ & 1 & 3 & A
\\ \hline 
 1 & $SO(3)\times \widehat{U(2)}$ & $(C,0_{NB})$ & $t^1$ & 1 &  3  
& C \\ \cline{3-7}
& & $(S,0_{NB})$ &  $t^0$ & 1 &  3  & D \\ \cline{2-7}
& $SO(5)\times \widehat{U(1)}$ & $(C,0_{NB})$ &  $t^1$ & 1 &  3  
& C \\ \cline{3-7}
& & $(C,0_{B})$ &  $t^1$ & 0 &  3  &  N\\ \cline{2-7}
& $SO(7)$ &  $(C)$ &  $t^1$ & 0 & 5 & N\\ \cline{2-7}
& $\widehat{U(3)}$  &  $(0_{NB})$ & $t^0$ & 1 & 5 & D 
\\ \hline 
 2 & $SO(3)\times \widehat{U(2)}$ & $(C,1_{NB})$ & $t^1$ & 1 &  1  
&  \\ \cline{3-7}
& & $(S,1_{NB})$ &  $t^0$ & 1 &  1  & E \\ \cline{3-7}
& & $(C,0_{NB})$ &  $t^1$ & 1 &  2  & F \\ \cline{3-7}
& & $(S,0_{NB})$ &  $t^0$ & 1 &  2  & G \\ \cline{3-7}
& & $(S,0_{B})$ &  $t^0$ & 0 &  1  &  \\ \cline{3-7}
& & $(C,0_{B})$ &  $t^1$ & 0 &  1  &  O\\ \cline{2-7}
& $SO(5)\times \widehat{U(1)}$ & $(C,1_{B})$ &  $t^1$ & 0 &  3  &  
\\ \cline{3-7}
& & $(C,0_{NB})$ &  $t^1$ & 1 &  3  & F \\ \cline{2-7}
& $SO(7)$ &  $(C)$ &  $t^1$ & 0 & 5 & O\\ \cline{2-7}
& $\widehat{U(3)}$  &  $(1_{NB})$ & $t^0$ & 1 & 2 & E \\ \cline{3-7}
&  &  $(0_{NB})$ & $t^0$ & 1 & 4 & G
\\ \hline 
 3 & $SO(3)\times \widehat{U(2)}$ & $(C,0_{NB})$ & $t^1$ & 1 &  1  
& H \\ \cline{3-7}
& & $(S,0_{NB})$ &  $t^0$ & 1 &  1  & I  \\ \cline{3-7}
& & $(C,1_{NB})$ &  $t^1$ & 1 &  1  &  \\ \cline{3-7}
& & $(S,1_{NB})$ &  $t^0$ & 1 &  1  & J \\ \cline{3-7}
& & $(C,1_{B})$ &  $t^1$ & 0 &  1  &  \\ \cline{3-7}
& & $(S,1_{B})$ &  $t^0$ & 0 &  1  &  \\ \cline{2-7}
& $SO(5)\times \widehat{U(1)}$ & $(C,0_{NB})$ &  $t^1$ & 1 &  3  & H 
\\ \cline{3-7}
& & $(C,1_{B})$ &  $t^1$ & 0 &  3  &  \\ \cline{2-7}
& $SO(7)$ &  $(C)$ &  $t^1$ & 0 & 5 & \\ \cline{2-7}
& $\widehat{U(3)}$  &  $(0_{NB})$ & $t^0$ & 1 & 3 & I \\ \cline{3-7}
& &  $(1_{NB})$ & $t^0$ & 1 & 3 & J \\ \cline{3-7}
&  &  $(0_{B})$ & $t^0$ & 0 & 1 & 
\\ \hline

\end{tabular}
\end{center} 
\caption{ \sl Summary of the phase structures of $SO(7)$ gauge group. 
Here again we use  capital letters in the last column to 
indicate  the various 
phases having a smooth transition. 
}
\label{tableso7}
\end{table}

%%%%%%%%%%%%%%%%%%%%%%%%%%%%%%%%%%%%%%%%%%%%%%%%%%

\begin{table}
\begin{center}
\begin{tabular}{|c|c|c|c|c|c|c|} \hline
 $N_f$  &  Group  & Branch & Power of  $t(=x^2)$ &  $U(1)$ 
& Number of vacua & Connection
 \\ \hline
4 & $SO(3)\times \widehat{U(2)}$ & $(C,1_{NB})$ & $t^1$ & 1 &  1 
 &  \\ \cline{3-7}
& & $(S,1_{NB})$ &  $t^0$ & 1 &  1  & K  \\ \cline{3-7}
& & $(C,0_{NB})$ &  $t^1$ & 1 &  2  & L \\ \cline{3-7}
& & $(S,0_{NB})$ &  $t^0$ & 1 &  2  & M \\ \cline{3-7}
& & $(C,2_{B})$ &  $t^1$ & 0 &  1  &  \\ \cline{3-7}
& & $(S,2_{B})$ &  $t^0$ & 0 &  1  &  \\ \cline{2-7}
& $SO(5)\times \widehat{U(1)}$ & $(C,1_{B})$ &  $t^1$ & 0 &  3  
&  \\ \cline{3-7}
& & $(C,0_{NB})$ &  $t^1$ & 1 &  3  & L \\ \cline{2-7}
& $SO(7)$ &  $(C)$ &  $t^1$ & 0 & 5 & \\ \cline{2-7}
& $\widehat{U(3)}$  &  $(1_{NB})$ & $t^0$ & 1 & 2 & K \\ \cline{3-7}
&  &  $(0_{NB})$ & $t^0$ & 1 & 3 & M \\ \cline{3-7}
&  &  $(2_{NB})$ & $t^0$ & 1 & 1 &  \\ \cline{3-7}
& &  $(1_{B})$ & $t^0$ & 0 & 1 &  
\\ \hline 
\end{tabular}
\end{center} 
\caption{ \sl Continued. 
Summary of the phase structures of $SO(7)$ gauge group.}
\label{tableso7-1}
\end{table}

%%%%%%%%%%%%%%%%%%%%%%%%%%%%%%%%%%%%%%%%%%%%%%%%%%%%%%%%%%%%%%%%%%%%%
\section{Quartic superpotential with massless flavors}
%%%%%%%%%%%%%%%%%%%%%%%%%%%%%%%%%%%%%%%%%%%%%%%%%%%%%%%%%%%%%%%%%%%%%%

\indent

Thus far we have discussed the phase structures of massive flavors.
In this section 
we will focus on {\it massless} flavors. Classically, under
the breaking pattern $SO(N_c)\to SO(N_0) \times \prod_{j=1}^n U(N_j)$, these
massless flavors will be charged under  $SO(N_0)$ since
$W'(x)|_{x=\pm m=0}=0$ is always true in our considerations
\footnote{For degenerated cases, we have $SO(N_c) \to 
\widehat{SO(N_c)}$ and $SO(N_c) \to U([N_c/2])$ for quartic 
superpotential.}.

To demonstrate the new features coming from the massless flavors
and compare them with the weak and strong coupling analyses, let
us start with the following quadratic superpotential first
\bea
W(\Phi)= {1\over 2} \mu \tr \Phi^2~.
\nonu
\eea
The curve is given by
\bea
y^2= P^2_{2[N_c/2]}(x) - 4\Lambda^{2(N_c-N_f-2)} x^{2(1+\epsilon
+N_f)}
\nonu
\eea
where $\epsilon=0$ for $N_c$ odd and $1$ for $N_c$ even and 
$N_f< N_c-2$ for asymptotic free theory. 
For the non-baryonic branch,
the curve should be factorized as
\bea
y^2= F_4(x) H^2_{2[N_c/2]-2}(x)
\label{quadcurve}
\eea
with $F_4(x)=x^2 F_2(x)$ where the appearance of $x^2$ in this case
is characteristic of the non-baryonic branch. 
See for example, (\ref{42swso2nb}) 
and (\ref{42swso2n+1b}).
 Furthermore, we have the  following 
relationship
\bea
F_2(x)= W'(x)^2 + d= x^2 +d~.
\nonu
\eea
Because of the special form of $F_4(x)$, we do not have
the $r$-th branch  in the $U(N_c)$ case. 

To show it more
clearly, let us consider the example of
$SO(2N_c)$ gauge theory with $2M$ flavors. If we use $t=x^2$, the
curve is given by
\bea
y^2=\prod_{j=1}^{N_c} (t-\phi_j^2)^2- 4\Lambda^{2(2N_c-2-2M)} 
t^{2+2M}.
\nonu
\eea
The non-baryonic vacua should be factorized as
\bea
y^2=F_2(t) H^2_{N_c-1}(t),~~~~~~F_2(t)=t(t+d)~.
\nonu
\eea
Due to  the
factor $t$ of $F_2(t)$, we should have, for example, $\phi_{N_c}=0$
and the curve becomes
\bea
y^2=t^2 \left[ \prod_{j=1}^{N_c-1} (t-\phi_j^2)^2- 
4\Lambda^{2(2N_c-2-2M)} t^{2+2(M-1)} \right]= t (t+d) H^2_{N_c-1}(t)
\nonu
\eea
which means that the function $H_{N_c-1}(t)$ 
must have a factor $t$. Cancelling out 
factor $t^2$ at both sides, we get
\bea
y_1^2=\prod_{j=1}^{N_c-1} (t-\phi_j^2)^2- 
4\Lambda^{2(2N_c-2-2M)} t^{2+2(M-1)}=t (t+d) H^2_{N_c-2}(t)
\nonu
\eea
which is the non-baryonic curve  of 
$SO(2(N_c-1))$ with $2(M-1)$ flavors. Since we
require that the number of flavors satisfies
$2M<2N_c-2$ for asymptotic free theory, the above operation 
can be iterated at most $(M+1)$ steps and  the
factorization problem is reduced to
\bea
y^2=t^{2(M+1)}[P^2_{N_c-(M+1)}(t)-4\Lambda^{2(2N_c-2-2M)}]=
t(t+a)t^{2(M+1)} H^2_{N-(M+1)-1}(t).
\nonu
\eea
The presence of  factor $t^{2(M+1)}$ indicates that no matter what factor
$t^{2r}$ with $r< M+1$ we start with, we always end up with 
$r=M+1$ branch. This particular branch is nothing but the
Chebyshev point emphasized in the weak and strong coupling
analyses given in section 3.1.2. 
It is easy to check that for other $N_c$ and $N_f$,
the factorized curves are also at the Chebyshev point.
For the Special branch, we have already written down the form of curves
in the strong coupling analysis  where the basic feature
of this branch is that the curve is a complete square.

Having obtained the quadratic superpotential, we
can progress 
to the general superpotential with degree $2(n+1)$ where $k=n$.
Again, for example,  the $SO(2N_c)$ with $2M$ flavors 
at the general non-baryonic branch should have the 
factorization, by generalizing  (\ref{quadcurve})
\bea
y^2=\prod_{j=1}^{N_c} (t-\phi_j^2)^2- 4\Lambda^{2(2N_c-
2-2M)} t^{2+2M}
=tF_{2n+1}(t) H^2_{N_c-(n+1)}(t).
\nonu
\eea
Similar to the above derivation, factor $t$ tells us that
at least one of $\phi_i$'s is vanishing, 
so that there is a factor $t^2$ on
the left hand side. Then either $H_{N_c-(n+1)}(t)$
or $F_{2n+1}(t)$ must be divided by  factor $t$. 

If $t$ is present in the $F_{2n+1}(t)$, we should have
$F_{2n+1}(t=x^2)=x^2 \left[ ({W'(x) \over x})^2+ f_{n-1}(x^2) \right]$
and the curve becomes
\begin{equation} 
\label{baryonic-infact}
y^2= F_{2n}(t) [ tH_{N-(n+1)}(t)]^2 
\end{equation}
with, (for example see (\ref{condition4n})),
\bea
 F_{2n}(t=x^2)= \left({W_{2n+1}'(x) \over x} \right)^2+ 
f_{n-1}(x^2)~.
\nonu
\eea
The factorization form (\ref{baryonic-infact}) is, in fact,
the expression for the Special branch because there is no
overall factor $t$ in the curve. Geometrically,
 the function 
$F_{2n+1}(t)$ divided by $t$ means that the cycle around
the origin $t=0$ 
on the reduced Riemann surface is degenerated and the
remaining non-trivial cycles are around $U(N_i)$ factors.
In other words, the non-baryonic solution found in this
case comes from the Special root discussed in the
weak and strong coupling analyses. 

Instead, 
%be to the factor of  $F_{2n+1}(t)$, 
if $t$ is present in 
the function $H_{N-(n+1)}(t)$, then we can  cancel the $t^2$ factor
on both sides of the curve and reduce the problem 
to $SO(2(N_c-1))$ with $2(M-1)$ massless flavors. 
Iterating this procedure, we can reach the Chebyshev point as
in the quadratic superpotential if it does not stop at
the above Special point. 

In summary, the above analysis shows that 
the factorization of curves will 
stop only  at either the Chebyshev point or the Special
point, which is
consistent with  the results from 
the weak and strong coupling analyses.

So considered, we can now focus on 
the following quartic superpotential
$W(\Phi)={1\over 4}\tr \Phi^4-{\alpha^2 \over 2}\tr \Phi^2$, so
that 
\begin{equation} 
\label{massless-W}
W'(x)=x(x^2-\alpha^2)~.
\end{equation}
We also consider only the case $N_f<N_c-2$ for asymptotic free theory.
For convenience of the following discussions, we would like to summarize the
essential points of the form of Seiberg-Witten curves:

%\begin{itemize}
a). At the Chebyshev point, the power of $t=x^2$ in the
curve is
$(N_f+3)$ for $(N_f,N)=(\mbox{even}, \mbox{even})$, $(N_f+1)$ for 
$(N_f,N)=(\mbox{even}, \mbox{odd})$, $(N_f+2)$ for $(N_f,N)=
(\mbox{odd}, \mbox{even})$,
and  $(N_f+2)$ for $(N_f,N)=(\mbox{odd}, \mbox{odd})$. Notice that for 
all four cases, the power is always an odd number which 
is consistent with the factorization form $t H^2(t)$; i.e.,
except one factor of $t$, the other power of $t$ can be considered
as  double roots. The number is given by $(N-N_f-2)$.

b). At the Special point which
exists only when $\widetilde{N}_c= 2N_f-N+4\geq 0$, the
characters of the curve are: (1) When $N_f<N-2$,
the power of $t$ is 
$(2N_f-N+4)$ for $N$ even and $(2N_f-N+3)$ for $N$ odd.
When $N_f\geq N-2$, the power of $t$ is $N$ for  $N$ even
or $(N-1)$ for $N$ odd. Notice that the power is always even,  so
it can be considered to belong into the double root in the
factorization form of curve; (2) the number of the double
root increases by one compared with that at the Chebyshev point.

c). From (a) and (b) the simplest way to distinguish between
the Chebyshev branch and the Special branch is to see whether the power of
$t$ is odd or even. 

d). Because of point (b), the Special branch has
a similar form of factorization as one of degenerated cases
and will be discussed together in the following discussion.

%\end{itemize}

%%%%%%%%%%%%%%%%%%%%%%%%%%%%%%%
\subsection{ $SO(4)$ case}
%%%%%%%%%%%%%%%%%%%%%%%%%%%%%%

\indent

Because of $N_f<N_c-2$ for asymptotic free theory, we only have 
two choices: $N_f=0$ and $N_f=1$. The $N_f=0$ case has been 
discussed in section 4 under massive flavors where we only need to
replace $m$ by $\alpha$. However, the $N_f=1$ will give {\it new}
results. There are three classical limits $SO(4)\to \widehat{SO(2)}
\times U(1)$, $SO(4)\to \widehat{SO(4)}$, and $SO(4)\to U(2)$. 
For $\widehat{SO(2)}$ 
with one flavor, there exists only the Special branch
which can also be seen by taking into consideration the 
counting of the Chebyshev
branch, $N-N_f-2=2-1-2<0$, which does not exist. 
For $\widehat{SO(4)}$ with one flavor,
we can have both the Chebyshev branch and the Special branch 
because it is possible that
$2N_f-N+4=2>0$ and $N-N_f-2=1 > 0$. 

%%%%%%%%%%%%%%%%%%%%%%%%%%%%%%%%%%%%%%
 {\bf 1 Non-degenerated case}
%%%%%%%%%%%%%%%%%%%%%%%%%%%%%%%%%%%%

\noindent
The curve should be
\bea
y^2= P^2_4(x)-4x^6\Lambda^{2}= x^2 F_6(x)
\nonu
\eea
by taking the $m \rightarrow 0$ limit into (\ref{example-1}).
Using $P_4(x)=x^2 (x^2-A)$, we get $F_6(x)=
x^2[(x^2-A)^2- 4x^2\Lambda^{2}]$. From the 
relationship between $F_6(x)$ and $W'(x)$ we
find that $A=\alpha^2$. In the classical limit $\Lambda \to 0$,
we get $SO(4)\to \widehat{SO(2)} \times U(1)$. Notice that there
is a $x^4=t^2$ factor in the $y^2$  indicating that  
the solution found
comes from the Special root
instead of the Chebyshev point.

%%%%%%%%%%%%%%%%%%%%%%%%%%%%%%%%%%%%%%%%%%%%%%%%%%%%% 
 {\bf  2. Degenerated case and Special branch}
%%%%%%%%%%%%%%%%%%%%%%%%%%%%%%%%%%%%%%%%%%%%%%%%%%%%%%

\indent

As previously noted, since the Special branch has one
extra double root and the degenerated case has one root
of $W'(x)$ without wrapping D5-branes, the starting 
factorization forms of the curve are the same and we should
consider
them simultaneously.
The curve should be
\bea
y^2= P^2_4(x)-4x^6\Lambda^{2}=F_4(x) H^2_2(x)
\nonu
\eea
or
\begin{equation} 
\label{massless-SO4-1-D}
y^2=(t^2-s_1 t+s_2)^2- 4 t^3\Lambda^{2}=(t^2+a t+b) 
(t+c)^2,~~~~~t=x^2. 
\end{equation}
Given the $a$ 
there are three solutions for $(s_1,s_2,b,c)$.
The first is $({-a-4 \Lambda^2 \over 2}, 0, 
({-a-4 \Lambda^2 \over 2})^2,0)$, where $c=0$ gives a factor
$t^2$ in the curve. Notice that there is no smooth 
transition  found in this solution.

To determine the classical 
limit, we need to determine the behavior of $a$. There are
two cases  to be analyzed. The first case is that 
there are some D5-branes wrapping around $x=\pm \alpha$. Then we 
get $a=-(2 \alpha^2 + 4\Lambda^2)$, which produces the
classical limit $SO(4)\to \widehat{SO(2)} \times U(1)$.
In fact, this solution is identical to the one given in 
the previous paragraph where we claimed that it belongs to
the Special branch. There is only one solution which
is consistent with the counting.
The second case is that all D5-branes wrap
around the origin $x=0$, so curve (\ref{massless-SO4-1-D})
describes the $SO(4)\to \widehat{SO(4)}$ classically.
With one flavor, we can have the Special branch with 
a factor $t^2$ by $(2N_f-N+4)=2$ or the Chebyshev branch with
$t^3$ where $N_f+2=3$. 
To get the Special branch we must set $a=-2\Lambda^2$
so that $(t^2+a t+b)=(t-\Lambda^2)^2$. There is only 
one solution. 
 To get the Chebyshev
branch, we need $b=0$ so that $a=-4\Lambda^2$. Notice the
counting of vacua matches the formula $(N-N_f-2=4-1-2=1)$.

The second and third cases have a somewhat complex expression
as a function of $a$ and neither $b$ or $c$ is zero for general $a$
\bean
s_1 & = & {-2a +\Lambda^2\mp 3 
\Lambda\sqrt{-2 a+\Lambda^2} \over 2}, \\
s_2 & = & {a^2-19 a \Lambda^2+16 \Lambda^4\pm 5a 
\Lambda\sqrt{-2 a+\Lambda^2}\mp 16 \Lambda^3\sqrt{-2 a+\Lambda^2}
\over 4}, \\
b & = & {a^2\over 4}+2\Lambda^4\mp 2\Lambda^3\sqrt{-2 a+\Lambda^2}
+a(-3\Lambda^2 \pm \Lambda\sqrt{-2 a+\Lambda^2}), \\
c & = & {a-5\Lambda^2\pm 3 \Lambda\sqrt{-2 a+\Lambda^2} 
\over 2}.
\eean
Under the limit $\Lambda\to 0$, $(s_1,s_2,b,c)\to
(-a,{a^2\over 4}, {a^2\over 4},{a\over 2})$ which gives
 $SO(4)\to U(2)$ with two vacua coming from the $U(2)$. 
Again by the relationship between $F_4(x)$ and $W'(x)/x$
we get $a=-(2 \alpha^2 + 4\Lambda^2)$. Notice that we 
cannot take the limit $\Lambda\to 0, a\to 0$ because
the curve form (\ref{massless-SO4-1-D}) has no 
factor $t$ before the limit is reached, whereas if there 
are D5-branes wrapping around origin $x=0$, the starting curve
must have a factor $t$. Because of this inconsistency, we
can not take that limit. Similar observations have been
presented in \cite{bfhn}.

The phase structure has been summarized in Table \ref{tableso4-massless}.

\begin{table}
\begin{center}
\begin{tabular}{|c|c|c|c|c|c|c|} \hline
 $N_f$  &  Group  & Branch &  Power of 
$t(=x^2)$ &  $U(1)$ & Number of vacua & Connection
 \\ \hline
1 &  $\widehat{SO(2)}
\times U(1)$ & $(S,0_{NB})$ & $t^2$ & 1 &  1  & \\ \cline{2-7}
  & $\widehat{SO(4)}$  & $(C)$ &  $t^3$ & 0 & 1 & \\\cline{3-7}
  &  & $(S)$ &  $t^2$ & 0 & 1 & \\\cline{2-7}
& $U(2)$ & $(0_{NB})$ & $t^0$ & 1 & 2 & \\ \hline     
\end{tabular} 
\end{center}
\caption{ \sl The phase structure of $SO(4)$ with one massless flavor.
There are no smooth transitions between these vacua because there is no 
vacua which has the same number of  the power of $t$ and the number of 
$U(1)$. }
\label{tableso4-massless}
\end{table}

%%%%%%%%%%%%%%%%%%%%%%%%%%%%
\subsection{ $SO(5)$ case}
%%%%%%%%%%%%%%%%%%%%%%%%%%%%%

\indent

For the $SO(5)$ gauge group, $N_f$ can be zero, one, or two. 
There are three breaking patterns $SO(5)\to \widehat{SO(3)}\times 
U(1)$, $SO(5)\to U(2)$, and $SO(5)\to \widehat{SO(5)}$. The 
$N_f=0$ case has been discussed in  
section 4 under massive flavors.
For the $N_f=2$ case, notice that the curve is
\bea
y^2=P^2_4(x)-4\Lambda^2 x^6
\nonu
\eea
which is exactly the same form as the curve of $SO(4)$ with
$N_f=1$. Thus,  the results are the same as  $SO(4)$ with
one flavor. In fact, it is an example of the addition map
previously discussed in Appendix B. 
For $N_f=1$, the curve
\bea
y^2=P^2_4(x)-4\Lambda^4 x^4
\nonu
\eea
is the same as that of $SO(4)$ without any flavors, so
the 
results can be applied  here. However, now we need to 
distinguish between Special vacua and Chebyshev vacua
for the case of massless flavors.
Because of this, we will repeat the $SO(5)$ example, but not discuss 
$SO(7)$ at all because everything in $SO(7)$ can be reduced
to the results of $SO(6)$. Additionally, we will use a somewhat different
method to redo the calculations. It will be interesting
to compare  these two methods. 

%%%%%%%%%%%%%%%%%%%%%%%%%%%%%%%%%%%%%%%%
 {\bf 1. Non-degenerated case}
%%%%%%%%%%%%%%%%%%%%%%%%%%%%%%%%%%%%%%%

\noindent
The curve should be
\bea
y^2= P^2_4(x)-4x^4\Lambda^{4}= x^2 F_6(x).
\nonu
\eea
Using $P_4(x)=x^2(x^2-u)$ and $F_6(x)=W'(x)^2+a x^2 +b$, we found 
that $b=0$, $u=\alpha^2$, and $a=-(\alpha^4-4 \Lambda^4)$,
which gives one vacuum of  $SO(5)\to \widehat{SO(3)}\times 
U(1)$. Again, the fact that $y^2=x^4 F_4(x)$ indicates that 
it is the Special vacuum. The reason that
the Chebyshev branch of $\widehat{SO(3)}\times U(1)$
does not exist can be seen from  the counting $N-N_f-2=3-1-2=0$
for the Chebyshev branch.

%%%%%%%%%%%%%%%%%%%%%%%%%%%%%%%%%%%%%%%%%%%%%%%%%%%%%%
 {\bf 2. Degenerated case and Special branch}
%%%%%%%%%%%%%%%%%%%%%%%%%%%%%%%%%%%%%%%%%%%%%%%%%%%%%

\noindent
The curve should be
\bea
y^2= P^2_4(x)-4x^4\Lambda^{4}=F_4(x) H^2_2(x)
\nonu
\eea
or
\bea 
%\label{massless-SO5-1-D}
y^2=(t^2-s_1 t+s_2)^2- 4 t^2\Lambda^{4}=
(t^2+a t+b) (t+c)^2, \qquad t=x^2. 
\nonu
\eea
There are three solutions for $(s_1,s_2,b,c)$. The first 
is $({-a\over 2},0,{a^2\over 4}-4 \Lambda^4,0)$. To determine
the classical limit we need to determine $a$. Similar to 
the discussion of $SO(4)$ with one flavor we need to consider
various situations. The first  is that
there are D5-branes wrapping
around the $x=\pm \alpha$, so  $a=-2 \alpha^2$, which
gives  the Special vacuum of $SO(5)\to \widehat{SO(3)}\times U(1)$
presented in the 
previous paragraph. The second is that all D5-branes
are wrapped around the origin $x=0$ 
so we have $SO(5)\to \widehat{SO(5)}$. 
Then we need a factor $t^3$ for
the Chebyshev vacua where $N_f+2=3$ and 
$t^0$ for Special vacua where $2N_f-N+3=0$. Since $c=0$ 
gives at least $t^2$, we can only get two Chebyshev vacua
by setting $b=0$ or $a=\pm 4 \Lambda^2$, which match the
counting $(N-N_f-2)=5-1-2=2$. The missing Special vacuum
will be discussed immediately.

The second and third solutions are 
$(-a\pm 2 \Lambda^2, ({a\mp 4 \Lambda^2 \over 2})^2, 
({a\mp 4 \Lambda^2 \over 2})^2, {a\mp 4 \Lambda^2 \over 2})$.
For nonzero D5-branes wrapping around the $x=\pm \alpha$, again
we have $a=-2 \alpha^2$, so there are two vacua for 
$SO(5)\to U(2)$. Beside that, if we set $a=2\Lambda^2$ for
the second solution or $a=-2\Lambda^2$ for the 
third solution, both
will give the same curve having  a complete square form,
 which is identical with the Special vacuum of
$SO(5)\to \widehat{SO(5)}$.

We want to emphasize again that the calculations here are
exactly the same calculations of $SO(4)$ with $N_f=0$. The 
only difference is how to explain and describe 
these results in the phase
structure of $SO(5)$ gauge group.

%%%%%%%%%%%%%%%%%%%%%%%%%%%%%%%
\subsection{ $SO(6)$ case}
%%%%%%%%%%%%%%%%%%%%%%%%%%%%%%

\indent

For the $SO(6)$ gauge group, the number of
flavors $N_f$ can be $0,1,2$, or $3$. There exist
the following four breaking patterns: $SO(6)\to \widehat{SO(2)}\times
U(2)$, $SO(6)\to \widehat{SO(4)}\times U(1)$, 
$SO(6)\to \widehat{SO(6)}$, and $SO(6)\to U(3)$. For the $SO(2)$ factor,
 only the Special branch exists because for the Chebyshev branch
there is a negative vacuum number $N-N_f-2 < 0$, which does not exist  
when  for $\widehat{SO(4)}$ and
$\widehat{SO(6)}$ factors, 
both the Special branch and Chebyshev branch  exist. 
The $N_f=0$ case
has been discussed in section 4 under massive flavors, so
we are left only with $N_f=1,2,3$.

%%%%%%%%%%%%%%%%%%%%%%%%%%%%%%%%%%%%
$\bullet$ { $N_f=1$ }
%%%%%%%%%%%%%%%%%%%%%%%%%%%%%%%%%%%%

%%%%%%%%%%%%%%%%%%%%%%%%%%%%%%%%%%%%%%%%
 {\bf  1. Non-degenerated case}
%%%%%%%%%%%%%%%%%%%%%%%%%%%%%%%%%%%%%%%%%

\noindent
The curve should be
\bea
y^2= P^2_6(x)-4x^6\Lambda^{6}= x^2 F_6(x) H^2_2(x).
\nonu
\eea
Using $F_6(x)=W'(x)^2+bx^2+d$, $H_2(x)=x^2+a$, and
$P_6(x)=x^2(x^2-t_1)(x^2-t_2)$, we found four
solutions for $(t=t_1t_2, u=t_1+t_2, a, b,d)$. 
The first is $(0, \alpha^2, 0, 0, -4 \Lambda^6)$,
which gives $y^2=x^6 (W'(x)^2-4 \Lambda^6)$ and
$P_6(x)=x^4(x^2-\alpha^2)$. Classically it
gives the Chebyshev branch of $SO(6)\to \widehat{SO(4)}\times U(1)$
with the counting $N-N_f-2=4-1-2=1$.
The factor $(x^2)^3$ is exactly the one we need for
$SO(4)$ with $N_f=1$ at the Chebyshev point by noticing
$(N_f+2)=3$.

The second solution has $d=0$, so the curve is
$y^2= \left[({W'(x)\over x})^2+b \right] \left[x^2(x^2+a) \right]^2$.
At the classical limit it becomes
$(0,\alpha^2, 0, 0,0)$ which gives 
$SO(6)\to \widehat{SO(4)}\times U(1)$. Because of the
factor $(x^2)^2$, it is the Special branch by 
noticing $(2N_f-N+4)=2$. The counting for  
the number of vacua is also
consistent.

The third and fourth solutions have $d=0$ and the classical
limit becomes $(\alpha^4, 2 \alpha^2, -\alpha^2,0,0)$, which
gives $SO(6)\to \widehat{SO(2)}\times U(2)$. Since $d=0$, they
are in fact the Special vacua with the counting
two coming from the $U(2)$ factor.  

To discuss the smooth transition, we can use the above
solutions directly by taking various limits. However, since
the solutions are so complex, it is not easy to see the results.
However, it is easy to see that the first solution {\it cannot}
be smoothly interpolated to the other three solutions. We will
discuss the relationship of the other three solutions 
immediately by another method.

%%%%%%%%%%%%%%%%%%%%%%%%%%%%%%%%%%%%%%%%%%%%%%%%%%%%%%%%%
 {\bf  2. Degenerated case and Special branch}
%%%%%%%%%%%%%%%%%%%%%%%%%%%%%%%%%%%%%%%%%%%%%%%%%%%%%%%

\noindent
The curve should be factorized as
\bea
y^2= P^2_6(x)-4x^6\Lambda^{6}=  F_4(x) H^2_4(x),
\nonu
\eea
or
\bea
\label{massless-SO6-1-D}
y^2=P^2_3(t)- 4 t^3 \Lambda^{6}=F_2(t) H^2_2(t), \qquad
t=x^2.
%\nonu
\eea
To solve the problem, we parameterize as follows:
\bea
F_2(t)=(t-a)^2+b, \qquad H_2(t)=t^2+c t+d, \qquad
P_3(t)=t^3-s_1 t^2
+s_2 t-s_3~.
\nonu
\eea
There are four solutions. The first  is given by
\bean
s_1 & = & {-3b\over 4 \Lambda^2}, \qquad s_2={3 b^2 \over 16 \Lambda^4}
+{3b\over 4}, \qquad
s_3= -{ (b+ 4 \Lambda^4)^3 \over 64 \Lambda^6}, 
\nonu \\
a & = & {-b \over 4 \Lambda^2}+\Lambda^2, \qquad c=
{b \over 2 \Lambda^2}+\Lambda^2,~~~d={ (b+ 4 \Lambda^4)^2 \over
 16 \Lambda^4}.
\nonu
\eean
There is only one sensible limit $b/ \Lambda^2=\beta \neq 0$
where $P_3(t)\to (t+{\beta \over 4})^3$, so it gives
$SO(6)\to \widehat{U(3)}$. The second  and third solutions
have
\bea
s_1=3(1\pm i\sqrt{3}){b\over 8 \Lambda^2},
\nonu
\eea  
which has also only one sensible limit $b/ \Lambda^2=\beta \neq 0$
and gives $SO(6)\to \widehat{U(3)}$. In fact, 
for all of these solutions
we should consider two cases: one is $b=0$ and the other, $b\neq 0$.
For $b=0$, all three solutions give the same curve 
(\ref{massless-SO6-1-D}) as a square form with
$P_3(t)= t^3+\Lambda^{3}$. This vacuum
is the Special vacuum of $SO(6)\to \widehat{SO(6)}$. For $b\neq 0$, 
these three solutions give the three vacua of $SO(6)\to U(3)$.
Since $SO(6)\to \widehat{SO(6)}$ has $b=0$ and  a factor $t$ at
the curve, there is no smooth connection between $\widehat{SO(6)}$
and $U(3)$.

The fourth solution has
\bean
s_1 & = & { b^2 \over 8 \Lambda^6}-{4 \Lambda^6 \over b}, \qquad
s_2={b\over 256}(64+{b^3\over  \Lambda^{12}}), \qquad
s_3=0,
\nonu \\
a & = & {b^2\over 16  \Lambda^6}-{4  \Lambda^6 \over b}, \qquad
c={- b^2 \over 16  \Lambda^6}, \qquad
d=0.
\nonu 
\eean
In fact, since $d=0$ and $s_3=0$, we can factorize out the $t^2$ 
factor and the curve
 (\ref{massless-SO6-1-D}) is reduced to the degenerated case of 
$SO(5)$ without flavor. Thus, we immediately get the following results:
(1) There is a smooth interpolation 
between $\widehat{SO(2)}\times U(2)$
and $\widehat{SO(4)}\times U(1)$; (2) There are two Special
vacua for $\widehat{SO(2)}\times U(2)$, one Special vacuum
for $\widehat{SO(4)}\times U(1)$, and three 
Chebyshev vacua $(N-N_f-2)=(6-1-2)=3$ for $SO(6)\to
\widehat{SO(6)}$.

%%%%%%%%%%%%%%%%%%%%%%%%%%%%%%%%%%%%
$\bullet$ {\bf $N_f=2$}
%%%%%%%%%%%%%%%%%%%%%%%%%%%%%%%%%%%%

%%%%%%%%%%%%%%%%%%%%%%%%%%%%%%%%%%%%%%
 {\bf 1. Non-degenerated case}
%%%%%%%%%%%%%%%%%%%%%%%%%%%%%%%%%%%%%

\noindent
For simplicity we will use $t=x^2$. 
The curve is given by
\bea
y^2=P^2_3(t)-4 \Lambda^4 t^4= t F_3(t) H^2_1(t).
\nonu
\eea
Due to the factor $t$ on the right hand side, 
we write $P_3(t)=t(t^2-s_1 t+ s_2)$,
$H_1(t)=t+a$, and $F_3(t)=t^3+b t^2+c t+d$. There are
three solutions for $(s_1,s_2,a,c,d)$ as a function of $b$.
The first  is $(-b/2,0,0,b^2/4-4 \Lambda^4,0)$
which gives $SO(6)\to \widehat{SO(4)}\times U(1)$. Since 
$F_3(t)=t(t^2+bt+(b^2/4-4 \Lambda^4))$, it is, in fact, the solution
of Special or degenerated case.
The other two solutions are
$(-b\pm 2\Lambda^2,{1\over 4}(b\mp 4\Lambda^2)^2,
{1\over 2}(b\mp 4\Lambda^2), {1\over 4}(b\mp 4\Lambda^2)^2, 0)$,
which give  $SO(6)\to \widehat{SO(2)}\times
U(2)$. Again, since $d=0$ they are the solutions
of the Special or degenerated case.
The counting number two comes from the $U(2)$ factor. Finally,
using the relationship of $W'(x)$ and $F_3(t)$ we get
$b=-2 \alpha^2$. There is no smooth transition between these
three solutions.
It is noteworthy that there is no Chebyshev vacua of
$SO(6)\to \widehat{SO(4)}\times U(1)$ because the counting
leads to $(N-N_f-2)=(4-2-2)=0$.

%%%%%%%%%%%%%%%%%%%%%%%%%%%%%%%%%%%%%%%%%%%%%%%%%%%%%
 {\bf 2. Degenerated case and Special branch}
%%%%%%%%%%%%%%%%%%%%%%%%%%%%%%%%%%%%%%%%%%%%%%%%%%%%%

\noindent
The curve should be
\begin{equation} 
\label{SO6-2-bary}
y^2=P^2_3(t)-4 \Lambda^4 t^4=  F_2(t) H^2_2(t).
\end{equation}
Let us consider the solutions case by case. First, if there is
a factor $t^4$, we have $P_3(t)=t^2(t-s)$, $H_2(x)=t^2$, and 
$F_2(t)=(t-s)^2-4 \Lambda^4$. There are two situations 
 to be considered. If there are D5-branes wrapping
 around the $x=\pm \alpha$,
we need to have $F_2(t)=(t-\alpha)^2+d$, so $s=\alpha^2$,
$d=-4 \Lambda^4$ and  $SO(6)\to \widehat{SO(4)}\times U(1)$.
It is the same solution presented in the previous paragraph.
The power $4=(2N_f-N+4)$ of $t$ is the expected character
of $SO(4)$ at the Special vacua.  
 If there are no
D5-branes wrapping around the $x=\pm \alpha$, we need a factor
$t$ from $F_2(t)$ which is equal to set  $s^2=4\Lambda^4$.
The curve has a factor $t^5$ which gives the Chebyshev vacua of 
$SO(6)\to \widehat{SO(6)}$. The counting two is also consistent
with the fact that $(N-N_f-2)=2$.

If there is only one $t^2$ factor in (\ref{SO6-2-bary}),
we write $P_3(t)=tP_2(t)$ and $H_2(t)=t(t+a)$. Canceling out the
factor $t^2$, we get $P^2_2(t)-4 \Lambda^4 t^2=F_2(t) (t+a)^2$.
It can be solved if we require $P_2(t)-2 \eta \Lambda^2 t=
(t+a)^2$, so $F_2(t)=(t+a)^2+4 \eta \Lambda^2 t$. If there are
D5-branes wrapping around $x=\pm \alpha$, we get $a=-\alpha^2$
and two Special vacua for  $SO(6)\to \widehat{SO(2)}\times U(2)$,
as found in the previous paragraph. 
If there is no D5-brane wrapping around
$x=\pm \alpha$, there are two choices: either $F_2(t)$ has
a factor $t$ so that $a=0$ or $F_2(t)$ is a square.
The former where $a=0$ gives a factor $t^5$, which is the one
discussed before. The latter is given by $P_2(t)=t^2+ \Lambda^4$,
which gives the one vacuum of $SO(6) \to \widehat{SO(6)}$
at the Special branch.

Finally we consider the case in which there is no $t^2$ factor
at all, so all D5-branes must wrap around $x=\pm \alpha$. To
achieve this, we must have
\bea
P_3(t)-2 \Lambda^2 t^2=(t+a)^2(t+b), \qquad
P_3(t)+2 \Lambda^2 t^2=(t+c)^2(t+d)~.
\nonu
\eea
This problem is the same problem of $U(3)$ with $N_f=4$ 
which has been solved in \cite{bfhn}
\bea
a=s_1+s_2,~~~c=s_1-s_2,~~~b=s_1-2\Lambda^2-2 s_2+{2 s_1 \Lambda^2
\over s_2},~~~d=s_1+2\Lambda^2+2 s_2+{2 s_1 \Lambda^2
\over s_2},
\nonu
\eea
with $s_2^2(s_2+\Lambda^2)=s_1^2 \Lambda^2$. Furthermore,
using $(t+b)(t+d)=(t-\alpha^2)^2+e$ we get
\bea
bd=-2\alpha^2  \to s_1 \left(1+{2\Lambda^2 \over s_2} \right)
=-\alpha^2.
\nonu
\eea  
For fixed $\alpha$, there are three solutions 
which have the classical limits
$(s_1,s_2)\to (-\alpha^2,0)$. These three vacua give
$SO(6)\to U(3)$.

It is worth remarking about the different behaviors. We
observe a smooth interpolation 
between $\widehat{SO(4)}\times U(1)$
and $\widehat{SO(2)}\times U(2)$ for $N_f=1$
at Special branch, but not for $N_f=2$. The reason for that
comes from  the power of $t$. The Special branch of $SO(2)$ always
has a factor $t^2$ in the curve. 
For $SO(4)$, it is $(2N_f-N+4)=2N_f$,
so only for the case with 
$N_f=1$  can we have a factor $t^2$. This also explains  why 
there is no smooth interpolation of $\widehat{SO(4)}\times U(1)$
or $\widehat{SO(2)}\times U(2)$ to $U(3)$ because there is
no $t$ factor for the curve of $U(3)$.

%%%%%%%%%%%%%%%%%%%%%%%%%%%%%%%%%%%%%
$\bullet$ { \bf $N_f=3$ }
%%%%%%%%%%%%%%%%%%%%%%%%%%%%%%%%%%%

%%%%%%%%%%%%%%%%%%%%%%%%%%%%%%%%%%%%%%
 {\bf  1. Non-degenerated case}
%%%%%%%%%%%%%%%%%%%%%%%%%%%%%%%%%%%%%%%

\noindent
Setting $t=x^2$, the curve of non-degenerated case should be
\bea
y^2=P^2_3(t)- 4\Lambda^2 t^5= t F_3(t) H^2_1(t).
\nonu
\eea
Writing $P_3(t)=t(t^2-s_1 t+ s_2)$,
$H_1(t)=t+a$, and $F_3(t)=t^3+b t^2+c t+d$, we found three
solutions. The first one has $(s_1,s_2,a,c,d)=({-b-4 \Lambda^2 
\over 2},0,0,{(b+4 \Lambda^2)^2 \over 4})$. It gives
 one vacuum of $SO(6)\to \widehat{SO(4)}\times U(1)$.
However, since $F_3(t)=t(t^2+bt+c)$, it is in fact 
at the Special branch. It is also noteworthy that
the power of $t$ is $4$, which is not equal to $(2N_f-N+4)=6$.
The reason is that  $SO(4)$ with three flavors is not asymptotically
free (recalling that $N_f<N-2$); thus, the power  stops at $N$.
In fact, we have a similar result that there is a factor
$t^2$ for $SO(2)$ with flavors.
The other two solutions have $d=0$ and the classical limits
$(-b, b^2/4, b/2,b^2/4,0)$, so give $SO(6)\to \widehat{SO(2)}\times
U(2)$. Again, since $d=0$, these two vacua are in fact 
at the Special branch. The counting number two comes from the
$U(2)$ factor. Finally, we can fix as follows:
$b=-2 \alpha^2-4\Lambda^2$.
The reason why
we did not find the Chebyshev branch in the non-degenerated
case is that the number of counting becomes
$(N-N_f-2)<0$ for both the $SO(2)$ and $SO(4)$ 
gauge groups.

%%%%%%%%%%%%%%%%%%%%%%%%%%%%%%%%%%%%%%%%%%%%%%%%%%%%%
{\bf  2. Degenerated case and Special branch}
%%%%%%%%%%%%%%%%%%%%%%%%%%%%%%%%%%%%%%%%%%%%%%%%%%%%%%%

\noindent
Let us now consider the Special branch or the degenerated
case with the following curve
\bea 
%\label{SO6-2-bary1}
y^2=P^2_3(t)-4 \Lambda^2 t^5=  F_2(t) H^2_2(t).
\nonu
\eea
We will discuss them case by case. First, if there is a $t^4$ factor in the
curve, we must take $P_3(t)=t^2(t-s)$, $H_2(t)=t^2$ and
$F_2(t)=(t-s)^2-4 \Lambda^2 t$. There are three possible values
for $s$. If $s=-\Lambda^2$, the curve is a complete square  and
gives the Special vacuum of $SO(6)\to \widehat{SO(6)}$. 
Another case is $s=\alpha^2$ which gives one Special vacuum
of $SO(6)\to \widehat{SO(4)}\times U(1)$. As we have remarked 
before, since $N_f=3>(N-2)=1$, the power of $t$ is $N=4$.
Finally, if $F_2(t)$ has just one factor $t$, or if $s=0$, it 
gives the one vacuum of  the Chebyshev branch
 $SO(6)\to \widehat{SO(6)}$.

Second, if there is only $t^2$ factor in the curve, we have
$P_3(t)=t(t^2-s_1 t+s_2)$, $H_2(t)=t(x+c)$. Cancelling the $t^2$
at both sides, we get the reduced curve
\bea
P^2_2(t)- 4 \Lambda^2 t^3= F_2(t) H^2_1(t).
\nonu
\eea
This is exactly the same curve (\ref{massless-SO4-1-D}) and we 
get two Special vacua of  $SO(6)\to \widehat{SO(2)}\times U(2)$
by keeping only those solutions with a factor $t^2$.

The last case is when  there is no factor $t$ in the curve and
all D5-branes wrap around the root $x=\pm \alpha$, so the gauge
group is broken to $SO(6)\to U(3)$. Using this fact, we know
immediately $F_2(t)=(t-\alpha^2)^2+d-4 \Lambda^2 t$. There are
three solutions
\footnote{It is very complicated to solve the 
factorization. However, notice that at the limit $\Lambda\to 0$,
there are $a\to \alpha^2, c\to -2\alpha^2, d\to \alpha^4$
and $b\to 0$. If we set $\alpha=1$
and $\Lambda=0.001$, numerically it is easy to see only three
solutions that satisfy this limit}.

The phase structure of $SO(6)$ with massless flavors has
been summarized in Table \ref{tableso6-massless}.

\begin{table}
\begin{center}
\begin{tabular}{|c|c|c|c|c|c|c|} \hline
 $N_f$  &  Group  & Branch &  Power of $t(=x^2)$ &  $U(1)$ & 
Number of vacua & Connection
 \\ \hline
 1 & $\widehat{SO(2)}
\times U(2)$ & $(S,0_{NB})$ & $t^2$ & 1 &  2  & A\\ \cline{2-7}
& $\widehat{SO(4)}
\times U(1)$ & $(C,0_{NB})$ &  $t^3$ & 1 &  1  & \\ \cline{3-7}
& & $(S,0_{NB})$ &  $t^2$ & 1 &  1  &  A\\ \cline{2-7}
& $\widehat{SO(6)}$ &  $(C)$ &  $t^3$ & 0 & 3 & \\ \cline{3-7}
&      & $(S)$ &  $t^0$ & 0 & 1 & \\ \cline{2-7}
& $U(3)$  &  $(0_{NB})$ & $t^0$ & 1 & 3 & 
\\ \hline 
2 & $\widehat{SO(2)}
\times U(2)$ & $(S,0_{NB})$ & $t^2$ & 1 &  2  & \\ \cline{2-7}
& $\widehat{SO(4)}
\times U(1)$ & $(S,0_{NB})$ &  $t^4$ & 1 &  1  &  \\ \cline{2-7}
& $\widehat{SO(6)}$ &  $(C)$ &  $t^5$ & 0 & 2 & \\ \cline{3-7}
&      & $(S)$ &  $t^2$ & 0 & 1 & \\ \cline{2-7}
& $U(3)$  &  $(0_{NB})$ & $t^0$ & 1 & 3 & 
\\ \hline 
 3 & $\widehat{SO(2)}
\times U(2)$ & $(S,0_{NB})$ & $t^2$ & 1 &  2  & \\ \cline{2-7}
& $\widehat{SO(4)}
\times U(1)$ & $(S,0_{NB})$ &  $t^4$ & 1 &  1  &  \\ \cline{2-7}
& $\widehat{SO(6)}$ &  $(C)$ &  $t^5$ & 0 & 1 & \\ \cline{3-7}
&      & $(S)$ &  $t^4$ & 0 & 1 & \\ \cline{2-7}
& $U(3)$  &  $(0_{NB})$ & $t^0$ & 1 & 3 & 
\\ \hline 
\end{tabular}
\end{center} 
\caption{ \sl Summary of the phase structures of $SO(6)$ gauge group
with massless flavors. It is worth comparing with the
the phase structures of $SO(6)$ with massive flavors.
Here again we use  capital letters in the last column
to indicate  the phases having smooth
transition. }
\label{tableso6-massless}
\end{table}

%%%%%%%%%%%%%%%%%%%%%%%%%%%%%%%%%%%%
\subsection{ $SO(7)$ case}
%%%%%%%%%%%%%%%%%%%%%%%%%%%%%%%%%%%%

The curve is
\bea
y^2=P^2_3(t)- 4\Lambda^{10-2N_f} t^{1+N_f}.
\nonu
\eea
For $N_f=0$, it has been discussed in the section 5 of massive
flavors. For $N_f\geq 1$, the curve can be written as
\bea
y^2=P^2_3(t)- 4\Lambda^{10-2N_f} t^2 t^{N_f-1}
\nonu
\eea
which is the same curve of $SO(6)$ with $(N_f-1)$ flavors.
Therfore, all the results of $SO(6)$ can be applied  here.

%%%%%%%%%%%%%%%%%%%%%%%%%%%%%%%%%%%
\subsection{The smooth interpolation}
%%%%%%%%%%%%%%%%%%%%%%%%%%%%%%%%%%%%

\indent

Now let us discuss in this subsection the general picture of 
smooth interpolations. The classical limit has three types:
$SO(N_c)\to \widehat{SO(N_0)}\times U(N_1)$,
$SO(N_c)\to U([N_c/2])$, and $SO(N_c)\to \widehat{SO(N_c)}$.
The first two have $U(1)$ left at IR while for the last one, 
there is no
$U(1)$ left at all. Because of this fact, only the first two types can have
smooth interpolations by realizing the presence of
the number of $U(1)$ left.

Now for the type  $SO(N_c)\to U([N_c/2])$, the curve 
has no factor $t$ in the factorization form, while for the type
$SO(N_c)\to \widehat{SO(N_0)}\times U(N_1)$, depending on the
Chebyshev branch or Special branch, it has a different power of $t$.

For the Chebyshev branch, the power of $t=x^2$ is always an 
{\sl odd} number and
mainly the function of $N_f$. Thus, the type 
$SO(N_c)\to \widehat{SO(N_0)}\times U(N_1)$ cannot be smoothly
connected to the type  $SO(N_c)\to U([N_c/2])$, but can have
smooth interpolations inside itself. For example, if
$SO(N_c)\to \widehat{SO(N_0)}\times U(N_1)$ and 
$SO(N_c)\to \widehat{SO(M_0)}\times U(M_1)$ with $N_f<N_0-2$ and
$N_f<M_0-2$ (so the Chebyshev branch exists for both 
$\widehat{SO(N_0)}$ and $\widehat{SO(M_0)}$),  the curves will
have the same power of  $t$ and can give the 
above two different classical
limits. To demonstrate the analysis, let us consider the $SO(8)$ with
one flavor.  The curve is
\bea
y^2=P^2_8(x)-4 x^6 \Lambda^{10}= x^2 F_6(x) H^2_4(x). 
\nonu
\eea 
The required $x^6$ factor forces us that $P_8(x)=x^4(x^4-s_1 x^2+s_2)$,
$H_4(x)=x^2(x^2+a)$. Writing $F_6(x)=x^6+b x^4+c x^2+d$, we 
have the following solutions
\bean
b & = & 2(a\mp {i \Lambda^5 \over a^{3/2}}), \qquad
c=a^2\mp  6 {i \Lambda^5 \over a^{1/2}}-{\Lambda^{10} \over a^3},
\qquad
   d=-4 {\Lambda^{10} \over a^2}, \\
s_1 & = & -2a\pm  {i \Lambda^5 \over a^{3/2}}, \qquad s_2=
a^2\mp 3 {i \Lambda^5 \over a^{1/2}}.
\nonu
\eean 
We can take the following limits: (1) $\Lambda\to 0$,
$a$ fixed, we get $SO(8)\to \widehat{SO(4)}\times U(2)$;
(2) $\Lambda\to 0$, $a\to 0$, but ${i \Lambda^5 \over a^{3/2}}\to
\beta\neq 0$, we get $SO(8)\to \widehat{SO(6)}\times U(1)$. We
have not shown that it is always true that 
$\widehat{SO(N_0)}\times U(N_1)$ is connected to 
 $\widehat{SO(M_0)}\times U(M_1)$ under the above conditions, but
we expect it to be true.

Now let us consider the Special branch in which the power of $t=x^2$
is always an {\sl even} number. 1) First, if the power is $zero$,
it is possible to connect $SO(N_c)\to \widehat{SO(N_0)}\times U(N_1)$
to $SO(N_c)\to U([N_c/2])$. Recalling that the power is
given by $(2N_f-N+4)/(2N_f-N+3)$ for $N$ is even/odd 
and $N_f<N-2$ and $N/(N-1)$ for $N_f\geq N-2$, 
we need $(2N_f-N+4)/(2N_f-N+3)=0$ or $N_f=(N-4)/2$ for
$N$ even and $N_f=(N-3)/2$ for $N$ odd. This has
been observed for $SO(5)$ with $N_f=0$ where there is, in fact,
 a smooth
interpolation in
$\widehat{SO(3)}\times U(1)\leftrightarrow U(2)$. 

2) Second, if
the power is not zero, we can only expect a smooth
interpolation inside the first type. Since the power is a function
of both $N_f$ and $N$, 
the symmetry breaking pattern
\bea
\widehat{SO(N_0)}\times U(N_1)
\nonu 
\eea 
is connected to  the following symmetry breaking pattern 
\bea
\widehat{SO(M_0)}\times U(M_1)
\nonu
\eea 
(assuming
$M_0<N_0$ and both are even numbers) only if $N_f\geq M_0-2$, 
$N_f<N_0-2$ and $(2N_f-N_0+4)=M_0$. One such  example
is $SO(6)$ with one flavor where $\widehat{SO(2)}\times U(2)$
is smoothly connected to $\widehat{SO(4)}\times U(1)$. In fact,
case 1) where the power of $t$ 
is zero and  case 2) where the power of $t$ is not zero
are related to each other by the addition map.

It is noteworthy that for the smooth transition in the
Special branch, the condition $(2N_f-N_0+4)=M_0$ is exactly
the condition of the Seiberg dual pair between $SO(N_0)$ and $SO(M_0)$.
Therefore, the smooth transition can be connected through the Seiberg
duality. However, for a smooth transition in the
Chebyshev branch, no such relationship exists and a smooth transition
cannot be understood from the Seiberg duality.

%%%%%%%%%%%%%%%%%%%%%%%%%%%%%%%%%%%%%%%%%%%%%%%%%%%%%%%%%%%%%%%%%%%%%%%%%%%%%
%\newpage
\vspace{1cm}
\centerline{\bf Acknowledgments}

BF wants to thank  Vijay Balasubramanian,
Freddy Cachazo, Oleg Lunin and Asad Naqvi for the discussions,
P.C. Argyres and K. Konishi for the helpful correspondences
and  the
High Energy Group at the University of Pennsylvania
for their generous hospitality. YO would like to 
thank Hiroaki Kanno for useful discussions.
This research of CA  was supported by Korea Research Foundation
Grant(KRF-2002-015-CS0006). This research of BF
is supported under the NSF grant PHY-0070928.

%%%%%%%%%%%%%%%%%%%%%%%%%%%%%%%%%%%%%%%%%%%%%%%%%%%%%%%%%%%%%%%%%%%%%%%%%%%%%
\appendix

\renewcommand{\thesection}{\large \bf \mbox{Appendix~}\Alph{section}}
\renewcommand{\theequation}{\Alph{section}\mbox{.}\arabic{equation}}

%%%%%%%%%%%%%%%%%%%%%%%%%%%%%%%%%%%%%%%%%%%%%%%%%%%%%%%%%%%%%%%%%%%%%%%%%%%%%
\section{\large \bf Strong gauge coupling approach: superpotential and 
a generalized Konishi anomaly equation for $SO(2N_c)$ case }
\setcounter{equation}{0}
%%%%%%%%%%%%%%%%%%%%%%%%%%%%%%%%%%%%%%%%%%%%%%%%%%%%%%%%%%%%%%%%%%%%%%%%%%%%%

In \cite{ao} the strong gauge coupling approach for $SO/USp$ pure gauge 
theories were studied generally. That analysis was an extension of 
\cite{eot}  to 
allow a more general superpotential 
in which the degree  can be arbitrary without any restrictions. 
We extend the analysis discussed in \cite{ao} to $SO(2N_c)$ gauge theory 
with $N_f$ flavors by using the method of \cite{aot,ahn98,Tera97}. 
%Although in this section for the simplicity we concentrate on the 
%massless case, the generalization to massive case is straightforward.  
Let us consider  superpotential regarded as a small perturbation of 
an ${\cal N}=2$ $SO(2N_c)$ gauge theory
\cite{ty,kty,as,aps,hanany,hms,dkp,aot,eot,feng1}
\begin{eqnarray}
W(\Phi)=\sum_{s=1}^{k+1}\frac{g_{2s}}{2s}\mbox{Tr}\Phi^{2s}\equiv
\sum_{s=1}^{k+1} g_{2s} u_{2s}, \qquad u_{2s} \equiv \frac{1}{2s}
\mbox{Tr} \Phi^{2s}
\label{treesup}
%\nonu
\end{eqnarray}
where $\Phi$ is an adjoint scalar chiral superfield and we denote
its eigenvalues by  $\pm   \phi_I(I=1,2,\cdots, N_c)$.
% that are purely
%imaginary values. 
The degree of the superpotential $W(\Phi)$ is
$2(k+1)$. Since $\Phi$ is an antisymmetric matrix we can transform
to the following simple form, 
%\footnote{
%In this case, we have $\mu_i=1$ for $i=1,2, \cdots, n$, $n=N_c-r$
%and $\mu_{n+1}=2r$ and we inert the imaginary $i$ 
%in the $\Phi$.}, 
\bea 
\Phi= \left( {0 \atop -1 }{ 1 \atop 0 } \right) \otimes
\mbox{diag} ( i \phi_1, i \phi_2, \cdots,  i \phi_{N_c-r}, 0, 0, 
\cdots, 0).
\label{Phi}
\eea
When we replace $\mbox{Tr} \Phi^{2j} $ with $\left\langle\mbox{Tr}
\Phi^{2j}\right\rangle$, the superpotential becomes an effective
superpotential. We introduce a classical $2N_c \times 2N_c$ matrix
$\Phi_{cl}$ such that $\left\langle\mbox{Tr}
\Phi^{2j}\right\rangle = \mbox{Tr} \Phi^{2j}_{cl}$ for
$j=1,2,\cdots, N_c$. Additionally, 
$u_{2j} \equiv \frac{1}{2j} \mbox{Tr}
\Phi^{2j}_{cl}$  are independent. However, for $2j > 2N_c$, both
$\mbox{Tr} \Phi^{2j} $ and $\left\langle\mbox{Tr}
\Phi^{2j}\right\rangle$ can be written as the $u_{2j}$ of $2j\leq 2N_c
$. Classical vacua can be obtained by putting all the
eigenvalues of $\Phi$ and $\Phi_{cl}$ equal to the roots of
$W^{\prime}(z)=\sum_{s=1}^{k+1} g_{2r}z^{2s-1}$. We will take the
degree of superpotential to be $2(k+1) \leq 2N_c$ first in which the
$u_{2j}$ are independent and $\left\langle\mbox{Tr}
\Phi^{2j}\right\rangle = \mbox{Tr} \Phi^{2j}_{cl}$. Then we will
take the degree of superpotential to be arbitrary. Until now we 
have reviewed the discussion given in \cite{ao}. Next we will study  some 
restriction that is an important idea in the discussion below.  

Let us consider the special $(N_c-r)$ dimensional submanifold of 
the Coulomb 
branch where some of the branching points of the moduli space collide. 
The index $r$ runs from $0$ to min$(N_c,N_f/2)$ \cite{aps, Tera97} 
and classifies the branch. On the $r$-th branch, the effective theory 
becomes $SO(2r)\times U(1)^{N_c-r}$ with $N_f$ massless flavors. Thus, 
after turning on the tree level superpotential since there exist 
$U(1)^n$ gauge groups with $2n \le 2k$, the remaining $(N_c-r-n)$ $U(1)$ 
factors are confined, which lead to $(N_c-r-n)$ massless monopoles or 
dyons. This occurs only at points where at $W^{\prime}=0$ the monopoles 
are massless on some particular submanifold $<u_{2r}>$. This can be 
done by including the $(N_c-r-n)$ monopole hypermultiplets in the 
superpotential. Then  the exact effective superpotential by adding 
(\ref{treesup}), near a point with $(N_c-r-n)$ massless monopoles, is given by
\bea 
W_{eff}=\sqrt{2} \sum_{l=1}^{N_c-n-r} M_{l}(u_{2s}) q_l
\widetilde{q}_l + \sum_{s=1}^{k+1} g_{2s} u_{2s}. 
\nonu 
\eea
By varying this with respect to $u_{2s}$,
we get an  equation of motion similar to  pure Yang-Mills theory
except that the extra terms on the left hand side since 
%on the non-baryonic root, 
the $u_{2s}$ with $2s > 2(N_c-r)$ are dependent on 
$u_{2s}$ with $2s \leq 2(N_c-r)$. 
%At a point in the nonbaryonic branch root, 
There exist $(N_c-n-r)$ equations 
for the $(k+1)$ parameters $g_{2s}$. 
Here $q_l$ and $\widetilde{q}_l$ are the monopole fields and 
$M_{l}(u_{2s})$ is the mass of $l$-th monopole as a function 
of the $u_{2s}$. The variation of $W_{eff}$ with respect to 
$q_l$ and $\widetilde{q}_l$ vanishes. However, variation 
of $W_{eff}$ with respect to $u_{2s}$ does not lead to the vanishing 
of $q_l\widetilde{q}_l$ and the mass of monopoles should vanish 
for $l=1,2, \cdots, (N_c-r-n)$ in a supersymmetric vacuum. Therefore, 
the superpotential in this supersymmetric vacuum becomes 
$W_{exact}=\sum_{r=1}^{k+1} g_{2s} <u_{2s}>$. The masses $M_i$ are
equal to the periods of some meromorphic one-form over some cycles
of the ${\cal N}=2$ hyperelliptic curve.

It is useful and convenient to consider a singular point in the 
moduli space where $(N_c-r-n)$ 
%mutually local 
monopoles are massless. 
Then the ${\cal N}=2$ curve of genus $(2N_c-2r-1)$ degenerates to a 
curve of genus $2n$ \cite{eot,feng1} and it is given by
\footnote{As we classified in previous section, there exist various 
curves (\ref{example-2}), (\ref{example-3}) and (\ref{example-4})
depending on the number of single and double roots. 
It is straightforward to
proceed to other cases so we restrict ourselves 
to here the particular case 
(\ref{curve}). }
\begin{eqnarray}
y^2 & = &
P_{2(N_c-r)}^2(x)-4\Lambda^{4(N_c-r)-2(N_f-2r+2)}x^{2(N_f-2r+2)}
\nonu \\
& = & x^2H_{2N_c-2r-2n-2}^2(x)F_{2(2n+1)}(x),
\label{curve}
\eea
where
\bea 
H_{2N_c-2r-2n-2}(x)=
%\prod_i^{2N_c-2n-2}(x-p_i)=
\prod^{N_c-n-r-1}_{i=1}(x^2-p_i^2), \quad F_{2(2n+1)}(x)
=\prod^{2n+1}_{i=1}(x^2-q_i^2).
\label{constraint}
\nonu
\eea
Here $H_{2N_c-2n-2r-2}(x)$ is a polynomial in $x$ of degree 
$(2N_c-2r-2n-2)$ that
gives $(2N_c-2r-2n-2)$ double roots and $F_{2(2n+1)}(x)$ is a 
polynomial in $x$ of degree $(4n+2)$ that is related to the 
deformed superpotential. Both functions are even functions 
in $x$. That is, a function of $x^2$ which is peculiar to the 
gauge group $SO(2N_c)$. Since we concentrate on the $(N_c-r)$ 
dimensional subspace of Coulomb phase the characteristic function 
$P_{2(N_c-r)}(x)$ is defined by $2(N_c-r)\times 2(N_c-r)$ matrix $\Phi_{cl}$ 
as follows:
\begin{eqnarray}
P_{2(N_c-r)}(x)&=&\det (x-\Phi_{cl})= \prod_{I=1}^{N_c-r}
(x^2-\phi^2_I). 
\label{P}
\end{eqnarray}
The degeneracy of the above curve can be checked by computing 
both $y^2$ and $\frac{\pa y^2}{\pa x^2}$ at the point $x=\pm p_i $
 and $x=0$ obtaining a zero. The factorization condition 
(\ref{curve}) can be described and encoded by Lagrange multipliers 
\cite{csw}. When the degree $(2k+1)$ of $W^{\prime}(x)$ is equal 
to $(2n+1)$, we obtain the result (\ref{newmatrix}).

$\bullet$ {\bf Field theory analysis for superpotential}

In \cite{bfhn} it was noted that when the degree of $W^{\prime}(x)$, 
$n$, was greater than or equal to
$(2N_c-N_f)$ the structure of the matrix model curve of 
$U(N_c)$ gauge theories was changed by the effect of the flavors. Thus, 
we first extended the discussion in \cite{bfhn} to $SO(2N_c)$ gauge theory 
and then later to the more general cases. 
The massless monopole constraint 
for $SO(2N_c)$ gauge theory with $N_f$ 
%massless 
flavors is described 
as follows:
\begin{eqnarray}
y^2&=&P_{2N_c}^2(x)-4\Lambda^{4N_c-4-2N_f}A(x)=x^2H_{2N_c-2n-2}^2(x)
F_{2(2n+1)}(x),  \nonu \\
&=&x^2\prod_{i=1}^{l-1} \left(x^2-p_i^2 \right)^2F_{2(2n+1)}(x),
\quad A(x)\equiv x^4 \prod_{j=1}^{N_f}(x^2-m_j^2)
\label{masslessSO}
\end{eqnarray}
where we used $l$ as the number of massless monopoles. From the 
equation (\ref{masslessSO}) we can find double roots at $x=0,\pm p_i 
\ i=1,\cdots, l-1$. We do not need take into account  
all points. Since $P_{2N_c}(x)$ and $A(x)\equiv x^4 \mbox{det}_{N_f}(x+m)$ 
are an even function of $x$, we have only to consider $x=0, x=+p_i$. For 
simplicity of equations we introduce $p_l=0$; thus, index $i$ runs from $1$ 
to $l$. With these conventions we have an effective superpotential with 
$l$ massless monopole constraints (\ref{masslessSO}),
\begin{eqnarray}
W_{{low}}&=&\sum_{t=1}^{k+1}g_{2t} u_{2t}  
+ \sum_{i=1}^{l} \left[L_i \left(P_{2N_c}(p_i)-2\epsilon_i \Lambda^
{2N_c-2-N_f}\sqrt{A(p_i)}\right) \right. \nonu \\
& + & \left. Q_i \frac{\partial }{\partial p_i}
\left( P_{2N_c}(p_i)-2\epsilon_i \Lambda^{2N_c-2-N_f}\sqrt{A(p_i)}  \right) 
\right] 
\nonumber
\end{eqnarray}
where $L_i$  and 
$Q_i$ are Lagrange multipliers and $\epsilon_i=\pm 1$. 
From the equation of motion for $p_i$ and $Q_i$ we obtain the following 
relations,
\begin{eqnarray}
Q_i=0, \qquad \frac{\partial }{\partial p_i}\left( P_{2N_c}(p_i)-
2\epsilon_i \Lambda^{2N_c-2-N_f}\sqrt{A(p_i)}\right)=0.
\nonu
\end{eqnarray}
The variation of $W_{low}$ with respect to $u_{2t}$ leads to 
\begin{eqnarray}
g_{2t}+\sum_{i=1}^{l}L_i\frac{\partial}{\partial u_{2t}}
\left(P_{2N_c}(p_i)-2\epsilon_i \Lambda^{2N_c-2-N_f}\sqrt{A(p_i)}\right)=0.
\nonu
\end{eqnarray}
Since $A(p_i)$ is independent of $u_{2t}$ the third term vanishes. 
By using $P_{2N_c}(p_i)=\sum_{j=0}^{N_c}s_{2j}p_i^{2N_c-2j}$ 
\footnote{The polynomial $P_{2N_c}(p_i)$ is given by 
 $\sum_{j=0}^{N_c}s_{2j}x^{2N_c-2j}=\prod_{I=1}^{N_c}
(x^2-\phi^2_I) $ where $s_{2j}$ and $u_{2j}$ are related each other by
so-called Newton's formula $ s_{2j} +\sum_{i=1}^j s_{2j-2i} u_{2i}=
0$ where $j=1,2, \cdots, N_c$ with $s_0=1$. From this recurrence 
relation we obtain $\frac{\pa s_{2j}}{\pa u_{2t}}=-s_{2j-2t}$ for $j \geq 
t$. }
we  obtain 
\begin{eqnarray}
g_{2t}=\sum_{i=1}^{l}L_i\frac{\partial}{\partial u_{2t}} P_{2N_c}(p_i)
=
\sum_{i=1}^{l}\sum_{j=0}^{N_c}L_ip_i^{2N_c-2j}s_{2j-2t}. 
\nonu
\end{eqnarray}
With this relation as in \cite{ookouchi} we can obtain the 
following relation
for $W^{\prime}(x)$.
\begin{eqnarray}
W^{\prime}(x)&=
&\sum_{i=1}^{l} \frac{xP_{2N_c}(x)}{x^2-p_i^2} L_i-x^{-1}
\sum_{i=1}^{l}2\epsilon_i L_i \Lambda^{2N_c-2-N_f}\sqrt{A(p_i)} +
{\cal O}(x^{-3}). 
\nonu
\end{eqnarray}
Defining a new polynomial $B_{2(l-1)}$ of order $2(l-1)$ 
as in \cite{civ}, 
\begin{eqnarray}
\sum_{i=1}^{l} \frac{L_i}{x^2-p_i^2} \equiv \frac{B_{2(l-1)}(x)}
{x^2H_{2(l-1)}(x)},
\nonu
\end{eqnarray}
the $W^{\prime}(x)$ is rewritten as 
\bea
W^{\prime}(x) =   \frac{xP_{2N_c}(x) B_{2(l-1)}(x)}
{x^2H_{2(l-1)}(x)} + {\cal O}(x^{-1}).
\nonu
\eea
Now we can compare them 
with both sides, in particular, for the power behavior of
$x$.  It is easy to see that 
the left hand side behaves like $(2n+1)$ while the right hand side behaves 
like
$2N_c-1-2(l-1)$ except the factor $B_{2(l-1)}(x)$ and therefore the condition
$l=(N_c-n)$ will give rise to the consistency and the polynomial 
$B_{2(l-1)}(x)$ becomes  constant.
By using this relation  and substituting the characteristic polynomial 
$P_{2N_c}(x)$ from the monopole constraints (\ref{masslessSO}),
we can obtain the following relation together with the replacement of
the polynomial $H_{2(l-1)}(x)$,
\begin{eqnarray}
F_{4n+2}(x)+\frac{4x^2\Lambda^{4N_c-2N_f-4}
\prod_{i=1}^{N_f}(x^2-m_i^2)}{\prod_{j=1}^{N_c-n-1}(x^2-p^2_j)^2}=
\frac{1}{g_{2n+2}^2}\left(W^{\prime}_{2n+1}(x)^2+
{\cal O}(x^{2n})\right). 
\label{newmatrix}
\end{eqnarray}
Thus, if $n>-N_f+2N_c-3$, the effect of flavor changes the geometry.
This is a new feature compared with  pure gauge theory with no 
flavors. This is due to the fact that the flavor-dependent part, $A(x)$
provides a new contribution.

The formula (\ref{newmatrix}) makes the assumption that
every root of $W'(x)$ has some D5-branes wrapping around it. If the
assumption
is not true, for example, for the breaking pattern 
$SO(N_c) \rightarrow \prod_{j=1}^n U(N_j)$,
instead of $SO(N_c) \rightarrow SO(N_0) \times \prod_{j=1}^n U(N_j)$,
the formula (\ref{newmatrix}) should be modified. The general form 
can be obtained as follows. Assuming a curve is factorized as
\bea 
\label{general-A(x)}
y^2=P^2_M(x)- A(x)= f_{2s}(x) H^2_{M-s}(x)
\eea
where $A(x)$ counts the contribution of flavors (massive or
massless) and $H_{M-s}(x)$ 
includes all the double roots (for example, the $x^2$ in [\ref{curve}]
should  take into account  $H_{M-s}[x]$), we will have 
\bea 
\label{general-relationship}
f_{2s}(x)+{A(x)\over H_{M-s}(x)^2}= W_s'(x)^2+{\cal O}(x^{s-1})
\eea
where $W'_s(x)=\prod_{j=1}^s (x-\lambda_j)$ and the $\lambda_j$'s are the
roots of $W'(x)$  with wrapping D5-branes. For example, for the
case considered in (\ref{newmatrix}), $\lambda_j=0, \pm \alpha_j$
so $W'_s(x)=W^{\prime}_{2n+1}(x)$. If among these roots
there is only the origin without
wrapping D5-branes, $W'_s(x)={W^{\prime}_{2n+1}(x) \over x}$, so
\bea
F_{4n}(x)+\frac{4x^4 \Lambda^{4N_c-2N_f-4}
\prod_{i=1}^{N_f}(x^2-m_i^2)}{\prod_{j=1}^{N_c-n}(x^2-p^2_j)^2}=
\frac{1}{g_{2n+2}^2}\left[ \left( \frac{W^{\prime}_{2n+1}(x)}{x} \right)^2+
{\cal O}(x^{2n-2})\right].
\label{condition4n}
\eea  
As soon as  $n>-N_f+2N_c-3$, 
the effect of flavor also changes the geometry. 

$\bullet$ {\bf Superpotential of degree $2(k+1)$ less than $2N$}

Now we generalize (\ref{newmatrix}) to $2n < 2k$ by introducing the
constraints (\ref{curve}). 
We follow the basic idea of \cite{csw,ao} and repeat
the derivations of (\ref{newmatrix}) and generalize to the arbitrary
degree of superpotential later. At first, in the range $2n+2 \le 2k+2\le
2N$, let us consider the superpotential  for 
$SO(2N_c)$ theory  with {\it massless} $N_f$ flavors
in the $r$-th branch 
%\footnote{We change the previous $r$ variable to make it
%even and so our $2r$ here corresponds to the previous $r$.} 
for simplicity (the massive flavors can be obtained by including the mass 
dependent parts appropriately. For massless case, as we remarked before
the $r$ is fixed to be some number for Chebyshev or Special branches.)
under these constraints (\ref{curve}) 
by starting from the curve in pure case and adding the
flavor dependent parts,
\begin{eqnarray}
W_{eff} & = & \sum_{s=1}^{k+1}g_{2s}u_{2s}  + \sum_{i=0}^{2N_c-2r-2n-2}
\left[L_i\oint \frac{P_{2(N_c-r)}(x)-2\epsilon_i x^{N_f-2r+2}
\Lambda^{2N_c-N_f-2}}{(x-p_i)}dx \right. \nonu \\
&+ & \left. B_i\oint \frac{P_{2(N_c-r)}(x)-2\epsilon_i x^{N_f-2r+2}
\Lambda^{2N_c-N_f-2}}{(x-p_i)^2}dx \right]
\label{eff}
\end{eqnarray}
where $L_i$ and $B_i$ are Lagrange multipliers imposing the
constraints and $\epsilon_i=\pm 1$. 
What we have done here is to include the $r$-dependence and $N_f$ dependence 
coming from the generalization of pure gauge theory \cite{ao}.   
For an equal massive case, 
we simply replace $ x^{N_f-2r+2}$ in the numerator with 
$ (x^2-m^2)^{N_f/2-r} x^2$.
The contour integration
encloses all $p_i$'s and the factor $1/2\pi i$ is absorbed in the
symbol of $\oint$ for simplicity. The $p_i$'s where $i=0,1,2,
\cdots, (2N_c-2r-2n-2)$ are the locations of the double roots of the curve 
$y^2=
P_{2(N_c-r)}^2(x)-4\Lambda^{4N_c-2N_f-4}x^{2N_f-4r+4}$ according to
the constraints (\ref{curve}) where the function $H_{2N_c-2r-2n-2}(x)$ has a 
factor $(x^2-p_i^2)$. The 
function $P_{2(N_c-r)}(x)$ depends on
$u_{2s}$. Note that the massless monopole points appear in pair
$(p_i,-p_i)$ where $i=1,2,\cdots, (N_c-r-n-1)$. Therefore, we denote half of
the $p_i$'s by $p_{N_c-r-n-1+i}=-p_i, i=1,2, \cdots, (N_c-r-n-1)$.
Moreover, we define $p_0=0$ and note that the summation index $i$ in the 
above starts from $i=0$. Since $P_{2(N_c-r)}(x)$ is an even function in $x$
and the property of $\epsilon_i$,
if the following constraints are satisfied at $x=p_i$ where $
i=1,2, \cdots, (N_c-r-n-1)$,
\begin{eqnarray}
\left(P_{2(N_c-r)}(x) -2x^{N_f-2r+2}\epsilon_i
\Lambda^{2N_c-N_f-2}\right)|_{x=p_i} & = & 0, \nonu \\
\frac{\partial }{\partial x}\left(P_{2(N_c-r)}(x)
-2x^{N_f-2r+2}\epsilon_i \Lambda^{2N_c-N_f-2} \right)|_{x=p_i} & = &
0, 
\label{condition}
\end{eqnarray}
then they also automatically are satisfied at $x=-p_i$. Then the
numbers of constraints that we should consider are $(N_c-r-n-1)$. Thus,
we denote the half of the Lagrange multipliers by
$L_{N_c-r-n-1+i}=L_i$ and $B_{N_c-r-n-1+i}=B_i$ where $i=1,2, \cdots,
(N_c-r-n-1)$. Due to the fact that the second derivative of $y$ 
with respect to $x$ at $x=p_i$
does not vanish, there are no  higher order terms such as
$(x-p_i)^{-a}, a=3,4,5, \cdots$ in the effective superpotential
(\ref{eff}). In other words, for given constraints (\ref{curve}), there
exist only a Lagrange multiplier $L_i$ with $(x-p_i)^{-1}$ and a Lagrange
multiplier $B_i$ with $(x-p_i)^{-2}$.

The variation of $W_{eff}$ with respect to $B_i$ leads to an integral 
expression which is the coefficient function of $B_i$ in (\ref{eff}) and  
by the
formula of contour integral, it is evaluated at $x=p_i$
and we have changed the derivative of $P_{2(N_c-r)}(x)$ with respect to
$x$ into $P_{2(N_c-r)}(x)$ and the trace of some quantity
\begin{eqnarray}
0&=&\oint \frac{P_{2(N_c-r)}(x) -2x^{N_f-2r+2} \epsilon_i
\Lambda^{2N_c-N_f-2}} {(x-p_i)^2}dx \nonu \\
& = & \left(P_{2(N_c-r)}(x) -2x^{N_f-2r+2}\epsilon_i
\Lambda^{2N_c-N_f-2}
\right)^{\prime} |_{x=p_i} \nonu \\
&=& \left(P_{2(N_c-r)}^{\prime}(x)-\frac{N_f-2r+2}{x}2\epsilon_i
x^{N_f-2r+2}
\Lambda^{2N_c-N_f-2} \right)|_{x=p_i}  \nonu \\
&=& \left(P_{2(N_c-r)}(x)\sum_{J=1}^{N_c-r}
\frac{2x}{x^2-\phi_J^2}-\frac{N_f-2r+2}{x}P_{2(N_c-r)}(x)\right)|_{x=p_i}
\nonu
\\
& = & P_{2(N_c-r)}(x) \left(\mbox{Tr}
\frac{1}{x-\Phi_{cl}}-\frac{N_f+2}{x} \right)|_{x=p_i}
\label{relation7}
\nonu
\end{eqnarray}
where we used the equation of motion for $L_i$
when we replace $2x^{N_f-2r+2}\epsilon_i \Lambda^{2N_c-N_f-2}$ 
with $P_{2(N_c-r)}(x)$ at $x=p_i$. \footnote{
The last equality
comes from the following relation, together with (\ref{Phi}),
$
\mbox{Tr} \frac{1}{x-\Phi_{cl}}=\sum_{k=0}^{\infty} x^{-k-1}
\mbox{Tr}\Phi_{cl}^k=\sum_{i=0}^{\infty}x^{-(2i+1)} \mbox{Tr}
\Phi^{2i}_{cl}
 = \sum_{i=0}^{\infty}x  \left(x^2 \right)^{-(i+1)}
\sum_{I=1}^{N_c} 2
\left(\phi_I^2\right)^i
= \left( \sum_{I=1}^{N_c-r} \frac{2x}{x^2-\phi_I^2} \right) + 
\frac{2r}{x} 
%\label{relation1}
$
where $\Phi_{cl}$ is antisymmetric matrix, and 
the odd power terms are
vanishing.
\label{note}}
The equation of motion for $B_i$ can be summarized as 
\footnote{
Through
the definition (\ref{P}),
the derivative $P_{2(N_c-r)}(x)$ with respect to $x$ is given by
$
P_{2(N_c-r)}^{\prime}(x)=
\left(\prod_{I}^{N_c-r}(x^2-\phi_I^2)\right)^{\prime}=2x\sum_
{J=1}^{N_c-r}
\prod_{I\ne J}^{N_c-r} (x^2-\phi^2_I)=P_{2(N_c-r)}(x) 
\sum_{J=1}^{N_c-r}
\frac{2x}{x^2-\phi_J^2}. 
%\nonu
$
Taking into account  the result of trace given in previous footnote 
\ref{note}, 
we can rewrite this result
as
$
\frac{P_{2(N_c-r)}^{\prime}(x)}{P_{2(N_c-r)}(x)}=
\mbox{Tr}\frac{1}{x-\Phi_{cl}}-\frac{2r}{x}$.
Therefore, we have
$
\frac{P_{2(N_c-r)}^{\prime}(x)}{P_{2(N_c-r)}(x)}=
\sum_{i=0}^{\infty}x^{-(2i+1)}
\mbox{Tr} \Phi^{2i}_{cl} =\sum_{i=0}^{\infty}x^{-(2i+1)} 2i
u_{2i}$. It is easy to check that for equal massive flavors, the 
corresponding relation leads to 
$
\left(\mbox{Tr} \frac{1}{x-\Phi_{cl}}-\frac{x N_f}{x^2-m^2}-\frac{2}{x}
\right)|_{x=p_i}=0, P_{2(N_c-r)}(x=p_0=0)=0, 
P_{2(N_c-r)}(x=p_i) \neq 0$.}
\bea 
\left(\mbox{Tr} \frac{1}{x-\Phi_{cl}}-\frac{N_f+2}{x}
\right)|_{x=p_i}=0,\quad P_{2(N_c-r)}(x=p_0=0)=0, \quad
P_{2(N_c-r)}(x=p_i) \neq 0 \nonu
\eea
where 
the characteristic function by solving the first order differential 
equation is given by 
\begin{eqnarray}
P_{2(N_c-r)}(x)= \left[x^{2(N_c-r)}\exp
\left(-\sum_{s=1}^{\infty}\frac{u_{2s}}{x^{2s}} \right)
\right]_{+}
\label{p2n}
\end{eqnarray}
and the polynomial part of a Laurent series inside the bracket 
is denoted here by
$+$. By putting the negative power of $x$ to zero, the 
$u_{2s}$ with $2s > 2(N_c-r)$ can be obtained the $u_{2s}$ with 
$2s \leq 2(N_c-r)$.

Next we consider the variation of $W_{eff}$ with respect to $p_i$,
\bea
0 = 2B_j \oint \frac{P_{2(N_c-r)}(x) -2x^{N_f-2r+2}\epsilon_i
\Lambda^{2N_c-N_f-2}}{(x-p_j)^3}dx
\nonu
\eea
where there is no $L_i$ term because 
we have used the equation of motion for $B_i$. In 
general since this integral does not vanish, we should 
have $B_j=0$ because the curve does not have more than cubic 
roots due to the fact that  $y^2$ 
contains a polynomial $H_{2N_c-2r-2n-2}^2(x)$ as we have discussed before. 

Let us consider 
variation of $W_{eff}$ with respect to $u_{2s}$, by 
using the relation 
$\frac{\pa P_{2(N_c-r)}(x)}{\pa u_{2s}} = 
-\left[ \frac{P_{2(N_c-r)(x)}}{x^{2s}} \right]_{+}$, which can be 
checked from the definition 
(\ref{p2n}) and remembering 
the fact that the 
function $P_{2(N_c-r)}(x)$ depends on
$u_{2s}$,
\begin{eqnarray}
0=g_{2s}-\sum_{i=0}^{2N_c-2r-2n-2}\oint \left[\frac{P_{2(N_c-r)}(x)}{x^{2s}}
\right]_{+} \frac{L_i}{x-p_i}dx,
\nonu
\end{eqnarray}
where we used $B_i=0$ at the level of equation of motion. 
Multiplying this with $z^{2s-1}$ and summing over $s$ where 
$z$ is inside the contour of integration,
we can obtain the first derivative $W^{\prime}(z)$,
\begin{eqnarray}
W^{\prime}(z)&=&\sum_{s=1}^{k+1}
g_{2s}z^{2s-1}=\sum_{i=0}^{2N_c-2r-2n-2}\oint
\sum_{s=1}^{k+1}z^{2s-1}\frac{P_{2(N_c-r)}(x)}{x^{2s}}
\frac{L_i}{(x-p_i)}dx.
\label{relation2}
\end{eqnarray}
Let us introduce a new polynomial $Q(x)$ defined as
\begin{eqnarray}
\sum_{i=0}^{2N_c-2r-2n-2}\frac{xL_i}{(x-p_i)}   
%L_0+
%\sum_{i=1}^{N_c-r-n-1}xL_i\left(\frac{1}{x-p_i}+\frac{1}{x+p_i}
%\right)
=L_0+\sum_{i=1}^{N_c-r-n-1}\frac{2x^2L_i}{x^2-p_i^2}
 \equiv  
\frac{Q(x)}{H_{2N_c-2r-2n-2}(x)}
\label{relation6}
\end{eqnarray}
where we used the fact that 
$L_{N_c-r-n-1+i}=L_i$ and $p_{N_c-r-n-1+i}=-p_i$ 
for $i=1,2, \cdots, (N_c-r-n-1)$.
By using this new function we can rewrite (\ref{relation2}) as
\begin{eqnarray}
W^{\prime}(z)=\oint \sum_{s=1}^{k+1}\frac{z^{2s-1}}{x^{2s}}
\frac{Q(x)P_{2(N_c-r)}(x)}{xH_{2N_c-2r-2n-2}(x)}dx.
\label{maruB}
\end{eqnarray}
Since $W^{\prime}(z)$ is a polynomial of degree $(2k+1)$, we found
the order of $Q(x)$ as $(2k-2n)$, so we denote it by
$Q_{2k-2n}(x)$. Thus, we have found the order of polynomial $Q(x)$
and therefore the order of integrand in (\ref{relation2}) is like
${\cal O}(x^{2k-2s+1})$. Thus, if $s\ge k+1$ it does not
contribute to the integral because the power of $x$ in this region
implies that the Laurent expansion around the origin vanishes. We
can replace the upper value of summation, $k+1$,  with the infinity and 
by computing the infinite sum over $s$ we get
\begin{eqnarray}
W^{\prime}(z)=\oint \sum_{s=1}^{\infty}\frac{z^{2s-1}}{x^{2s}}
\frac{Q_{2k-2n}(x)P_{2(N_c-r)}(x)}{xH_{2N_c-2r-2n-2}(x)}dx=\oint z
\frac{Q_{2k-2n}(x)P_{2(N_c-r)}(x)}{x(x^2-z^2)H_{2N_c-2r-2n-2}(x)}dx.
\label{prires}
\nonu
\end{eqnarray}
From (\ref{curve}) we have the relation,
\begin{eqnarray}
P_{2(N_c-r)}(x)=x\sqrt{F_{2(2n+1)}(x)}H_{2N_c-2r-2n-2}(x)+{\cal
O}(x^{-2N_c+2N_f-2r+4}).
\label{prires1}
\end{eqnarray}
Let us emphasize the presence of the second term which will play the 
role and {\it cannot} 
be ignored because the power of $x$ can be greater than or 
equal to $-1$  depending on the number of flavor $N_f$ which can vary 
in various regions. For pure cases, there were no 
contributions from the second terms. 
Therefore, we have
\begin{eqnarray}
W^{\prime}(z)=\oint z\frac{y_m(x)}{x^2-z^2}dx+\oint 
{\cal O}\left(x^{-4N_c+2N_f+2k+3} \right)dx,
\label{effNf}
\end{eqnarray}
where we have defined $y_m$ as follows,
\begin{eqnarray}
y_m^2(x)=F_{2(2n+1)}(x)Q_{2k-2n}^2(x)
\end{eqnarray}
corresponding to the matrix model curve. 
In (\ref{effNf}), the second term does contribute if the condition 
$-4N_c+2N_f+2k+3 \geq -1$ holds. 
Then we get an expected and generalized result: 
\begin{eqnarray}
y_m^2(x)&=&F_{2(2n+1)}(x)Q^2_{2k-2n}(x)= \left\{
				\begin{array}{ll}
				{W_{2k+1}^{\prime}}^2(x)+
{\cal O}\left(x^{2k} \right) & k\geq 2N_c-N_f-2 \\
				{W_{2k+1}^{\prime}}^2(x)+
{\cal O}\left(x^{4N_c-2N_f-4} \right) & k < 2N_c-N_f-2
				\end{array}
				\right. 
		\nonu	\\	&\equiv& 
{W_{2k+1}^{\prime}}^2(x)+f_{2M}(x),\qquad 2M= \mbox{max} (2k,4N_c-2N_f-4)
\label{matrixcurveform}
\end{eqnarray}
where both $F_{2(2n+1)}(x)$ and $Q_{2k-2n}(x)$ are functions of
$x^2$, then $f_{2M}(x)$ is  also a function of $x^2$. We put the
subscript $m$ in the $y_m$ in order to emphasize the fact that
this corresponds to the matrix model curve. When $2k=2n$, we
reproduce (\ref{newmatrix}) with $Q_0= g_{2n+2}$. 
The second term on the left hand side of (\ref{newmatrix}) behaves like as
$x^{2+2N_f-4(N_c-n-1)}= x^{2N_f-4N_c+4n+6}$. Depending on whether the power of
this is greater 
than or equal to $2n=2k$, the role of flavor is effective or not.  
When $n=k > -N_f +2N_c-3$, the flavor dependent part will contribute to
the 
$W^{\prime}(x)$. 
The above relation
determines a polynomial $F_{2(2n+1)}(x)$ in terms of $(2n+1)$
unknown parameters by assuming the leading coefficient of $W(x)$ to be
normalized by 1. These parameters
can be obtained from both $P_{2(N_c-r)}(x)$ and $H_{2N_c-2r-2n-2}(x)$ through
the factorization condition (\ref{curve})
\footnote{ One can proceed the case where there are no D5-branes
wrapping around the origin similarly. Our corresponding monopole constraint
equation (\ref{prires1}) becomes
$
P_{2(N_c-r)}(x)=\sqrt{F_{4n}(x)}H_{2N_c-2r-2n}(x)+{\cal
O}(x^{-2N_c+2N_f-2r+4})$. 
Using this relation, it is straightforward to 
arrive the following matrix model curve in this case: 
\begin{eqnarray}
y_m^2(x)&=&F_{4n}(x)Q^2_{2k-2n}(x)= \left\{
				\begin{array}{ll}
				\left(\frac{W_{2k+1}^{\prime}(x)}{x}\right)^2+
{\cal O}\left(x^{2k-2} \right) & k\geq 2N_c-N_f-2 \\
		\left(\frac{W_{2k+1}^{\prime}(x)}{x}\right)^2		+
{\cal O}\left(x^{4N_c-2N_f-6} \right) & k < 2N_c-N_f-2
				\end{array}
				\right. 
		\nonu	\\	&\equiv& 
\left(\frac{W_{2k+1}^{\prime}(x)}{x}\right)^2+
f_{2M}(x),\qquad 2M= \mbox{max} (2k-2,4N_c-2N_f-6).
\nonu
\end{eqnarray}}.

When $k$ is arbitrary large, 
we refer to the version one in the hep-th archive 
for details.

$\bullet$ {\bf A generalized Konishi anomaly}

%%%%%%%%%%%%%%%%%%%%%%%%%%%%%%%%%%%%%%%%%%%%%%%%%%%%%%%%%%%%%%%%%%%%%
%\subsection{CSW Appendix B for SO(2N_c) case}
%%%%%%%%%%%%%%%%%%%%%%%%%%%%%%%%%%%%%%%%%%%%%%%%%%%%%%%%%%%%%%%%%%%%%

Now we are ready to study the derivation of the generalized
Konishi anomaly equation based on the results of previous section.
As in \cite{csw}, we will restrict ourselves 
to the case with $\left\langle
\mbox{Tr} W^{\prime}(\Phi) \right\rangle=\mbox{Tr}
W^{\prime}(\Phi_{cl})$ and assume that the degree of
superpotential $(2k+2)$ is less than $2N_c$. By substituting
(\ref{relation6}) into (\ref{maruB}) we can write the derivative
of  superpotential $W^{\prime}(\phi_I)$
\begin{eqnarray}
W^{\prime}(\phi_I)=\sum_{i=0}^{2N_c-2r-2n-2} \oint
\phi_I\frac{P_{2(N_c-r)}(x)}{(x^2-\phi^2_I)} \frac{L_i}{(x-p_i)}dx
%\label{koni2}
\nonu
\end{eqnarray}
where we varied $W(\phi_I)$ with respect to $\phi_I$ rather than
$u_{2s}$ and used the result of $B_i=0$. Note that the characteristic 
function is given by  $P_{2(N_c-r)}(x)=
\prod_{I=1}^{N_c-r}(x^2-\phi^2_I)$. Using the above expression, 
one obtains the
following relation 
\footnote{When we change the summation index from $k$ to $i$, the only
odd terms appear because effectively the product of $\Phi_{cl}$
and $W^{\prime}(\Phi_{cl})$ does contribute only under  that
condition
$\mbox{Tr} \frac{W^{\prime}(\Phi_{cl})}{z-\Phi_{cl}}=\mbox{Tr}
\sum_{k=0}^{\infty}z^{-k-1}\Phi_{cl}^kW^{\prime} (\Phi_{cl})  =
\sum_{i=0}^{\infty}z^{-(2i+1)-1}2
\sum_{I=1}^{N_c-r} \phi_I^{2i+1}W^{\prime}(\phi_I) 
=2\sum_{I=1}^{N_c-r} \phi_I W^{\prime}(\phi_I)
\frac{1}{(z^2-\phi_I^2)}$.
The even terms do not contribute. Here $z$ is outside
the contour of integration.
We recognize that the
following factor can be written as, by simple manipulation between
the property of the trace we have seen before,
$
\frac{2\phi_I^2}{(z^2-\phi_I^2)(x^2-\phi_I^2)}=
\frac{1}{(x^2-z^2)}\left(z\mbox{Tr}\frac{1}{z-\Phi_{cl}}-x\mbox{Tr}
\frac{1}{x-\Phi_{cl}} \right)$.
},
\begin{eqnarray}
\mbox{Tr} \frac{W^{\prime}(\Phi_{cl})}{z-\Phi_{cl}} & = &
%\mbox{Tr}
%\sum_{k=0}^{\infty}z^{-k-1}\Phi_{cl}^kW^{\prime} (\Phi_{cl})  =
%\sum_{i=0}^{\infty}z^{-(2i+1)-1}2
%\sum_{I=1}^{N_c-r} \phi_I^{2i+1}W^{\prime}(\phi_I) \nonu \\
2\sum_{I=1}^{N_c-r} \phi_I W^{\prime}(\phi_I)\frac{1}{(z^2-\phi_I^2)}
\nonu \\
& = & 
\sum_{I=1}^{N_c-r}
\frac{2\phi_I^2}{(z^2-\phi_I^2)}\sum_{i=0}^{2N_c-2r-2n-2} \oint
\frac{P_{2(N_c-r)}(x)}{(x^2-\phi_I^2)}\frac{L_i}{(x-p_i)}dx \nonu \\
&= & \oint
\sum_{i=0}^{2N_c-2r-2n-2}\frac{P_{2(N_c-r)}(x)L_i}{(x^2-z^2)(x-p_i)}
\left(z\mbox{Tr}
\frac{1}{z-\Phi_{cl}}-x\mbox{Tr}\frac{1}{x-\Phi_{cl}} \right)dx.
\label{koni4}
\end{eqnarray}
As in the case of \cite{csw}, we can rewrite outside contour
integral in terms of two parts as follows:
\begin{eqnarray}
\oint_{z_{out}}=
\oint_{z_{in}}-\oint_{C_z+C_{-z}}
\label{contour}
\end{eqnarray}
where $C_z$ and $C_{-z}$ are the small contour around $z$ and $-z$
respectively. Thus, the first term in (\ref{koni4}) (corresponding to the
second term in $[B.3]$ of \cite{csw}) can be written as, by exploiting the
relation (\ref{relation6}) to write in terms of 
$Q_{2k-2n}(x)$ and $H_{2N_c-2r-2n-2}(x)$
\begin{eqnarray}
\left( \mbox{Tr}\frac{1}{z-\Phi_{cl}} \right)
\oint_{z_{out}}
\frac{zQ_{2k-2n}(x)P_{2(N_c-r)}(x)}{x H_{2N_c-2r-2n-2}(x)(x^2-z^2)}dx. 
\nonu
\eea
Let us emphasize
that in this case also we cannot drop the terms of order ${\cal
O}(x^{-2N_c+2N_f-2r+4})$ in the characteristic function $P_{2(N_c-r)}(x)$. 
In order to take into account of this,  we have to use the
above change of integration, then the first term of (\ref{koni4}) 
is given by
\bea
&&
\left( \mbox{Tr}\frac{1}{z-\Phi_{cl}}\right)
\left( \oint_{z_{in}}
\frac{zQ_{2k-2n}(x)P_{2(N_c-r)}(x)}{xH_{2N_c-2r-2n-2}(x)(x^2-z^2)}dx-
\oint_{C_{z}+C_{-z}}
\frac{zQ_{2k-2n}(x)P_{2(N_c-r)}(x)}{xH_{2N_c-2r-2n-2}(x)(x^2-z^2)}dx 
\right) \nonu \\
&& = \left( \mbox{Tr}\frac{1}{z-\Phi_{cl}} \right) 
\left(W^{\prime}(z)- \frac{y_m(z)
P_{2(N_c-r)}(z)}{\sqrt{P_{2(N_c-r)}^2(z)-4z^{2N_f-4r+4}\Lambda^{4N_c-2N_f-4}}} 
\right), 
\label{firstterm}
\end{eqnarray}
where the first term was obtained by the method done previously
and the second term was calculated at the poles and we used
\begin{eqnarray}
H_{2N_c-2r-2n-2}(z)=
\frac{\sqrt{P_{2(N_c-r)}^2(z)-4\Lambda^{4N_c-2N_f-4}z^{2N_f-4r+4}}}
{z\sqrt{F_{2(2n+1)}(z)}},  y_m^2(z) =
F_{2(2n+1)}(z) Q^2_{2k-2n}(z).
\nonu
\end{eqnarray}

The crucial difference between $U(N_c)$ case and $SO(2N_c)$ case comes
from the second term of (\ref{koni4}) (corresponding to the first
term in $(B.3)$ of \cite{csw}), which vanishes in $U(N_c)$ case. 
\footnote{ Let us write it as, after an integration over $x$ and 
using the (\ref{condition}) to change the trace part into 
$1/p_i$ term 
$
-
%\sum_{i=0}^{2N_c-2r-2n-2} 
\oint \frac{L_iP_{2(N_c-r)}(x)x}
{(x-p_i)(x^2-z^2)}\mbox{Tr}\frac{1}{x-\Phi_{cl}}dx
=-
%\sum_{i=0}^{2N_c-2r-2n-2} 
\frac{L_i p_i P_{2(N_c-r)}(x=p_i)}
{(p_i^2-z^2)}\mbox{Tr}\frac{1}{p_i-\Phi_{cl}}= 
-
%\sum_{i=0}^{2N_c-2r-2n-2}
(N_f+2)\frac{L_i P_{2(N_c-r)}(x=p_i)}{(p_i^2-z^2)}$. For equal massive
flavors, this becomes $-
\left(\frac{N_f}{1-\frac{m^2}{p_i^2}}+
2 \right)\frac{L_i P_{2(N_c-r)}(x=p_i)}{(p_i^2-z^2)}$. Due to the mass
dependent term containing $p_i$ dependent term, 
we can not simplify  the second term (\ref{koni4}) (as we go to the contour
integral around $x$ again, when
we consider massive flavors) further
like as
(\ref{secondterm}) for the massless case.
\label{foot}} 
Now we use the result of the equation of motion for $B_i$
(the equation just above (\ref{relation7})) 
in order to change the trace part and arrive at
the final contribution of the second term (\ref{koni4}) 
as follows: 
\bea
&&  -\sum_{i=0}^{2N_c-2r-2n-2}
(N_f+2)\frac{L_i P_{2(N_c-r)}(x=p_i)}{(p_i^2-z^2)} \nonu \\
&& =  -\sum_{i=0}^{2N_c-2r-2n-2}
\oint \frac{(N_f+2)P_{2(N_c-r)}(x)}{(x^2-z^2)}
\frac{L_i}{(x-p_i)}dx \nonu \\ 
&& =-(N_f+2)\frac{W^{\prime}(z)}{z}
+\frac{(N_f+2)}{z}
\frac{y_m(z) P_{2(N_c-r)}(z)}{\sqrt{P_{2(N_c-r)}^2(z)-4z^{2N_f-4r+4}
\Lambda^{4N_c-2N_f-4}}}. 
\label{secondterm}
\end{eqnarray}
%where once again we substituted (\ref{relation6}) into
%(\ref{prires}). 
We used here again the property of contour integration (\ref{contour}).
Therefore, we obtain the (\ref{koni4}) by combining the two contributions
(\ref{firstterm}) and (\ref{secondterm})
\bea 
\mbox{Tr}
\frac{W^{\prime}(\Phi_{cl})}{z-\Phi_{cl}}
&=& \left( \mbox{Tr}\frac{1}{z-\Phi_{cl}} \right)
\left(W^{\prime}(z)- \frac{y_m(z)
P_{2(N_c-r)}(z)}{\sqrt{P_{2(N_c-r)}^2(z)-4z^{2N_f-4r+4}\Lambda^{4N_c-2N_f-4}}}
\right)\nonu \\
&&-(N_f+2)\frac{W^{\prime}(z)}{z}+\frac{(N_f+2)}{z}
\frac{y_m(z) P_{2(N_c-r)}(z)}{\sqrt{P_{2(N_c-r)}^2(z)-4z^{2N_f-4r+4}
\Lambda^{4N_c-2N_f-4}}}. 
\nonu 
\eea
Then the second term of above expression can be written
as \footnote{
One can multiply $z^{2r}$ both in the numerator and denominator
and then the orders of the characteristic polynomial are changed into $2N_c$. 
In other words, $z^{2r} P_{2(N_c-r)}(z)=P_{2N_c}(z)$. Remember that 
$ \mbox{Tr}\frac{1}{z-\Phi_{cl}}=
\frac{\left(z^{-{N_f+2}} P_{2N_c}(z) \right)^{\prime}}
{z^{-{N_f+2}} P_{2N_c}(z)}
+\frac{(N_f+2)}{z}$ and the quantum mechanical expression 
$
\left\langle \mbox{Tr}\frac{1}{x-\Phi} \right\rangle=\frac{d}{dx}
\log \left(P_{2(N_c-r)}(x)+\sqrt{P_{2(N_c-r)}^2(x)-4x^{2N_f-4r+4}
\Lambda^{4N_c-2N_f-4}} \right)$.} 
\bea 
 -\left( \mbox{Tr}  \frac{1}{z-\Phi_{cl}} \right) 
\!\!\!\!\!\!\!\!\!\!\!\!\!\! 
&& \frac{y_m(z) P_{2(N_c-r)}(z)}{\sqrt{P_{2(N_c-r)}^2(z)-4z^{2N_f-4r+4}
\Lambda^{4N_c-2N_f-4}}}\nonu \\
&=&  -\left( \mbox{Tr}  \frac{1}{z-\Phi_{cl}} \right) 
%\!\!\!\!\!\!\!\!\!\!\!\!\!\!
\frac{y_m(z) P_{2N_c}(z)}{\sqrt{P_{2N_c}^2(z)-4z^{2N_f+4}
\Lambda^{4N_c-2N_f-4}}}\nonu \\
 & = & - \left(\frac{\left(\frac{P_{2N_c}(z)}{z^{N_f+2}}\right)^{\prime}}
{\left(\frac{P_{2N_c}(z)}{z^{N_f+2}}\right)}
+\frac{N_f+2}{z}\right)
\frac{y_m(z) P_{2N_c}(z)}{\sqrt{P_{2N_c}^2(z)-4z^{2N_f+4}\Lambda^
{4N_c-2N_f-4}}} \nonu \\
& = & -\left( \left\langle \mbox{Tr}\frac{1}{z-\Phi}
\right\rangle - \frac{\left(N_f+2 \right) }{z} \right) y_m(z) \nonu \\
& - &
\frac{(N_f+2)}{z}
\frac{y_m(z) P_{2N_c}(z)}{\sqrt{P_{2N_c}^2(z)-4z^{2N_f+4}
\Lambda^{4N_c-2N_f-4}}}.
%\frac{P_{2N_c}^{\prime}(z)}{\sqrt{P_{2N_c}^2(z)-4z^4\Lambda^{4N_c-4} }}
\nonu
 \eea
Therefore, we obtain (\ref{koni4})
as
\bea
\mbox{Tr}
\frac{W^{\prime}(\Phi_{cl})- W^{\prime}(z)}{z-\Phi_{cl}}  = 
-(N_f+2)\frac{\left(W^{\prime}(z)-y_m(z)\right)}{z} 
-  \left\langle \mbox{Tr}\frac{1}{z-\Phi}
\right\rangle y_m(z). 
%-\frac{(N_f+2)}{z}
%\frac{y_m(z) P_{2N_c}(z)}{\sqrt{P_{2N_c}^2(z)-4z^{2N_f+4}
%\Lambda^{4N_c-2N_f-4}}}. 
\nonu
\eea
Taking into account  the relation
\begin{eqnarray}
\tr \frac{W^{\prime}(\Phi_{cl})-W^{\prime}(z)}{z- \Phi_ {cl}}=
\left\langle \mbox{Tr}
\frac{W^{\prime}(\Phi)-W^{\prime}(z)}{z-\Phi} \right \rangle 
\nonu
\end{eqnarray}
we can write  the quantum mechanical expression as follows:
\begin{eqnarray}
 \left\langle \tr \frac{W^{\prime}(\Phi)}{z-\Phi} \right\rangle &
=& 
\left\langle \mbox{Tr} \frac{1}{z-\Phi} \right \rangle W^{\prime}(z) 
+\tr \frac{W^{\prime}(\Phi_{cl})-W^{\prime}(z)}{z- \Phi_ {cl}}
\nonu \\
&=&
\left( \left\langle \tr \frac{1}{z-\Phi} \right \rangle
-\frac{\left(N_f+ 2\right)}{z} 
\right) \left(W^{\prime}(z)-y_m(z)\right) 
%\nonu
%\\
%& - & %2\frac{W^{\prime}(z)}{z} +
% \frac{(N_f+2) }{z}
%\frac{y_m(z) P_{2N_c}(z)}{\sqrt{P_{2N_c}^2(z)-4z^{2N_f+4}
%\Lambda^{4N_c-2N_f-4}}}  
%\nonu
\label{Konequation}
\end{eqnarray}
which is the generalized Konishi anomaly equation for the $SO(2N_c)$
case. The resolvent of the matrix model $R(z)$ is related to
$W^{\prime}(z)-y_m(z)$.
When $N_f=0$, this result reproduces the result in \cite{ao} exactly.
It does not depend on the $r$ with which we start from. 
For massive flavors, one can generalize the above description similarly, but 
as we mentioned in the footnote \ref{foot}, we would get rather involved 
expressions. 

For $SO(2N+1)$ case, 
we refer to the version one in the hep-th archive 
for details.

%%%%%%%%%%%%%%%%%%%%%%%%%%%%%%%%%%%%%%%%%%%%%%%%%%%%%%%%%%%%%%%%%%%%%%%%%%%%%
\section{\large \bf Addition and multiplication maps  }
\setcounter{equation}{0}
%%%%%%%%%%%%%%%%%%%%%%%%%%%%%%%%%%%%%%%%%%%%%%%%%%%%%%%%%%%%%%%%%%%%%%%%%%%%%

\indent

In \cite{csw} it was noted that all the confining vacua of
higher rank gauge groups can be  
constructed from the Coulomb vacua with lower rank  gauge groups by using
the Chebyshev polynomial through the multiplication map. 
Under this map, which was called multiplication
map, the vacua with the classical gauge group $\prod_{i=1}^n U(N_i)$ for a 
given superpotential are mapped to 
the vacua with the gauge group $\prod_{i=1}^n U(KN_i)$ for the {\it same} 
superpotential where $K$ is a multiplication index. 
This multiplication map was extended 
to the $SO/USp$ gauge theories in \cite{ao} and $U(N_c)$ gauge theory
with flavors in \cite{bfhn}. In addition to this multiplication map, 
another map (called the addition map) was introduced in \cite{bfhn}. The 
addition map reduces many analyses to  simpler cases. Thus,
we also introduce these two maps, the addition map and 
the multiplication map for the $SO(N_c)$ gauge theory with flavors.

%%%%%%%%%%%%%%%%%%%%%%%%%%%
$\bullet$ {\bf Addition map}
%%%%%%%%%%%%%%%%%%%%%%%%%

Unlike the unitary gauge group, for $SO(N_c)$ with flavors, there
are two kinds of addition maps: one is for massive flavors and the other
 is for massless flavors. They should be treated separately.
As we will show shortly, the addition
map with massive flavors can only connect the $SO(N_c)$ and
$SO(M_c)$ for both $N_c$ and $M_c$ are an even numbers 
(or they are odd at the same time), 
but the addition map
with massless flavors can connect the $SO(2N_c)$ to 
$SO(2N_c+1)$.

Let us start with  {\it massive} flavors first by using the 
$SO(2N_c)$ gauge group
as an example.  On assumption that we have two 
theories, $SO(2N_c)$ with $N_f$ flavors in the $r$-th branch and 
$SO(2N_c^{\prime})$ with $N_f^{\prime}$ flavors in the $r^{\prime}$-th 
branch. On these branches, the SW curves are described by 
\begin{eqnarray}
y^2_1&=&(x^2-m^2)^{2r}\left[P^2_{2(N_c-r)}(x)-4x^4\Lambda^{4N_c-2N_f-4}
(x^2-m^2)^{N_f-2r} \right], \nonumber \\
y^2_2&=&(x^2-m^2)^{2r^{\prime}}\left[P^2_{2(N_c^{\prime}-
r^{\prime})}(x)-4x^4\Lambda^{4N_c^{\prime}-2N_f^{\prime}-4}
(x^2-m^2)^{N_f^{\prime}-2r^{\prime}} \right].
\nonumber
\end{eqnarray}
For the two curves, if we have the following relations, we 
find that the polynomials in the bracket are identical.
\begin{eqnarray}
2(N_c-r)=2(N_c^{\prime}-r^{\prime}),\qquad 4N_c-2N_f=4N_c^{\prime}-
2N_f^{\prime},\qquad N_f-2r=N_f^{\prime}-2r^{\prime}.
\label{solu}
\end{eqnarray}
These relations are exactly the same as the ones for the $U(N_c)$ case. 
We expect that certain vacua of $SO(2N_c)$ gauge theory with $N_f$
flavors in the $r$-th branch are related to the one of 
 $SO(2N_c^{\prime})$ gauge theory with $N_f^{\prime}$
flavors in the $r^{\prime}$-th branch.
For the $SO(2N_c+1)$ case we have the same relation, because the only 
difference comes from the factor of $x^4\Lambda^{-4}$ which does 
not matter for the relation. They are a common factor in both SW curves.
Namely, we have only to replace $x^4
\Lambda^{4N_c-2N_f-4}$ with $x^2\Lambda^{4N_c-2N_f-2}$. 
It is also noteworthy that starting with $SO(N_c)$ for
$N_c$ even (or odd), we can only get $SO(M_c)$ for $M_c$ even 
(or odd).

Thus, if on the $r$-th branch, the $SO(N_c)$ theory with $N_f$ 
massive flavors has a classical limit $SO(N_0)\times U(N_1)$ with 
the {\it effective} $N_f$ massless flavors charged under 
the $U(N_1)$ factor in the quartic superpotential, 
then a $SO(N_c+2d)$ theory  with $(N_f+2d)$ massive flavors will 
have a classical limit $SO(N_0)\times U(N_1+d)$ with the {\it effective}
$(N_f+2d)$ massless flavors charged under $U(N_1+d)$ 
on the $(r+d)$-th branch, according to (\ref{solu}). 
Notice that for massive flavors, the counting
of vacua of the $U(N_1)$ factor is $2N_1-N_f=2(N_1+d)-(N_f+2d)$,
which is invariant under the above addition map, so it is consistent. 
In fact, for massive flavors, it is the factor $U(N_i)$ with
the {\it effective} massless flavors so we get the 
same results as in \cite{bfhn}.

Now let us discuss the addition map of {\it massless} flavors. 
For
$SO(2N_c+1)$ with $N_f$ massless flavors, the curve is
characterized by
\bea 
%label{add-SO-1}
y^2=P^2_{2N_c}(x)-4\Lambda^{4N_c-2N_f-2} x^2 (x^2)^{N_f}
\nonu
\eea
which can be rewritten as
\bea 
\label{add-SO-2}
y^2=P^2_{2N_c}(x)-4\Lambda^{4N_c-2(N_f-1)-4} x^4 (x^2)^{N_f-1}.
\eea
However, the form of (\ref{add-SO-2}) can be interpreted as
the curve of $SO(2N_c)$ with $(N_f-1)$ massless flavors. In
other words, $SO(2N_c+1)$ gauge group with $N_f$ massless
flavors is equivalent to the $SO(2N_c)$ gauge group with
$(N_f-1)$ massless flavors. It is obvious that the above
interpretation can only be done for  massless flavors
where $SO(2N_c)$ is related to $SO(2N_c+1)$. We call it the
``special addition maps''. In fact, this special 
addition map comes from a well known result
in the field theory: giving one flavor in
$SO(2N+1)$ non-zero vacuum expectation value,
it can be higgsed down to the $SO(2N_c)$ gauge group.

It seems that
except the reinterpretation of $SO(2N_c+1)$ curve to
$SO(2N_c)$ curve, we should have a similar addition map of
massive flavors (we will call it the ``general addition 
map'') by just setting $m=0$. However, we need
to be careful in applying the general addition map. 
As we 
did in the weak and strong coupling analyses, for the $SO(N_c)$
gauge group with massless flavors, there is no concept of
$r$-th branch and what we have is the Chebyshev branch or
the Special branch where the power of $x$ is fixed for a given
$SO(N_c)$ factor. The general addition map is used to 
relate the Chebyshev branches (or the Special branches) of
two different gauge groups. 
%More details can be found
%in later section and 
We will give one explicit example to
demonstrate the general idea. For $SO(2N_c)$ with $2M$ massless
flavors, the Chebyshev branch requires a factorization of the curve
to be
\bea
y^2=(x^2)^{2M+2} \left
[P^2_{2N_c-(2M+2)}(x)-4 \Lambda^{4N_c-2(2M-2)} \right].
\nonu
\eea  
Similarly, for  $SO(2N_c')$ with $2M'$ massless flavors, the curve 
at the Chebyshev branch is
\bea
y^2=(x^2)^{2M'+2} \left
[P^2_{2N_c'-(2M'+2)}(x)-4 \Lambda^{4N_c'-2(2M'-2)} \right].
\nonu
\eea
Therefore, if
\bea
N_c-(M+1)=N_c'-(M'+1),
\nonu 
\eea
we can reduce the Chebyshev branch of $SO(2N_c')$ with $2M'$ massless
flavors to the Chebyshev branch of $SO(2N_c)$ with $2M$ massless
flavors. Notice that there is no  index $r$ in the above relationship.

%%%%%%%%%%%%%%%%%%%%%%%%%%%%%%%%%%%
$\bullet$ {\bf Multiplication map}
%%%%%%%%%%%%%%%%%%%%%%%%%%%%%%%%%%%%%%

Now we will discuss  the multiplication map. The method
we used is from the bottom to the top, i.e., starting from the 
known factorization of gauge group with lower rank to the
unknown factorization of gauge group with higher rank. 
First let us assume that the factorization of curve is given by
\bea 
\label{SO-general}
y^2= P^2_{2N_c}(x)- 4x^{2s} \Lambda_0^{4N_c-2s-2N_f} A(x)=
f_{2p}(x) H^2_{2N_c-p}(x)
\nonu
\eea
where $s=2$ for the $SO(2N_c)$ group and $s=1$ for $SO(2N_c+1)$ group
and $A(x)$ the contribution of flavors (it can be very general with
different masses). Next let us divide out the common factor
between $P^2_{2N_c}(x)$ and $x^{2s} A(x)$. For example, if
one of the eigenvalues of $\Phi$ in the $SO(2N_c)$ gauge theory is zero, 
we can divide out the $x^4$ factor
at the two sides of the curve. Or if $A(x)$ has factor $(x^2-m^2)^2$
and we are at the $r=1$ branch, the factor $(x^2-m^2)^2$ can also be divided
out. After dividing out all the common factors, the remaining
curve is given by
\bea 
\label{reduced}
P^2_{M}(x)-4 \Lambda_0^{4N_c-2s-2N_f} \widetilde{A}(x)=
\widetilde{f}_{2\widetilde{p}}(x) H^2_{M-\widetilde{p}}(x) 
\nonu
\eea
where $\widetilde{p}$ is some number.
Now let us define
\bea 
\label{widetilde-x}
\widetilde{x}={P_{M}(x) \over 2\eta \Lambda_0^{2N_c-s-N_f}
\sqrt{\widetilde{A}(x)}}
%\nonu
\eea
and
\bea 
\label{P_KM}
P_{KM}(x)=2 \left(\eta \Lambda_0^{2N_c-s-N_f}\sqrt{\widetilde{A}(x)} 
\right)^K
{\cal T}_{K}(\widetilde{x})
%\nonu
\eea
with $\eta^{2K}=1$. 
It is worth to notice that although $\widetilde{x}$ has
$\sqrt{\widetilde{A}(x)}$ in the denominator, the $P_{KM}(x)$ is 
a perfect polynomial of $x$. The reason is very simple:
${\cal T}_{K}(t)$ has only an even (odd) power of $t$ in the
polynomial if $K$ is an even (odd) number, so 
$\sqrt{\widetilde{A}(x)}^{K-q}=\widetilde{A}(x)^{{K-q\over 2}}$ is
a polynomial of $x$ because $(K-q)$ is always an even number
\footnote{
We want to thank Freddy Cachazo and Oleg Lunin for discussing
this general situation of the multiplication map.}. Using
(\ref{widetilde-x}) and (\ref{P_KM}) it is easy to see
that one obtains
\bean
&& P^2_{KM}(x)-4 \Lambda^{(4N_c-2s-2N_f)K} \widetilde{A}(x)^K \nonu \\
&& =  4 \Lambda^{(4N_c-2s-2N_f)K} \widetilde{A}(x)^K
\left[ \left(\frac{P_{KM}(x)}{ 2 \left(\eta 
\Lambda_0^{2N_c-s-N_f}\sqrt{\widetilde{A}(x)} \right)^K}
\right)^2-1 \right] \nonu \\
&& =  4 \Lambda^{(4N_c-2s-2N_f)K} 
\widetilde{A}(x)^K \left[{\cal T}^2_{K}(\widetilde{x})-1 \right] \nonu \\
&& =  \left[2 \left(\eta \Lambda_0^{2N_c-s-N_f}
\sqrt{\widetilde{A}(x)}\right)^K
{\cal U}_{K-1}(\widetilde{x}) \right]^2 \left( \widetilde{x}^2-1 
\right) \nonu  \\
&& =  \left[\left(\eta \Lambda_0^{2N_c-s-N_f}\sqrt{\widetilde{A}(x)}
\right)^{K-1}
{\cal U}_{K-1}(\widetilde{x}) \right]^2 
\left[P^2_{M}(x)-4 \Lambda_0^{4N_c-2s-2N_f} 
\widetilde{A}(x) \right] \nonu \\
&& =  \widetilde{f}_{2\widetilde{p}}(x) \left[ H_{M-\widetilde{p}}(x)
\left(\eta \Lambda_0^{2N_c-s-N_f}\sqrt{\widetilde{A}(x)}\right)^{K-1}
{\cal U}_{K-1}(\widetilde{x}) \right]^2. \nonu 
\eean
Notice that a similar reason guarantees that
$\sqrt{\widetilde{A}(x)}~^{K-1}{\cal U}_{K-1}(\widetilde{x})$
is a polynomial of $x$. The above calculation shows that we have
the solution of factorization of $P_{KM}(x)$ with matter
dependent part $\widetilde{A}(x)^K$. Finally to recover to the 
$SO(M_c)$ gauge
group, we need to multiply back some factor of $x^p$ to make
sure that $(KM+p/2)$ is even and $x^p \widetilde{A}(x)^K$ 
has $x^2$ or $x^4$ factor.

To demonstrate the above general results, let us consider 
several concrete
examples:
%\begin{itemize}

1). The first example is that we consider the 
breaking pattern $SO(2N_c)\to \prod_{j=1}^n U(N_j)$
at the $r=0$ branch, so the $P_{2N_c}(x)$ does {\it not} have a common
factor with $x^4 A(x)=x^4 (x^2-m^2)^{N_f}$. \footnote{For simplicity,
we have assumed that all $N_f$ flavors have the same mass.
Generalization to different masses is obvious.} 
Putting $M=2N_c$
and $\widetilde{A}(x)=x^4 (x^2-m^2)^{N_f}$, we get
$ \widetilde{A}(x)^K=(x^2-m^2)^{KN_f} x^{4K}$. This means that
the multiplication map lifts the original  solutions to $SO(2KN_c)$ with
$KN_f$ {\it massive} flavors and $2(K-1)$ {\it massless} flavors (or
$SO(2KN_c+1)$ with $KN_f$ {\it massive} flavors and 
$2(K-1)+1$ {\it massless} flavors). It is somewhat 
surprising that we get the {\it massless} flavors 
after lifting. The reason is that the original breaking
$SO(2N_c)\to \prod_{j=1}^n U(N_j)$ can be effectively considered
as $U(N_c)\to \prod_{j=1}^n U(N_j)$ with $N_f$ massive flavors plus
two massless flavors which come from a factor $x^4=(x^2)^2$ by looking at
the SW curve. 
From this new point of view, we  have {\it massless} flavors
at the beginning.

2). For the breaking $SO(2N_c)\to SO(2N_0) \times \prod_{j=1}^n 
U(N_j)$,
we can divide out at least the factor $x^4$ from the SW curve
because the characteristic
function $P_{2N_c}(x)$ and $4x^{4} \Lambda_0^{4N_c-4-2N_f} A(x)$
possess a factor $x^2$, so $M=(2N_c-2)$
and $\widetilde{A}(x)=(x^2-m^2)^{N_f}$. The multiplication 
map will give the factorization of $SO(K(2N_c-2)+2)$ with
$KN_f$ flavors. It is significant that the $+2$
of the rank $K(2N_c-2)+2$ comes from putting back the $x^4$ factor
to $P^2_{KM}(x)-4 \Lambda^{(4N_c-2s-2N_f)K} \widetilde{A}(x)^K$.
Since we can not multiply a factor $x^2$ back ($(KM+p/2)$ is not even), 
it is impossible to
obtain the map from $SO(2N_c)$ to $SO(2N_c+1)$ which 
reveals the same phenomena observed in \cite{ao} for the pure
$SO(2N_c)$ gauge theory. 
However, if there are massless flavors, one gets from $SO(2N_c)$
to $SO(2M_c+1)$ through the special addition map.
The multiplication map has another application.  We have seen that 
the relation between 
$ f_{2p}(x)$ and $W^{\prime}(x)$ is changed in certain situation 
under the presence of the flavors. For $SO(2N_c)$ gauge theory 
with $N_f$ flavors at the $n$-th branch, 
when $n > -N_f +2N_c-3$,  the relation between 
$ f_{2p}(x)$ and $W^{\prime}(x)$ is modified. On the other hand, 
for  $SO(2KN_c-2K+2)$ gauge theory with $K N_f$ flavors at $n$-th
branch, this happens when 
$ n > -KN_f +(2KN_c-2K+2)-3$. Since we are considering the particular 
case $N_f < 2N_c-2$,
for asymptotic free theory,
the relation  between 
$ f_{2p}(x)$ and $W^{\prime}(x)$ will be different for $SO(2N_c)$ gauge and 
$SO(2KN_c-2K+2)$ gauge theories. If $K$ is large enough, the presence of
flavors will not modify the relationship between 
$ f_{2p}(x)$ and $W^{\prime}(x)$ and the geometric proofs 
\cite{cv,feng1,ookouchi} can go 
through \cite{bfhn}. In other words, the geometric picture is really for
a large $N_c$ limit.

3).  To break $SO(2N_c+1)\to SO(2N_0+1) \times \prod_{j=1}^n
 U(N_j)$,
we can divide out at most the factor $x^2$ if all masses are not
zero, so $M=(2N_c-1)$ and $\widetilde{A}(x)=(x^2-m^2)^{N_f}$.
The multiplication map will give us $KM=K(2N_c-1)$. 
Notice that when we multiply back $x^p$ we require 
$(KM+p/2)$ is even. If $K$ is even,
we can only multiply back $x^4$ factor so it is
$SO(K(2N_c-1)+2)$ with $KN_f$ flavors. If $K$ is odd,
we can only multiply back $x^2$ factor so it is
$SO(K(2N_c-1)+1)$ with $KN_f$ flavors. Notice that
this conclusion is made under the assumption that there are
no massless flavors at the beginning. If there are
massless flavors, by the ``special addition maps'',
we can jump from $SO(2N_c)$ gauge theory to $SO(2N_c+1)$ gauge theory.

%\end{itemize}

\bibliographystyle{JHEP}

\end{document}